\documentclass[12pt]{article}
\usepackage[dvips]{graphicx}
\usepackage{amssymb}
\usepackage{amsmath}
\usepackage{epsfig}
\usepackage{cite}
\usepackage{hyperref}
\usepackage{breakurl}
\usepackage{bbold}
\usepackage{multirow}
\usepackage{tabu}

\numberwithin{equation}{section}
\numberwithin{table}{section}

\def\g{{\mathfrak{g}}}
\def\f{{\mathfrak{f}}}
\def\k{{\mathfrak{k}}}
\def\h{{\mathfrak{h}}}
\def\u{{\mathfrak{u}}}
\def\x{{\mathfrak{x}}}
\def\a{{\mathfrak{a}}}
\def\U{{\mathrm{U}}}
\def\su{{\mathfrak{su}}}
\def\SU{{\mathrm{SU}}}
\def\so{{\mathfrak{so}}}
\def\SO{{\mathrm{SO}}}

\def\SL{{\mathrm{SL}}}
\def\usp{{\mathfrak{usp}}}
\def\USp{{\mathrm{USp}}}

\def\E{{\mathrm{E}}}

\DeclareRobustCommand{\SkipTocEntry}[4]{}
\RequirePackage{color}
\newcommand{\cO}{\mathcal{O}}
\newcommand{\cT}{\mathcal{T}}

\newcommand{\cP}{\mathcal{P}}

\newcommand{\cD}{\mathcal{D}}
\newcommand{\cL}{\mathcal{L}}

\newcommand{\cK}{\mathcal{K}}
\newcommand{\cM}{\mathcal{M}}
\newcommand{\cN}{\mathcal{N}}

\newcommand{\cH}{\mathcal{H}}

\newcommand{\cF}{\mathcal{F}}

\newcommand{\cR}{\mathcal{R}}
\newcommand{\cV}{\mathcal{V}}
\newcommand{\cQ}{\mathcal{Q}}
\newcommand{\cX}{\mathcal{X}}

\newcommand{\dd}{\mathrm{d}}

\newcommand{\cc}{\mathrm{C}}

\newcommand{\id}{{\mathbb1}}    
\def\AdS{\textrm{AdS}}

\newcommand{\al}[1]{{\alpha_{#1}}}
\newcommand{\hal}[1]{{\hat\alpha_{#1}}}
\newcommand{\tal}[1]{{\tilde\alpha_{#1}}}
\newcommand{\be}[1]{{\beta_{#1}}}
\newcommand{\hbe}[1]{{\hat\beta_{#1}}}
\newcommand{\tbe}[1]{{\tilde\beta_{#1}}}
\newcommand{\ga}[1]{{\gamma_{#1}}}
\newcommand{\hga}[1]{{\hat\gamma_{#1}}}
\newcommand{\tga}[1]{{\tilde\gamma_{#1}}}

\newcommand{\hde}[1]{{\hat\delta_{#1}}}

\DeclareMathOperator{\tr}{tr}

\begin{document}

\begin{titlepage}
\begin{center}
\rightline{\small ZMP-HH/17-27}
\rightline{\small CPHT-RR054.102017}
\rightline{\small EMPG-17-20}

\vskip 0.4cm

{\Large \bf Maximally Supersymmetric AdS Solutions \\[0.2cm] and their Moduli Spaces}
\vskip 0.7cm

{\bf  Severin L\"ust,$^{a}$ 
Philipp R\"uter$^{b}$ 
and Jan Louis$^{c,d}$}

\vskip 0.5cm

$^{a}${\em Centre de Physique Th\'eorique, \'Ecole Polytechnique, CNRS,
91128 Palaiseau, France}
\vskip 0.2cm

$^{b}${\em Maxwell Institute for Mathematical Sciences and Department of Mathematics, \\ Heriot--Watt University, % \\ Colin Maclaurin Building, Riccarton, 
Edinburgh EH14 4AS, U.K.}
\vskip 0.2cm

$^{c}${\em Fachbereich Physik der Universit\"at Hamburg, \\
Luruper Chaussee 149, 22761 Hamburg, Germany}
\vskip 0.2cm

$^{d}${\em Zentrum f\"ur Mathematische Physik,
Universit\"at Hamburg, \\
Bundesstrasse 55, 20146 Hamburg, Germany}

\vskip 0.3cm

{\tt severin.luest@polytechnique.edu, pr26@hw.ac.uk, jan.louis@uni-hamburg.de}

\end{center}

\vskip 0.6cm

\begin{center} {\bf ABSTRACT } \end{center}

\noindent

We study maximally supersymmetric AdS$_D$ solutions of gauged supergravities in dimensions $D \geq 4$.
We show that such solutions can only exist if the gauge group after spontaneous symmetry breaking is a product of two reductive groups \(H_R \times H_\mathrm{mat}\), where \(H_R\) is uniquely determined by the 
%number of space-time dimensions and supersymmetries
dimension \(D\) and the number of supersymmetries \(\cN\) 
while \(H_\mathrm{mat}\) is unconstrained.
This resembles the structure of the global symmetry groups of the holographically dual SCFTs, where \(H_R\) is interpreted as the R-symmetry and \(H_\mathrm{mat}\) as the flavor symmetry.
Moreover, we discuss possible supersymmetry preserving continuous deformations,
%The moduli spaces spanned by these deformations 
which correspond to the conformal manifolds of the dual SCFTs.
Under the assumption that the scalar manifold of the supergravity is a symmetric 
%homogeneous 
space we derive general group theoretical conditions on these moduli.
Using these results we determine the AdS solutions of all gauged supergravities with more than 16 real supercharges.
We find that almost all of them do not have supersymmetry preserving deformations
with the only exception being the maximal supergravity in five dimensions with a moduli space given by \(\SU(1,1)/\U(1)\).
Furthermore, we determine the AdS solutions of four-dimensional \(\cN = 3\) supergravities and show that they similarly do not admit supersymmetric moduli.

%\vfill

%\today

%September 2017

\end{titlepage}

%%%%%%%%%%%%%%%%%%%%%%%%%%%%%%%%%%%%%%%%%%%%%%%%%%%%

\tableofcontents

%%%%%%%%%%%%%%%%%%%%%%%%%%%%%%%%%%%%%%%%%%%%%%%%%%%%

%%%%%%%%%%%%%%%%%%%%%%%%%%%%%%%%%%%%%%%%%%%%%%%%%%%%
\section{Introduction}

Anti-de Sitter backgrounds of supergravity theories play an important role in the context of the AdS/CFT correspondence \cite{Maldacena:1997re} and therefore have been intensely studied in the past.
In particular, supersymmetric solutions of the form \(\AdS_D \times Y_{11/10-D}\) of ten or eleven dimensional supergravity have attracted lots of attention.
%\Snote{Add refs here?}
An alternative strategy, however, is to study AdS solutions directly in \(D\)-dimensional gauged supergravity without referring to an explicit higher-dimensional origin.
This approach allows us to derive constraints on the existence and on the geometry of the moduli space of an AdS background.
The moduli space considered here is the manifold which is spanned by the supersymmetry preserving deformations of the supergravity solution and is the holographically dual object to the conformal manifold of the boundary superconformal field theory (SCFT).
This analysis has been explicitly performed for four-dimensional \(\cN = 1,2\) and \(4\) supergravities in \cite{deAlwis:2013jaa, Louis:2014gxa},
for five-dimensional \(\cN=2\) and \(\cN = 4\) theories in \cite{Tachikawa:2005tq, Louis:2016qca, Louis:2015dca},
for six-dimensional \(\cN = (1,1)\) theories \cite{Karndumri:2016ruc} and for half-maximal seven-dimensional supergravity in \cite{Louis:2015mka}.
Moreover, the moduli spaces of AdS solutions of various maximal supergravities have been discussed in \cite{Rueter}.

In this paper we aim for a systematic analysis of AdS$_D$ solutions with unbroken supersymmetry of all gauged supergravity theories in  \(D \geq 4\) space-time dimensions.
Similarly to \cite{Louis:2016tnz}, where a complete classification of all background space-times of maximally supersymmetric solutions of gauged supergravities was given, we perform most parts of the analysis in a general framework independent of the number of dimensions and supercharges.

A maximally supersymmetric solution is characterized by the vanishing of all supersymmetry variations of the fermions.
These variations are generically scalar field dependent and 
therefore such solutions can only exist at specific points or submanifolds of the scalar field space.
% where the fermionic shift matrices take specific values \Snote{More explicit?}.
%The shift matrices in turn
The scalar field dependent part of the fermion variations (i.e.~the fermionic shift matrices)
 are parametrized by the Killing vectors and moment maps of the gauge group \(G^g\).
Consequently, a maximally supersymmetric AdS solution is not possible for arbitrary gaugings.
The most characteristic feature of the gauge group in an AdS background is that it always contains a subgroup \(H^g_R\) which is solely generated by the vector fields in the gravitational multiplet, i.e.~the graviphotons.
\(H^g_R\) is uniquely determined to be the maximal subgroup of the R-symmetry group \(H_R\) such that it can be gauged by the graviphotons and such that the gravitino mass matrix is invariant with respect to \(H^g_R\).
Furthermore, in the vacuum the gauge group \(G^g\) must be spontaneously broken to a subgroup 
\begin{equation}\label{eq:AdSgaugegroup}
H^g = H^g_R \times H^g_\mathrm{mat} \,,
\end{equation}
where the second factor \(H^g_\mathrm{mat}\) is unconstrained by the conditions on the shift matrices but can only be gauged by vector multiplets.
Moreover, \(H^g_R\) and \(H^g_\mathrm{mat}\) are products of abelian and compact semi-simple Lie groups.

This resembles the structure of the holographically dual SCFTs.
A gauge symmetry of the AdS background translates via the AdS/CFT dictionary \cite{Gubser:1998bc, Witten:1998qj} to a global symmetry of the boundary CFT.\footnote{If we denote the conserved current of a global symmetry of the boundary CFT by \(J\) it couples via \(\int_{\partial AdS} A \wedge \ast J\) to the gauge field \(A\) of a local symmetry in the bulk.}
The first factor \(H^g_R\) in \eqref{eq:AdSgaugegroup} corresponds to the R-symmetry of the SCFT.
As a subgroup of the full superconformal group the R-symmetry must always be present and cannot be chosen freely.\footnote{Note, however, that there are SCFTs without an R-symmetry, as for example three-dimensional \(\cN=1\) theories. In this case also the gauge group factor \(H^g_R\) of the dual supergravity solution is trivial.}
Moreover, many SCFTs display additional global symmetries which commute with the R-symmetry.
They are called flavor symmetries and correspond to the second factor \(H^g_\mathrm{mat}\).

In the subsequent part of this paper we 
%turn to our main object of interest and discuss 
focus on the supersymmetric deformations of AdS solutions, i.e.~their moduli spaces.
A necessary condition for a scalar field to be a supersymmetric modulus is that the first order variations of the fermionic shift matrices with respect to this scalar field vanish.
This implies that the scalar field is massless.
However, consistency requires that there is one massless scalar field per spontaneously broken gauge group generator.
We show
% from the generic structure of gauged supergravities 
that generically there is 
%indeed
one such massless field (the Goldstone boson) for each broken generator of the gauge group \(G^g\).
%These fields are Goldstone bosons and constitute the additional degrees of freedom of those gauge fields which obtain a mass during spontaneous symmetry breaking.
These can therefore not be counted as candidates for (supersymmetric) moduli.

To obtain concrete results we focus on theories where the scalar field space is a symmetric space \(\cM = G/H\) for a Lie group \(G\) and its maximally compact subgroup \(H\).
This is necessarily the case for all supergravity theories with more than 8 real supercharges, e.g.~for half-maximal and higher supersymmetric theories.
Moreover, if there are more than sixteen supersymmetries, the field content of the theory is completely fixed and there is no freedom left in the choice of \(G\) and \(H\).
However, even for theories with 8 and less supercharges there exist various examples of symmetric spaces which are admissible as scalar manifolds.
Restricting to symmetric scalar geometries allows us to discuss the moduli spaces in a group theoretical and algebraic language.
We obtain explicit constraints on the existence of supersymmetric moduli and derive conditions under which the moduli space is symmetric again.
An essential prerequisite for this analysis is the knowledge of the gauged R-symmetry group \(H^g_R\) in \eqref{eq:AdSgaugegroup}. 

As a first important class of examples we apply these results to all theories with more than 16 supercharges which allow for maximally supersymmetric AdS$_D$ vacua.
These theories are characterized by the absence of any other supermultiplets than the gravity multiplet.
As a consequence, the entire gauge group must be given by \(H^g_R\) and is uniquely determined.
Supersymmetric moduli must necessarily be uncharged (i.e.~singlets) with respect to \(H^g_R\), so they can be found (or excluded) by a simple group theoretical decomposition.
This allows us to show that almost all of these solutions do not admit supersymmetric moduli,
in accordance with the recent classification of marginal deformations of SCFTs \cite{Cordova:2016xhm}.
The only exception occurs for maximal gauged supergravity in five dimensions.
Here the moduli space is given by \(\SU(1,1)/\U(1)\) and has a well-known holographic interpretation as the complex gauge coupling of four-dimensional \(\cN = 4\) super Yang-Mills theory.

For less supersymmetric theories further complications can arise.
In particular, if there are vector multiplets the gauge group \(G^g\) can be larger than just \(H^g_R\) and also non-compact.
However, as long as the scalar field space is symmetric our general results are still applicable.
This is for example the case for all half-maximal supergravities with \(q = 16\) real supercharges.
An explicit analysis of their AdS solutions can be found in \cite{Louis:2014gxa, Louis:2015dca, Karndumri:2016ruc, Louis:2015mka}.
Of similar type are the gauged \(\cN = 3\) supergravities in four dimensions with \(q = 12\) real supercharges.
We pick these theories as a second example and discuss their maximally supersymmetric AdS solutions.
As most of the cases with more than 16 real supercharges also these solutions do not admit for supersymmetric moduli.
%As a second example we discuss their AdS solutions and show that they do not admit for supersymmetric moduli.

%As a second example we discuss the AdS solutions of gauged \(\cN = 3\) supergravities in four dimensions.
%Due to the possible existence of vector multiplets their analysis is slightly more complicated.
%As most of the cases with more than 16 real supercharges also these solutions do not admit for supersymmetric moduli.

This paper is organized as follows:
In section~\ref{sec:prelim} we review basic concepts of gauged supergravities and set the stage for our analysis.
%In particular, our objective is to develop a unifying notation.
In section~\ref{sec:ads} we study maximally supersymmetric AdS solutions. 
We develop general properties of their gauge groups and discuss how to determine the moduli space for theories where the scalar manifold is a symmetric space.
%In sections~\ref{sec:examples} and \ref{sec:N=3} we show how to apply our previous results to specific examples.
In section~\ref{sec:examples} we compute the moduli spaces of AdS solutions for all theories with more than 16 real supercharges.
In section~\ref{sec:N=3} we discuss the AdS solutions of \(\cN = 3\) gauged supergravity in four dimensions.
In appendix~\ref{app:conventions} we summarize our notations and conventions.
In appendix~\ref{app:susy} we collect the general form of the supergravity Lagrangian and the supersymmetry transformation laws of the involved fields. Furthermore, we compute explicit expressions for the Killing vectors and their moment maps in terms of the fermionic shift matrices.
In appendix~\ref{app:representationtheory} we discuss properties of the gauged R-symmetry group.
In appendix~\ref{app:symP} and appendix~\ref{app:Tmoduli} we provide technical details needed in section~\ref{sec:adsmoduli}.

\section{Preliminaries}\label{sec:prelim}

In this section we discuss some basic concepts and properties of (gauged) supergravity theories.
We try to be as general as possible and do not focus on a specific space-time dimension or number of supercharges.
The main purpose of this section is to set the stage for the analysis in the subsequent sections and to introduce a unifying notation which allows us to discuss all cases more or less simultaneously, avoiding a cumbersome case-by-case analysis.%
\footnote{For a review of gauged supergravities see e.g.~\cite{Weidner:2006rp, Samtleben:2008pe, Trigiante:2016mnt}.
For a more detailed discussion of the geometrical structures underlying supergravities see e.g.~\cite{Cecotti:2015wqa}.}

\subsection{Supergravity}
A supergravity theory in \(D\) space-time dimensions always contains a gravitational multiplet.
The generic field content of this multiplet includes the metric \(g_{MN}\)  (\(M,N = 0, \dots, D-1\)), \(\cN\) gravitini \(\psi^i_M\) (\(i = 1, \dots, \cN\)), a set of \((p-1)\)-form fields or gauge potentials \(A^{(p-1)}\), a set of spin-\(\frac12\) fermions \(\chi^{\hat a}\) as well as a set of scalar fields \(\phi\).
Note that not all of these component fields
necessarily have to be
part of a given gravitational multiplet but we 
gave the most general situation.
Moreover, the theory might be coupled to additional multiplets, for example vector, tensor or matter multiplets.
If they are present, these multiplets always contain some spin-$\frac12$ fermions which we collectively call \(\chi^{\tilde a}\).
On the bosonic side they can have additional \((p-1)\)-form fields \(A^{(p-1)}\) among their components, as well as scalar fields which we universally call $\phi$.

We denote all form-fields from the gravitational multiplet as well as those from the other multiplets collectively by \(A^{I_p}\), where the index \(I_p\) labels all fields of the same rank $(p-1)$.
The reason for this is that there often exist duality transformations which mix fields from different multiplets and make it therefore impossible to distinguish from which multiplet a certain bosonic field originates.
%originally came from.
Moreover, we need to introduce the corresponding field strengths \(F^{I_p}\) which are differential forms of rank \(p\).
In some situations it will prove convenient to consider also the scalar fields \(\phi\) as $0$-form fields, so we often denote them by \(A^{I_1}\), and their field strengths by \(F^{I_1}\).

We collectively denote all spin-$\frac12$ fermions as \(\chi^a\), but we often want to distinguish the fermions which are part of the gravity multiplet from all the other fermions by calling the former \(\chi^{\hat a}\) and the latter \(\chi^{\tilde a}\).
This is possible because there is no symmetry or duality relating fermions from different types of multiplets.
The fermions \(\psi^i_M\) and \(\chi^a\) can always be arranged in representations of a group \(H\),
\begin{equation}\label{eq:H}
H = H_R \times H_\mathrm{mat} \,,
\end{equation}
where \(H_R\) is the R-symmetry group.
%i.e.~the automorphism group of the supersymmetry algebra,
% and \(H_\mathrm{mat}\) is a compact group which -- loosely speaking -- rotates multiplets of the same kind into each other.
Notice that all fields from the gravitational multiplet (i.e.~the gravitini \(\psi^i_M\) and the \(\chi^{\hat a}\)) are necessarily inert under \(H_\mathrm{mat}\) transformations, they can only transform non-trivially under \(H_R\).

Using these ingredients the general bosonic Lagrangian takes a relatively simple form and reads
\begin{equation}\label{eq:bosonicaction}
e^{-1} \cL_B = -\frac{R}{2} 
%+ \frac{e}{2} g_{rs}(\phi) D\phi^r \wedge \ast D\phi^s
- \frac{1}{2} \sum_{p \geq 1} M^{(p)}_{I_{p} J_{p}}\!\left(\phi\right)\, F^{I_p} \wedge \ast F^{J_p} + e^{-1} \cL_\mathrm{top} \,.
\end{equation}
The last part \(\cL_\mathrm{top}\) does not depend on the space-time metric and is therefore topological, a common example for such a term is a Chern-Simons term.
It is not necessarily part of every supergravity theory.
The matrices \(M^{(p)}_{I_pJ_p}(\phi)\) depend generically on all scalar fields and have to be symmetric and positive definite.
Therefore, they can be diagonalized by introducing vielbeins \(\cV^{\alpha_p}_{I_p}\), i.e.
\begin{equation}\label{eq:kinmatrix}
M^{(p)}_{I_pJ_p} = \delta_{\alpha_p\beta_p} \cV^{\alpha_p}_{I_p}  \cV^{\beta_p}_{J_p}  \,.
\end{equation}
Of course the vielbeins \(\cV^{\alpha_p}_{I_p}\) are scalar dependent as well.
We can use them to convert between the indices \(I_p\) and \(\alpha_p\).
It is convenient to introduce the abbreviations
\begin{equation}
F^{\alpha_p} = F^{I_{p}} \cV^{\alpha_p}_{I_p} \,.
\end{equation}
The benefit of working in this frame is that it allows us to couple the bosonic fields to the fermions, which is crucial for supergravity.
In fact the \(F^{\alpha_p}\) now transform under the same group \(H\) as the fermions but possibly in different representations.
Moreover, the invariance of the theory with respect to such \(H\)-transformations requires that \(\delta_{\alpha_p\beta_p}\) is \(H\)-invariant.
This means that if \({J_\al{p}}^\be{p}\) denotes an element of the Lie algebra \(\h\) of \(H\) in the respective matrix representation it needs to satisfy
\begin{equation}\label{eq:Hdelta}
{J_{(\al{p}}}^\ga{p} \delta_{\be{p})\ga{p}} = 0 \,.
\end{equation}

Later on it will be important to distinguish which of the form fields enter the supersymmetry variations of the gravitini.
For this purpose we go one step further and split the indices \(\alpha_p\) according to
\begin{equation}\label{indexsplit}
\alpha_p = \left(\hat \alpha_p, \tilde \alpha_p\right) \,,
\end{equation}
in the same way as we split the index \(a = \left(\hat a, \tilde a\right)\) labelling the spin-$\frac12$ fermions. 
We then denote by \(F^{\hat\alpha_p}\) 
the field strengths in the gravitational multiplet 
(e.g.\ the graviphotons for \(p = 2\)) and by  \(F^{\tilde\alpha_p}\)
the field strengths which arise in all other multiplets that might be present.
%Equivalently
Also \(F^{\hat\alpha_p}\) do not transform under \(H_\mathrm{mat}\) but only non-trivially under the R-symmetry \(H_R\). 
Note that this split depends on the scalar fields via the vielbeins $\cV$
and thus is background dependent.

In the general bosonic Lagrangian \eqref{eq:bosonicaction} we have written the kinetic term of the scalar fields on equal footing with all other %$p$-
form fields.
However, the scalar field sector is of particular relevance for the construction of supergravities, it is therefore appropriate to introduce a separate notation for its description.
Therefore, we often denote the scalar fields by \(\phi^r\) instead of \(A^{I_1}\) and their kinetic matrix by \(g_{rs}(\phi)\) instead of \(M^{(1)}_{I_1 J_1}(\phi)\).
Moreover, their field strengths \(F^{I_1}\) are given by the derivatives \(\dd \phi^r\), so their kinetic term can be expressed as
\begin{equation}\label{eq:sigmamodel}
\cL_\mathrm{kin,scal} = -\frac{e}{2} g_{rs}(\phi) \dd\phi^r \wedge \ast \dd\phi^s \,.
\end{equation}
%This is the Lagrangian of a (non-linear) sigma model.
The scalar fields can be interpreted as maps from the space-time manifold \(\Sigma\) into some target-space manifold \(\cM\) with Riemannian metric \(g\), i.e.
\begin{equation}
\phi \colon \Sigma \rightarrow \cM \,.
\end{equation}
From the discussion above it follows that the other fields (besides being space-time differential forms) must be sections of some vector bundles over \(\cM\) with bundle metrics \(M^{(p)}_{I_pJ_p}\) and structure group \(H\).
Using this language, the \(\cV^{\alpha_p}_{I_p}\) are nothing but local orthonormal frames on these bundles.
Sometimes we also want to introduce a local frame \(e^{\alpha_1}\) on \(\cM\), i.e.~\(
g_{rs} = \delta_{\alpha_1\beta_1} e^{\alpha_1}_r e^{\beta_1}_s %\,,
\),
such that \eqref{eq:sigmamodel} reads
\begin{equation}\label{eq:scalarvielbeins}
\cL_\mathrm{kin,scal} = \frac{e}{2} \delta_{\alpha_1\beta_1} \cP^{\alpha_1} \wedge \ast \cP^{\beta_1} \,, \qquad \text{with} \qquad \cP^{\alpha_1} = \phi^\ast e^{\alpha_1} = e^{\alpha_1}_r \dd \phi^r \,,
\end{equation}
where \(\phi^\ast\) denotes the pullback with respect to \(\phi\).

In a supersymmetric theory 
%bosonic and fermionic fields are mapped into each other via supersymmetry transformations, so 
also the fermions 
%should be 
are
sections of some vector bundles over \(\cM\).
In many cases these bundles correspond to the tangent bundle \(T\cM\) or are subbundles of \(T\cM\).
Let us make this more specific for the example of the gravitini.
%, which are the fermions that are present in every supergravity theory.
They are sections of a vector bundle %with structure group \(H_R\),
%so there exists the associated principle \(H_R\)-bundle
\begin{equation}
\cR \rightarrow \cM \,,
\end{equation}
with structure group \(H_R\).
On this bundle (or better on the associated principal bundle) there exists a local connection form \(\theta\),
%\(\theta \in \Omega^1(\cM, \h_R)\),
i.e.~a \(\h_R\)-valued 1-form on \(\cM\), where \(\h_R\) denotes the Lie-algebra of \(H_R\).
The corresponding curvature 2-form \(\Omega\)
%\(\Omega \in \Omega^2(\cM, \h_R)\)
is given by
\begin{equation}\label{eq:Rcurvature}
\Omega = \dd \theta + \theta \wedge \theta \,.
\end{equation}
This induces a covariant derivative \(\cD_M \psi^i_N\) which transforms covariantly under scalar-depedent \(H_R\)-transformations,
\begin{equation} \label{eq:covariantderiv}
\cD_M \psi^i_N = \nabla\!_M \psi^i_N - \left(\cQ^R_M\right)^i_j \psi^j_N \,,
\end{equation}
where \(\nabla_M\) is the space-time Levi-Cevita connection and \(\left(\cQ_M\right)^i_j\) is the pullback of the connection form \(\theta\), expressed in the appropriate \(\h_R\)-representation, i.e.
\begin{equation}\label{eq:Rconnection}
\cQ^R = \phi^\ast \theta \,.
\end{equation}
The corresponding curvature or field strength is obtained from the commutator of two covariant derivatives.
Explicitly, we have
\begin{equation}\label{eq:Dcomm}
\left[\cD_M, \cD_N \right] \epsilon^i = \tfrac{1}{4} R_{MNPQ}\Gamma^{PQ}\, \epsilon^i - \left(\cH^R_{MN}\right)^i_j \epsilon^j \,,
\end{equation}
where \(R_{MNPQ}\) is the space-time Riemann curvature tensor and \(\cH^R\) is the pullback of the curvature form \(\Omega\), i.e.~\(\cH^R = \phi^\ast \Omega\).

In a similar way we can introduce covariant derivatives for the other fermionic fields.
They transform in general not only under \(H_R\) but also under \(H_\mathrm{mat}\), or in other words they are sections of a vector bundle \(\cX \rightarrow \cM\) with structure group \(H\).
Analogous to our previous construction, we define
\begin{equation}\label{eq:chicovderiv}
\cD_M \chi^a = \nabla\!_M \chi^a - (\cQ_M)^a_b \chi^b= \nabla\!_M \chi^a - (\cQ^R_M)^a_b \chi^b - (\cQ^\mathrm{mat}_M)^a_b \chi^b \,,
\end{equation}
where \((\cQ_M)^a_b\) is the pull-back of the connection form on \(\cX\), expressed in the appropriate \(H\)-representation.
Since \(H\) is the product of \(H_R\) and \(H_\mathrm{mat}\) it splits into \(\cQ_M^R\) and \(\cQ_M^\mathrm{mat}\), where the former agrees with \eqref{eq:Rconnection}.
This indicates that in general \(\cR\) is a subbundle of \(\cX\). % unless \(\chi^a\) does not transform under the R-symmetry.
We finally want to note that according to the split \(a = (\hat a, \tilde a)\) we have \((\cQ_M)^{\hat a}_{\tilde a} = (\cQ_M)^{\tilde a}_{\hat a} = 0\) and \((\cQ_M^\mathrm{mat})^{\hat a}_{\hat b} = 0\).
The last identity is due to the fact that the components of the gravity multiplet do not transform with respect to \(H_\mathrm{mat}\).

We are now in the position to give the supersymmetry variations of the fermions.%
\footnote{The supersymmetry variations of the bosons (as well as of the fermions) are summarized in appendix~\ref{app:susyvariations}.}
They are of special importance in the following section, where we study maximally supersymmetric solutions.
In general they also contain terms of higher order in the fermionic fields.
However, we omit these terms as they vanish identically for the purely bosonic solutions we are interested in.
Under an infinitesimal supersymmetry transformation, described by the spinorial parameter \(\epsilon^i = \epsilon^i(x^M)\), the gravitini transform as
%The transformation of the gravitini takes the generic form
\begin{equation}\label{eq:ungaugedgravitinovariation}
\delta \psi^i_M = \cD_M  \epsilon^i  + \left(\cF_M\right)^i_j \epsilon^j \ ,
\end{equation} 
where \(\cD_M\) is the covariant derivative introduced in \eqref{eq:covariantderiv}.
The second term in \eqref{eq:ungaugedgravitinovariation} contains the various field strengths and is given by
\begin{equation}\label{eq:cFM}
\big(\cF_M\big)^i_j  =  \tfrac{1}{2D-4}\sum_{p \geq 2} \big(B_{ \hat\alpha_p}\big)^i_j\,
%F^{(p)} \cdot T_M = \alpha_{(p)} 
F^{\hat\alpha_p}_{{N_1}\dots {N_p}} {T_{(p)}^{{N_1}\dots {N_p}}}{}_M \,,
\end{equation}
where the \(B_{ \hat\alpha_p}\) are constant matrices correlating the different \(H_R\)-representations.
(See appendix~\ref{app:susyvariations} for a more detailed discussion of their properties.)  
The matrices \({T^{{N_1}\dots {N_p}}}{}_M\) are a specific combination of \(\Gamma\)-matrices and are defined in \eqref{eq:T}.

The supersymmetry variations of the spin-$\frac12$ fermions are even simpler and take the generic form
\begin{equation}
\delta \chi^a = \cF^a_i \epsilon^i \,,
\end{equation}
where \(\cF^a_i\) contains the various field strengths.
%The crucial observation is that
The variations of the fermions \(\chi^{\hat a}\) which are part of the gravity multiplet can contain only the field strengths \(F^{\hat \alpha_p}\), while the variations of the \(\chi^{\tilde a}\) depend only on \(F^{\tilde \alpha_p}\).
Explicitly \(\cF^a_i\) is given by
\begin{equation}\label{eq:cFhat}
\cF^{\hat a}_i = \sum_{p \geq 1}
%\sum_{\hat\alpha_p}
\big(C_{ \hat\alpha_p}\big)^{\hat a}_i \, F^{\hat\alpha_p}_{N_1\dots N_p} \Gamma^{N_1\dots N_p} \epsilon^i \,,
\end{equation}
and
\begin{equation}\label{eq:cFtilde}
\cF^{\tilde a}_i = \sum_{p \geq 1}
%\sum_{\tilde\alpha_p}
\big(C_{ \tilde\alpha_p}\big)^{\tilde a}_i \, F^{\tilde\alpha_p}_{N_1\dots N_p} \Gamma^{N_1\dots N_p} \epsilon^i \,.
\end{equation}
As in the gravitino variations the \(C_\al{p}\) are constant matrices. 
Contrary to \eqref{eq:cFM}, the sums in \eqref{eq:cFhat} and
\eqref{eq:cFtilde} start already at \(p = 1\)
and thus include  the fields strengths  of the scalar fields
\(F^{\alpha_1}_M = \cP^{\alpha_1}_M\) which
do not enter the gravitino variations \eqref{eq:ungaugedgravitinovariation}.

\subsection{Gauged Supergravity}\label{sec:gauging}

A generic supergravity theory is often invariant under a global symmetry group \(G\).
Let us denote the generators of \(G\) by \(t_\rho\), with \(\rho = 1, \dots , \dim(G)\).
They satisfy 
\begin{equation}\label{eq:Gstrconst}
\bigl[t_\rho, t_\sigma] = {f_{\rho\sigma}}^{\tau} t_\tau \,,
\end{equation}
where \({f_{\rho\sigma}}^{\tau}\) are the structure constants of the Lie algebra \(\g\) of \(G\).

We now want to gauge a subset of these symmetries, corresponding to a subgroup \(G^g \subseteq G\), i.e.~convert them from global to local symmetries.
Making a symmetry local is only possible if there exist appropriately transforming gauge fields, i.e.~1-forms or vector fields \(A^I\),% 
\footnote{For the sake of simplicity, from now on we often write for gauge fields \(A^I\) instead of \(A^{I_2}\).}
so that we can replace ordinary derivatives \(\partial_M\) by covariant derivatives \(D_M\),
\begin{equation}\label{eq:gencovderiv}
D_M = \partial_M - A^I_M X_I \,,
\end{equation}
whereas the \(X_I\) generate the respective subalgebra \(\g^g \subseteq \g\).
However, in supergravity the presence of gauge fields as well as their transformation behavior with respect to the global symmetry group \(G\) cannot be chosen freely but is usually restricted by supersymmetry.
This obstruction makes the gauging procedure more subtle.
To be more specific, let us denote the \(\g\)-representation of the gauge fields corresponding to the index \(I\) by \(\mathbf{v}\).
Clearly, the gauging can only be successful if the adjoint representation of \(\g^g\) can be found in the decomposition of \(\mathbf{v}\) into \(\g^g\)-representations.
 
The problem of finding a gaugeable subgroup \(G^g\) of \(G\) can be tackled systematically by means of the embedding tensor formalism \cite{Nicolai:2000sc,Nicolai:2001sv,deWit:2002vt} (see e.g.~\cite{Samtleben:2008pe} for a review).
Here one describes the embedding of \(\g^g\) into \(\g\) in terms of a constant map \(\Theta \colon \mathbf{v} \rightarrow \g\).
Explicitly, this embedding reads
\begin{equation}\label{eq:gaugegenerators}
X_I = {\Theta_I}^\rho t_\rho \,,
\end{equation}
where \({\Theta_I}^\rho\) is called the embedding tensor.
If we denote the generators of \(\g\) in the gauge field representation \(\mathbf{v}\) by \({(t_\rho)_I}^J\) and accordingly introduce \({X_{IJ}}^K = {(X_I){}_J}^K = {\Theta_I}^\rho {(t_\rho)_J}^K\), the condition that the \(X_I\) span a closed subalgebra of \(\g\) reads
\begin{equation}\label{eq:quadconstr}
\bigl[X_I, X_J] = -{X_{IJ}}^K X_K \,.
\end{equation}
Note that \({X_{IJ}}^K\) can only be regarded as the structure constants of \(\g^g\) under the above contraction with \(X_K\), on its own they do not even have to be antisymmetric in their lower indices.
This is the case because the \(X_I\) are not necessarily all linearly independent since the rank of \(\g^g\) might be smaller than the dimension of \(\mathbf{v}\).
The condition \eqref{eq:quadconstr} is equivalent to the \(\g^g\)-invariance of \(\Theta\), or explicitly
\({\Theta_I}^\rho \bigl( {(t_\rho)_J}^K {\Theta_K}^\sigma + {f_{\rho\tau}}^\sigma {\Theta_J}^\tau\bigr) = 0\).
Hence, it is called the quadratic constraint.

However, not every embedding which is actually compatible with the quadratic constraint can be realized in a given supergravity.
Supersymmetry imposes a second condition on the embedding tensor, called the linear constraint.
By construction \(\Theta\) transforms under \(\g\) in the product representation \(\mathbf{\overline{v}} \otimes \g\), which can be decomposed into a direct sum of irreducible \(\g\)-representations.
Not all of these irreducible representations describe a gauging which can be consistently realized in a supergravity theory.
Some of the irreducible representations in \(\mathbf{\overline{v}} \otimes \g\) are therefore not allowed and have to be set equal to zero.
Schematically, the linear constraint reads
\begin{equation}
\mathbb{P}\!_{lc} \, \Theta = 0 \,,
\end{equation}
where \(\mathbb{P}\!_{lc}\) is an operator that projects onto the forbidden \(\g\)-representations.
In a similar fashion one could also write the quadratic constraint as
\begin{equation}
\mathbb{P}\!_{qc} \, \Theta \otimes \Theta = 0 \,,
\end{equation}
with some appropriate projection operator \(\mathbb{P}\!_{qc}\).

A generic object \(\cO\) transforms under a local and infinitesimal gauge transformation parametrized by \(\lambda^I(x)\) according to
\begin{equation}\label{eq:gengaugetransf}
\delta \cO = \lambda^I X_I \cO = \lambda^I {\Theta_I}^\rho t_\rho \cO\,,
\end{equation}
where \(t_\rho\) are here the generators of \(G\) in the respective representation of \(\cO\).
In order for the covarariant derivative \(D_\mu \cO\) \eqref{eq:gencovderiv} to transform in the same way (i.e.~covariantly) the gauge fields \(A^I\) need to transform according to
\begin{equation}\label{eq:Atransf}
\delta A^I = D \lambda^I = \dd \lambda^I + {X_{JK}}^I A^J \lambda^K \,.
\end{equation} 
This transformation behavior requires an appropriate modification of the corresponding field strength 2-forms \(F^I\) such that they transform covariantly as well, i.e.
\begin{equation}\label{eq:Ftransf}
\delta F^I = - \lambda^J {X_{JK}}^I F^K \,.
\end{equation}
Note that this is precisely the same as \eqref{eq:gengaugetransf} for an object transforming in the gauge field representation \(\mathbf{v}\).
Due to the fact that the \({X_{IJ}}^K\) are not in one-to-one correspondence with the structure constants of \(\g^g\), finding covariantly transforming field strengths \(F^I\) is more subtle than in standard Yang-Mills theory.
The precise form of \(F^I\), however, is not important for the following discussion, so we do not need to comment further on this point.
Analogously, also the field strengths \(F^{I_p}\) of the other higher-rank form fields (if present) need to be modified appropriately.

Let us now turn to a discussion of the scalar field sector.
The sigma model Lagrangian \eqref{eq:sigmamodel} is invariant under all transformations of the scalar fields which leave the metric \(g_{rs}\) invariant.
In other words the global symmetry group \(G\) must be contained in the isometry group \(\mathrm{Iso}(\cM)\) of \(\cM\).
To be more specific, an infinitesimal transformation \(\phi^r \rightarrow \phi^r + \lambda^\rho k^r_\rho\) leaves \eqref{eq:sigmamodel} invariant if the \(k^r_\rho\) are Killing vectors of \(g_{rs}\), i.e. \(\nabla_{(r} k_{s)\rho} = 0\), and if the \(k_\rho\) generate a subgroup \(G\) of \(\mathrm{Iso}(\cM)\), i.e. \(\left[k_\rho, k_\sigma\right] = - {f_{\rho\sigma}}^\tau k_\tau\), where \({f_{\rho\sigma}}^\tau\) are the structure constants of the Lie algebra \(\g\) of \(G\), cf.~\eqref{eq:Gstrconst}.
%We now want
To gauge some of these symmetries
%, so according to our above considerations 
we select a subgroup \(G^g \subset G\) via
\begin{equation}
k_I = {\Theta_I}^\rho k_\rho \,,
\end{equation}
such that
\begin{equation}\label{eq:killingcommutator}
\bigl[k_I, k_J\bigr] = {X_{IJ}}^K k_K \,,
\end{equation}
where \({X_{IJ}}^K\) is defined
% in the same way
as in \eqref{eq:quadconstr}.
In the end we want to construct a Lagrangian which is invariant under local \(G^g\) transformations
\begin{equation}\label{eq:scalargaugetransf}
%\phi^r(x) \rightarrow \phi^r(x) + \lambda^I(x) k^r_I(\phi) \,,
\delta\phi^r = \lambda^I(x) k^r_I(\phi) \,,
\end{equation}
%where the infinitesimal parameters \(\lambda^I(x)\) are explicitly allowed to depend on the space-time coordinates.
%Such transformations induce additional terms in the derivative \(\dd\phi^r\) which have to be compensated by the introduction of
The corresponding covariant derivatives read
\begin{equation}\label{eq:scalarcovderiv}
D\phi^r = \dd \phi^r - A^I k^r_I \,,
\end{equation}
where the \(A^I\) transform according to \eqref{eq:Atransf}.
%The form of \eqref{eq:scalargaugetransf} and \eqref{eq:scalarcovderiv} indicates again that
The Killing vectors take the role of the general gauge group generators \(X_I\) in the scalar field sector.
Analogously the vielbeins \(\cP^\al{1}\) get replaced by
\begin{equation}\label{eq:gaugedP}
\hat \cP^\al{1} = \cP^\al{1} + A^I \cP^\al{1}_I \,,\qquad \cP^\al{1}_I  = k^r_I e^\al{1}_r \,.
\end{equation}
It is often beneficial to use \(\cP^\al{1}_I\), which are the Killing vectors expressed in the local frame \(e^\al{1}\)\eqref{eq:sigmamodel}, instead of working directly with \(k^r_I\).
%where the insertion map \(\iota_I\) is defined as \(\iota_I e^\al{1} = e^\al{1}(k_I) = k^r_I e^\al{1}_r\).

The complete supersymmetric Lagrangian consists not only of the sigma model part \eqref{eq:sigmamodel}, but also features all the other fields living in vector bundles over \(\cM\).
Therefore, a symmetry of the complete theory must be more than just an isometry of the scalar manifold \(\cM\).
We need to demand that Killing vectors are compatible with the various bundle structures.
For the R-symmetry bundle \(\cR\) these conditions read%
\footnote{Our discussion follows \cite{DAuria:1990qxt, Andrianopoli:1996cm}.}
\begin{equation}\label{eq:W}
\cL_I \Omega = \bigl[\Omega, W_I\bigr] \,, \qquad \cL_I \theta = \cD W_I \equiv \dd W_I + \bigl[\theta, W_I\bigr] \,.
\end{equation}
Here \(\cL_I\) denotes the Lie derivative in the direction of \(k_I\), i.e. \(\cL_I = \cL_{k_I}\),\footnote{
The Lie derivative describes how a scalar field dependent object varies under a variation of the scalar fields.
%The Lie derivative along Killing directions describes how an object varies under gauge transformations.
For example, under an infinitesimal gauge transformation \eqref{eq:scalargaugetransf} parametrized by \(\epsilon^I(x)\), a geometrical object \(T\) defined on \(\cM\) transforms according to \(\delta_\epsilon T = \epsilon^I \cL_I T\).}
 and \(\theta\) and \(\Omega\) are the connection and curvature form on \(\cR\), see \eqref{eq:Rcurvature}.
The \(W_I\) are (local) \(\h_R\)-valued functions on \(\cM\) which are required to satisfy the condition
\begin{equation}\label{eq:Wcocycle}
\cL_I W_J - \cL_J W_I + \bigl[W_I,W_J\bigr] = {X_{IJ}}^K W_K \,.
\end{equation}
To find the correct modification of the covariant derivative \eqref{eq:covariantderiv} of the gravitini and supersymmetry parameters one needs to introduce the generalized moment maps \(\cQ^R_I\) which are locally defined by
\begin{equation}\label{eq:genmomentmap}
\cQ^R_I = \iota_I \theta - W_I \,,
\end{equation}
where \(\iota_I \theta = k^r_I \theta_r\) denotes the insertion of the Killing vector \(k_I\) into \(\theta\).
It follows directly from the definition of the curvature form \(\Omega\) \eqref{eq:Rcurvature} and from \eqref{eq:Wcocycle} that
\begin{equation}\label{eq:Qderiv}
\cD \cQ^R_I = - \iota_I \Omega \,,
\end{equation}
which is often taken as the definition of \(\cQ^R_I\).
Moreover, it follows from \eqref{eq:Wcocycle} that the Lie derivative of the moment maps with respect to the Killing directions is given by
\begin{equation}\label{eq:momentmapliederiv}
\cL_I \cQ^R_J = - \bigl[W_I, \cQ^R_J \bigr] + {X_{IJ}}^K \cQ^R_K \,,
\end{equation}
which implies that they satisfy the equivariance condition
\begin{equation}\label{eq:equivariance}
\bigl[\cQ^R_I, \cQ^R_J\bigr] = - {X_{IJ}}^K \cQ^R_K + \Omega\!\left(k_I, k_J\right) \,.
\end{equation}
The transformation property \eqref{eq:momentmapliederiv} shows that \(\cQ^R_I\) is the correct object to build a gauged version \(\hat\cD\) of the covariant derivative \(\cD\) introduced in \eqref{eq:covariantderiv}.
Explicitly, we define
\begin{equation}\label{eq:gaugedQ}
\hat\cD_M \epsilon^i = \nabla\!_M \epsilon^i - (\hat\cQ^R_M)^i_j \epsilon^j \,,\qquad\text{with}\qquad \hat\cQ^R = \cQ^R + A^I \cQ^R_I \,.
\end{equation}
This covariant derivative transforms properly if the \(\epsilon^i\) transform under a gauge transformation as
\begin{equation}
\delta \epsilon^i = -\lambda^I (W_I)^i_j \epsilon^j \,,
\end{equation}
where \((W_I)^i_j\) is the \(\h_R\)-compensator \eqref{eq:W} expressed in the appropriate representation of \(H_R\). 
Analogously to \eqref{eq:Dcomm}, the commutator of two gauged covariant derivatives \(\hat\cD\) is given by
\begin{equation}\label{eq:Dcommgauged}
\bigl[\hat\cD_M, \hat\cD_N \bigr] \epsilon^i = \tfrac{1}{4} R_{MNPQ}\Gamma^{PQ}\, \epsilon^i - \bigl(\hat\cH^R_{MN}\bigr)^i_j \epsilon^j \,,
\end{equation}
where the field strength \(\hat\cH^R\) now also contains a term that depends on the field strengths \(F^I\) of the gauge fields \(A^I\),
\begin{equation}\label{eq:hrgenerators}
\hat\cH^R = \cH^R + F^{I} \cQ^R_{I} \,.
\end{equation}
Let us finally mention that even in the absence of scalar fields, i.e.~if \(\cM\) is degenerated to a point, it is still often consistent to assign a non-trivial (constant) value to \(\cQ^R_I\), known as a Fayet-Iliopoulos term \cite{Fayet:1974jb, Fayet:1975yi}.

In a similar fashion to the construction above we need to modify the \(H\)-covariant derivative \eqref{eq:chicovderiv} of the other fields and introduce
\begin{equation}\label{eq:chigaugedcovderiv}
\hat\cD_M \chi^a = \nabla\!_M \chi^a - (\hat\cQ_M)^a_b \chi^b = \nabla\!_M \chi^a - (\cQ_M)^a_b \chi^b - A^I (\cQ_I)^a_b \chi^b \,.
\end{equation}
Notice that the moment maps \(\cQ_I\) split in general according to
\begin{equation}
\cQ_I = \cQ^\cR_I + \cQ_I^\mathrm{mat} \,,
\end{equation}
where \(\cQ^R\) is the R-symmetry moment map which we have constructed above.

%We want to illustrate these concepts for
The tangent bundle \(T\cM\)
%which 
is by construction an \(H\)-bundle as well.
Here the connection form \(\theta\) is given by the Levi-Civita connection.
With respect to the local frame \(e^\al{1}\) it is defined as the solution of
\begin{equation}\label{eq:spinconnection}
\dd e^\al{1} + {\theta_\be{1}}^\al{1} \wedge e^\be{1} = 0 \,.
\end{equation}
Accordingly the covariant derivative of the Killing vectors \(\cP^\al{1}_I\) reads
\begin{equation}
\cD \cP^\al{1}_I = \dd \cP^\al{1}_I - {\theta_\be{1}}^\al{1} \cP^\be{1}_I \,,
\end{equation}
and the moment maps \(\cQ_I\) in the respective \(H\)-representation are given by \cite{Bandos:2016smv}
\begin{equation}\label{eq:TMQ}
{\bigl(\cQ_I\bigr)_\al{1}}^\be{1} = - \cD_\al{1} \cP^\be{1}_I \,.
\end{equation}
A moment map introduced in this way indeed satisfies the defining property \eqref{eq:Qderiv}.
This follows from the general fact that the second covariant derivative of a Killing vector is given by a contraction of the same Killing vector with the Riemann tensor (see e.g. \cite{Weinberg:1972kfs}).
Moreover, \eqref{eq:TMQ} implies that 
\begin{equation}\label{eq:Pcovderiv}
\cD_I \cP^\al{1}_J = {X_{IJ}}^K \cP^\al{1}_K - {\bigl(\cQ_I\bigr)_\be{1}}^\al{1} \cP^\be{1}_J \,,
\end{equation}
which in turn shows in combination with \eqref{eq:genmomentmap} that \(\cP^\al{1}_I\) transforms under a gauge transformation in the appropriate way, i.e.
\begin{equation}
\cL_I \cP^\al{1}_J = {X_{IJ}}^K \cP^\al{1}_K + {\bigl(W_I\bigr)_\be{1}}^\al{1} \cP^\be{1}_J \,.
\end{equation}

Let us again come back to the gauge field sector.
As we have seen above the field strengths \(F^I\) are not inert under gauge transformations but transform according to \eqref{eq:Ftransf}.
Therefore the gauge invariance of the kinetic term in \eqref{eq:bosonicaction} demands an analogous transformation law for the matrix \(M^{(2)}_{IJ}(\phi)\), i.e.
\begin{equation}
\cL_I M^{(2)}_{JK} = 2 {X_{I (J}}^L M^{(2)}_{K) L} \,,
\end{equation}
consistent with \(M^{(2)}_{IJ}\) transforming in the \(\left(\mathbf{\overline{v}} \otimes \mathbf{\overline{v}}\right)_\mathrm{sym}\) representation.
Correspondingly, the vielbeins \(\cV^\al{2}_I\) transform according to
\begin{equation}\label{eq:generalvielbeinvariation}
\cL_I \cV^\al{2}_J = {X_{IJ}}^K \cV^\al{2}_K + {(W_I)_\be{2}}^\al{2} \cV^\be{2}_J \,. 
\end{equation}
The additional term featuring the \(H\)-compensator \(W_I\) is due to the fact that \(\cV^\al{2}_I\) live in an \(H\)-bundle over \(\cM\).
%where \((W_I)^\al{2}_\be{2}\) is the \(H\)-compensator \eqref{eq:W} expressed in the corresponding representation.
Similar considerations hold for the other \(p\)-form fields.

In addition to the replacement of \(\cD\) with \(\hat\cD\) the gauging of the theory requires the modification of the fermionic supersymmetry variations by shift matrices \(A^i_{0\,j}\) and \(A^a_{1\,i}\).
These matrices in general depend on the scalar fields and the specific form of the gauging.
We will derive some explicit relations between \(A_0\) and \(A_1\) and the Killing vectors and moment maps in appendix~\ref{app:susycalculations}.
Altogether, the supersymmetry variations of the fermions read
\begin{subequations}\label{eq:generalfermionicvariations}
\begin{equation}\label{eq:gravitinovariation}
\delta \psi^i_M = \hat\cD_M  \epsilon^i  + \left(\cF_M\right)^i_j \epsilon^j + A^i_{0\,j} \epsilon^j\ ,
\end{equation} 
\begin{equation}\label{eq:spin12variation}
\delta \chi^a = \cF^a_i \epsilon^i + A^a_{1\,i} \epsilon^i \,,
\end{equation}
\end{subequations}
where \(\cF_M\) and \(\cF\) are the same objects as defined in \eqref{eq:cFM}, \eqref{eq:cFhat} and \eqref{eq:cFtilde}, depending on the gauge covariant field strengths.
In addition, the shift matrices also act as fermionic mass-matrices and we give their explicit form in \eqref{eq:appfermionmass}.
Moreover, supersymmetry requires the existence of a non-trivial scalar potential which can be expressed in terms of \(A_0\) and \(A_1\).
It is given by
\begin{equation}\label{eq:generalpotential}
%V = - \tfrac{2(D-1)(D-2)}{\cN} \tr(A_0^\dagger A_0) + \tfrac{2}{\cN}\tr(A_1^\dagger A_1) \,.
\delta^i_j V = - 2(D-1)(D-2) \bigl(A^\dagger_0\bigr)^i_k A^k_{0\,j} + 2 \bigl(A_1^\dagger\bigr)^i_a A^a_{1\,j} \,.
\end{equation}
%and \(V\) can be obtained by taking the trace on both sides.
Of course, for the gauging procedure to be consistent the potential must be invariant with respect to local \(G^g\) transformations, i.e.~\(\cL_I V = 0\).

We finally need to mention that in some cases there exist deformations %of a supergravity theory
which can not be expressed as the gauging of a global symmetry.
These deformation can give rise to fermion shift matrices and to a scalar potential as well.
Prominent examples are the superpotential of four-dimensional \(\cN = 1\) supergravity or massive type IIA supergravity in ten dimensions.

\subsection{Coset geometry}\label{sec:coset}
In this section we discuss the application of the previously introduced concepts to theories where the target space \(\cM\) is a symmetric space.\footnote{We follow the discussion of \cite{Bandos:2016smv, Trigiante:2016mnt}.}
This is necessarily the case for all theories with more than 8 real supercharges.
For these theories we can write \(\cM\) as a coset
\begin{equation}
\cM = \frac{G}{H} \,,
\end{equation}
where \(G\) is a non-compact Lie group and \(H\) its maximally compact subgroup.
%, such that \(\cM\) is non-compact and of definite signature.
\(H\) coincides with the group introduced in \eqref{eq:H}.
The points of \(\cM\) are the 
%equivalence classes in \(G\) with respect to the right multiplication of \(H\), i.e. \(g \sim g h\) for some \(h \in H\), and thus the
left-cosets \(gH\) with \(g \in G\).
Note that the map \(g \mapsto gH\) induces on \(G\) a natural structure as an \(H\)-principal bundle over \(G/H\), which is precisely the kind of structure needed for supergravity.

The Lie algebra \(\g\) of \(G\) can be decomposed as
\begin{equation}\label{eq:gdecomp}
\g = \h \oplus \k \,,
\end{equation}
where the direct sum is to be understood only as a direct sum of vector spaces.
Here \(\h\) denotes the Lie algebra of \(H\) and \(\k\) spans the remaining directions of \(\g\).
Since \(\h\) is a subalgebra of \(\g\) it is by definition closed with respect to the Lie-bracket, i.e.~\(\left[\h,\h\right] \subseteq \h\).
If \(\g\) is a reductive Lie algebra (this means it is the direct sum of only simple or abelian Lie-algebras) we can always find a decomposition of \(\g\) such that
\begin{equation}\label{eq:reductive}
\left[\h,\k\right] \subseteq \k \,.
\end{equation}
In this case also the coset space \(G/H\) is called reductive.
In particular, this means that \(\k\) transforms in an \(\h\)-representation with respect to the adjoint action.
Moreover, \(G/H\) is called symmetric if it is reductive and
\begin{equation}\label{eq:symmetric}
\left[\k,\k\right] \subseteq \h \,.
\end{equation}
All coset spaces that we encounter will be symmetric.
It is sometimes convenient to give an explicit basis for \(\h\) and \(\k\).
In this case we denote the generators of \(\h\) by \(J^A\) and the generators of \(\k\) by \(K^\alpha\).
In this basis the conditions \eqref{eq:reductive} and \eqref{eq:symmetric} in terms of the structure constants read
\begin{equation}
{f_{\alpha A}}^B = {f_{\alpha\beta}}^\gamma = 0 \,.
\end{equation}

Let \(\phi: \Sigma \rightarrow \cM \) be the scalar fields describing a sigma model on \(\cM\), and let \(\phi^r\) be the scalar fields in local coordinates.
Each value of \(\phi\) corresponds to a coset and can be therefore described by a coset representative \(L(\phi) \in G\).
Acting on \(L(\phi)\) from the left with some element \(g \in G\) yields another element in \(G\) that generically lies in a different coset, represented by \(L(\phi')\).
As \(g L(\phi)\) and \(L(\phi')\) are in the same \(H\)-coset, they must only differ by the right action of some \(h(\phi,g) \in H\) and therefore
\begin{equation}\label{eq:gcosettransf}
g L(\phi) = L(\phi') h(\phi, g) \,.
\end{equation}
To formulate the sigma model action we introduce the Maurer-Cartan form
\begin{equation}\label{eq:MCform}
\omega = L^{-1} \dd L \,,
\end{equation}
which takes values in \(\g\) and satisfies the Maurer-Cartan equation \(\dd \omega + \omega \wedge \omega = 0\).
We split \(\omega\) according to the decomposition \eqref{eq:gdecomp} of \(\g\),
\begin{equation}\label{eq:maurercartan}
\omega = \cP + \cQ \,,%\qquad\text{such that}\qquad \cP \in \k\,, \cQ \in \h \,,
\end{equation}
such that \(\cP\) takes values in \(\k\) and \(\cQ\) takes values in \(\h\)
or explicitly \(\cP = \cP^\alpha K_\alpha\) and \(\cQ = \cQ^A J_A\).
\(\cP\) is used to formulate the kinetic term of a sigma model on \(\cM\).
Its Lagrangian reads
\begin{equation}\label{eq:cosetlagrangian}
\cL_\mathrm{kin,scal} = -\frac{e}{2} \tr\left(\cP\wedge \ast \cP\right) = -\frac{e}{2} g_{\alpha \beta} \cP^\alpha \wedge \ast \cP^\beta \,,
\end{equation}
where \(g_{\alpha\beta} = \tr\left(K_\alpha K_\beta\right)\) is the restriction of the Killing form of \(\g\) on \(\k\).
Notice that it is always possible to find a basis of generators \(K_\alpha\) such that \(g_{\alpha\beta} = \delta_{\alpha\beta}\).
In this frame the \(\cP^\alpha\) directly correspond to the vielbeins introduced in \eqref{eq:scalarvielbeins}.
This Lagrangian is invariant under a global \(G\)-transformation \eqref{eq:gcosettransf}.
Indeed, \(\cP\) and \(\cQ\) transform as
\begin{equation}\begin{aligned}
\cP(\phi') &= h \cP(\phi) h^{-1} \,, \\
\cQ(\phi') &= h \cQ(\phi) h^{-1} + h \dd h^{-1} \,,
\end{aligned} \end{equation}
which shows the invariance of \eqref{eq:cosetlagrangian}.
Moreover, \(\cQ\) has the transformation behavior of a \(H\)-connection, it is the local connection form of the principal \(H\)-bundle over \(\cM\) and can be used to define an \(H\)-covariant derivative.
The action of this covariant derivative on the coset representative \(L\) is given by
\begin{equation}\label{eq:cosetreprderiv}
\cD L = d L - L \cQ = L \cP \,,
\end{equation}
where the second equality follows from the definition of \(\cP\) and \(\cQ\), see \eqref{eq:MCform} and \eqref{eq:maurercartan}.
The Maurer-Cartan equation expressed in terms of \(\cP\) and \(\cQ\) reads
\begin{equation}\begin{aligned} 
\cD \cP &= \dd \cP + \cQ \wedge \cP + \cP \wedge \cQ = 0 \,, \\
\cH &= \dd \cQ + \cQ \wedge \cQ = - \cP \wedge \cP \,.
\end{aligned}\end{equation}
The first equation can be rewritten as \(\dd \cP^\alpha + {f_{A\beta}}^\alpha \cQ^A \wedge \cP^\beta = 0\). %, where we expressed \(\cQ\) explicitly in the adjoint representation.
This is Cartan's structure equation for the vielbein \(\cP^\alpha\) and shows that \({\cQ_\beta}^\alpha = \cQ^A {f_{A\beta}}^\alpha\) is a connection on the tangent bundle \(T\cM\) compatible with the metric \(g_{\alpha\beta}\).
%The second equation is nothing but \eqref{eq:hscalar}, as can be seen by expressing it as \(\cH = - {f_{\alpha\beta}}^A J_A \cP^\alpha \wedge \cP^\beta\).
%Moreover, it shows that the holonomy group of \(\cM\) is given by \(\mathrm{Hol}(\cM) = H\).

Let us finally discuss the isometries of \(\cM\) and the gauged version of the above construction.
The metric \(g_{\alpha\beta}\) is invariant under the left action of \(G\), therefore every element of \(G\) (acting on \(\cM\) from the left) corresponds to an isometry of \(\cM\). 
%and therefore \(G\) must be (at least contained in) the isometry group of \(\cM\).
We start with a discussion of the action of an infinitesimal isometry on the coset representative \(L\), described by the left action of 
\begin{equation}
g = 1 + \epsilon^\rho t_\rho \,,
\end{equation}
where \(t_\rho\in \g\),
This induces a transformation of the scalars \(\phi\) along the corresponding Killing vector \(k_\rho\),
\begin{equation}
\phi' = \phi + \epsilon^\rho k_\rho \,.
\end{equation}
According to \eqref{eq:gcosettransf} we need a compensating \(H\)-transformation
\begin{equation}
h(\phi, g) = 1 - \epsilon^\rho W_\rho \,,
\end{equation}
where \(W_\rho \in \h\).
Inserting this into \eqref{eq:gcosettransf} and collecting all terms at linear order in the parameter \(\epsilon^\rho\) yields
\begin{equation}
L^{-1} t_\rho L = \iota_\rho \cP + \cQ_\rho \,,
\end{equation}
where the moment map \(\cQ_\rho\) is given by
\begin{equation}\label{eq:cosetmomentmap}
\cQ_\rho = \iota_\rho \cQ - W_\rho \,.
\end{equation}
Notice that this agrees precisely with the general form of the moment map as defined in \eqref{eq:genmomentmap}.
To describe a gauged sigma model on \(\cM\) 
%where some of its isometries are gauged 
we proceed along the lines of the general discussion and select a subalgebra \(\g^g\) of \(\g\) using the embedding tensor formalism.
The generators \(X_I\) of \(\g^g\) are given in terms of \(t_\rho\) by \eqref{eq:gaugegenerators}.
We then introduce the gauged version of the Maurer-Cartan form \eqref{eq:maurercartan}
\begin{equation}
\hat\omega = L^{-1} \left(d + A^I X_I\right) L \,.
\end{equation}
It is by construction invariant under a local transformation of the form \(\delta L = \epsilon^I(x) X_I L\) if we demand \(A^I\) to transform according to \eqref{eq:Atransf}.
We learn from our previous considerations that for the gauged versions of the vielbein \(\cP\) and the connection \(Q\) this yields
\begin{equation}\begin{aligned}
\hat\cP &= \cP + A^I \cP_I \,, \\
\hat\cQ &= \cQ + A^I \cQ_I \,,
\end{aligned}\end{equation}
where \(\cP_I = \iota_I \cP\) and \(\cQ_I\) is given in \eqref{eq:cosetmomentmap}.
This is exactly the same as \eqref{eq:gaugedP} and \eqref{eq:gaugedQ}, so \(\hat\cP\) and \(\hat\cQ\) indeed are the correct quantities to describe the gauged sigma model on \(\cM = G/H\).

Instead of working with the generators \(X_I\) themselves, it is often more convenient to work with their contracted or dressed version
\begin{equation}\label{eq:dressedgenerator}
\cT_I = L^{-1} X_I L = \cP_I + \cQ_I \,,
\end{equation}
and \(\cP_I\) and \(\cQ_I\) are the \(\k\)-part and \(\h\)-part of \(\cT_I\).
Since the coset representative \(L\) is invertible, \(\cT_I\) carries the same amount of information as \(X_I\) and clearly satisfies the same commutator algebra.
One can go one step further and also dress the remaining index \(I\) with the vielbein \(\cV^I_\al{2}\) to obtain
\begin{equation}\label{eq:Ttensor}
\cT_\al{2} = \cV^I_\al{2} \cT_I \,.
\end{equation}
This object is often called the T-tensor \cite{deWit:1981sst, deWit:1982bul}.
In the same way as the embedding tensor \(\Theta\) decomposes into irreducible representation of \(\g\), the T-tensor can be decomposed into irreducible representations of \(\h\).
Again, the linear constraint restricts which representations can appear in a consistently gauged supergravity.
The allowed representations for \(\cT_\al{2}\) can be obtained by branching the allowed \(\g\)-representation of \(\Theta\) into \(\h\)-representations.

The T-tensor -- or equivalently its components \(\cQ_\al{2}\) and \(\cP_\al{2}\) -- features in the construction of the fermionic shift matrices \(A_0\) and \(A_1\).
Denoting the \(\h\)-representations of the gravitini and the spin-1/2 fermions by \(\mathbf{s}\) and \(\mathbf{x}\), respectively, \(A_0\) and \(A_1\) a priori transform in the tensor product representations \(\mathbf{s} \otimes \mathbf{\overline s}\) and \(\mathbf{x} \otimes \mathbf{\overline s}\).%
%where \(\mathbf{s^\ast}\) denotes the dual representation.
\footnote{The \(\h\) representation of \(A_0\) is furthermore often restricted since the gravitino mass term \(A^i_{0\,j} \bar\psi_{M i} \Gamma^{MN} \psi^j_N\) can impose an (anti-)symmetry property on \(A_0\).}
The components of \(\cT_\al{2}\) that transform in a representation which appears in these tensor products agree with the respective components of \(A_0\) and \(A_1\).
In appendix~\ref{app:susycalculations} we further elaborate on the relation between the T-tensor and the fermionic shift matrices and give explicit expressions for \(\cQ_\al{2}\) and \(\cP_\al{2}\) in terms of \(A_0\) and \(A_1\).

We finally want to point out that also the vielbeins \(\cV^\al{p}_{I_p}(\phi)\) of the kinetic matrices \(\cM^{(p)}_{I_p J_p}\) \eqref{eq:kinmatrix} are nothing but the coset representative \(L(\phi)\) taken in the respective representations of \(G\) and \(H\).
In this sense we can express any scalar field dependence solely in terms of the coset representative (and its derivatives). 

\section{AdS solutions and their moduli spaces}\label{sec:ads}

\subsection{The gauge group of AdS solutions}\label{sec:adsgaugings}

In this section we derive conditions on the gauge group \(G^g\) such that maximally supersymmetric AdS$_D$ solutions exist.
In particular we discuss the spontaneous breaking of \(G^g\) to a reductive subgroup \(H^g\).

A maximally supersymmetric solution is characterized by the vanishing of all fermionic supersymmetry variations, i.e.
\begin{equation}\label{eq:susyvacuum}
\delta \psi^i_M = \delta \chi^a = 0 \,,
\end{equation}
where explicit expressions for \(\delta \psi^i\) and \(\delta \chi^a\) are given in \eqref{eq:gravitinovariation} and \eqref{eq:spin12variation}.
%From the general expressions \eqref{eq:gravitinovariation} and \eqref{eq:spin12variation} it follows that the fermionic shift matrices \(A_0\) and \(A_1\) need to satisfy \cite{Louis:2016tnz}
As discussed in \cite{Louis:2016tnz}, an AdS$_D$ solution is only possible in the absence of any background fluxes, entering the supersymmetry variations via the terms \(\cF_M\) and \(\cF\).
Such fluxes would not be compatible with the \(\SO(D-1, 2)\) isometries of the AdS background.
Therefore, for unbroken supersymmetry the conditions \eqref{eq:susyvacuum} imply
\begin{equation}\label{eq:adsconditions}
(A_0)^2 = - \frac{\Lambda}{2(D-1)(D-2)} \id \,,\quad
A_1 = 0 \,,
\end{equation}
where \(\Lambda\) is the negative cosmological constant.
Note that it follows from the general form of the scalar potential V given in \eqref{eq:generalpotential} that \(A_1 = 0\) already implies \((A_0)^2 \sim \id\).
Therefore demanding \(A_0 \neq 0\) and \(A_1 = 0\) is enough to guarantee that also the first equation in \eqref{eq:adsconditions} is solved for some value of \(\Lambda\).

The conditions \eqref{eq:adsconditions} in turn enforce constraints on the possible gauge groups of the theory.
Let us introduce the dressed moment maps and Killing vectors, 
%defined in \eqref{eq:gaugedP} and \eqref{eq:genmomentmap},
\begin{equation}\label{eq:dressedQP} 
\cQ^R_\al{2} = \cV^I_\al{2} \cQ^R_I \,,\qquad \cP_\al{2} = \cV^I_\al{2} \cP_I \,,
\end{equation}
where \(\cQ^R_I\) and \(\cP_I\) are defined in \eqref{eq:genmomentmap} and \eqref{eq:gaugedP} and
 \(\cV^I_\al{2}\) are the vielbeins of the vector field kinetic matrix \eqref{eq:kinmatrix}.\footnote{Note the similarity with the definition of the T-tensor in \eqref{eq:Ttensor}.}
In appendix~\ref{app:susycalculations} we derive expressions for \(\cQ_\al{2}^R\) and \(\cP_\al{2}\) in terms of \(A_0\) and \(A_1\), see \eqref{eq:QRA0A1} and \eqref{eq:PA1}.
For vanishing \(A_1\) they
%the resulting equations \eqref{eq:QRA0A1} and \eqref{eq:PA1}
read
\begin{equation}
\cQ^R_\al{2} = (D-3) \bigl\{A_0, B_\al{2}\bigr\} \,,\qquad\text{and}\qquad \cP_\al{2} B_\be{2} \delta^{\al{2}\be{2}} = 0 \,,
\end{equation}
where \(B_\al{2}\) are the same matrices as appearing in the supersymmetry variations of the gravitini \eqref{eq:cFM}.
As introduced in the previous section we want to employ the split of \(\al{2}\) into \(\hal{2}\) and \(\tal{2}\) \eqref{indexsplit},
where \(\hal{2}\) labels those fields strengths which enter the gravitini variations and \(\tal{2}\) their orthogonal complement.
Consequently, the \(B_\hal{2}\) are a set of linearly independent matrices, while on the other hand \(B_\tal{2} = 0\)
and we find the following general conditions for a maximally supersymmetric AdS solution in terms of the dressed moment maps and Killing vectors, 
\begin{equation}\begin{gathered}\label{eq:AdSconditionsQP}
\cQ^R_{\hal{2}} = (D-3) \bigl\{A_0, B_{\hal{2}}\bigr\} \,, \\
\cQ^R_{\tal{2}} = \cP_\hal{2} = 0 \,.
\end{gathered}\end{equation}
Of course these equations are only to be understood as restrictions on the background values of \(\cQ^R_\al{2}\) and \(\cP_\al{2}\), at an arbitrary point of the scalar manifold they do not need to be satisfied.

Let us analyze the implications of the equations \eqref{eq:AdSconditionsQP} on the gauge group \(G^g\). 
As discussed in section~\ref{sec:gauging} the generators of \(G^g\) are denoted by \(X_I\) \eqref{eq:gaugegenerators} and their action on the scalar manifold is described in terms of the Killing vectors \(\cP_I\) or equivalently by the dressed Killing vectors \(\cP_\al{2}\) defined in \eqref{eq:dressedQP}.
Contrary to \(\cP_\hal{2}\) the background values of the Killing vectors \(\cP_\tal{2}\) are unrestricted by \eqref{eq:AdSconditionsQP}, none the less some (or all) might also be vanishing.
For this reason we again split the index \(\tal{2}\) into \(\tal{2}'\) and \(\tal{2}''\) such that the background values of \(\cP_{\tal{2}'}\) are all non-vanishing and linearly independent and such that in the background \(\cP_{\tal{2}''} = 0\).
Let us furthermore collectively denote all Killing vectors with vanishing background value by \(\cP_{\alpha_2^0} = (\cP_\hal{2}, \cP_{\tal{2}''})\).

The Killing vectors \(\cP_{\alpha_2^0}\) with vanishing background value (or equivalently the generators \(X_{\alpha_2^0}\)) generate a subgroup 
\begin{equation}
H^g \subseteq G^g
\end{equation} 
of the gauge group.
To see this we express the commutator \eqref{eq:killingcommutator} of Killing vectors \(\cP_{\alpha_2^0}\) according to our split of indices as 
\begin{equation}\label{eq:P0comm}
\bigl[\cP_{\alpha_2^0}, \cP_{\beta_2^0}\bigr] = {X_{\alpha_2^0 \beta_2^0}}^{\gamma_2^0} \cP_{\gamma_2^0} + {X_{\alpha_2^0 \beta_2^0}}^{\gamma_2'} \cP_{\gamma_2'} \,,
\end{equation}
where \({X_{\al{2}\be{2}}}^\ga{2} = \cV^I_\al{2} \cV^J_\be{2} \cV^\ga{2}_K {X_{IJ}}^K\).
In the background only \(\cP_{\gamma_2'}\) on the right hand side of \eqref{eq:P0comm} does not vanish, which enforces \({X_{\alpha_2^0 \beta_2^0}}^{\gamma_2'} = 0\).
Moreover, inserting \(\cP_{\alpha_2^0} = 0\) into \eqref{eq:genmomentmap} gives \(\cQ_{\alpha_2^0} = - W_{\alpha_2^0}\) and from \eqref{eq:generalvielbeinvariation} we find
\begin{equation}\label{eq:backgroundQ}
{X_{\alpha_2^0 \be{2}}}^\ga{2} = {\bigl(\cQ_{\alpha_2^0}\bigr)_\be{2}}^\ga{2} \,.
\end{equation}
However, since \(\cQ_{\alpha_2^0}\) is an element of \(\h\) it satisfies \({\bigl(\cQ_{\alpha_2^0}\bigr)_\tbe{2}}^\hga{2} = {\bigl(\cQ_{\alpha_2^0}\bigr)_\hbe{2}}^\tga{2} = 0\)
%\begin{equation}
%{X_{\hal{2}\hbe{2}}}^\tga{2} = {X_{\tal{2}'' \tbe{2}}}^\hga{2} = {X_{\tal{2}''\hbe{2}}}^\ga{2} = 0 \,,
%\end{equation}
and therefore we find for the commutators \eqref{eq:quadconstr} of the corresponding gauge group generators
\begin{equation}\begin{aligned}\label{eq:Hcomm}
\bigl[X_\hal{2}, X_\hbe{2}\bigr] &= -{\bigl(X_\hal{2}\bigr)_\hbe{2}}^\ga{2} X_\ga{2} = 
- {\bigl(\cQ_\hal{2}\bigr)_\hbe{2}}^\hga{2} X_\hga{2}  \,, \\
\bigl[X_{\tal{2}''}, X_{\tbe{2}''}\bigr] &= - {\bigl(X_{\tal{2}''}\bigr)_{\tbe{2}''}}^\ga{2} X_\ga{2} = 
- {\bigl(\cQ_{\tal{2}''}\bigr)_{\tbe{2}''}}^\tga{2} X_\tga{2}  \,.
\end{aligned}\end{equation}
Moreover \eqref{eq:AdSconditionsQP} implies that \(\cQ_{\tal{2}''}\) cannot have any hatted indices and thus
\begin{equation}\label{eq:HRHmatcomm}
\bigl[X_{\tal{2}''}, X_\hal{2}\bigr] = - {\bigl(X_{\tal{2}''}\bigr)_\hal{2}}^\be{2} X_\be{2} = 
0 \,.
\end{equation}
(Note that equations \eqref{eq:backgroundQ} - \eqref{eq:HRHmatcomm} are understood to be evaluated in the background.)
Together \eqref{eq:Hcomm} and \eqref{eq:HRHmatcomm} show that \(H^g\) factorizes into two mutually commuting subgroups, i.e.
\begin{equation}
H^g = H^g_R \times H^g_\mathrm{mat} \,,
\end{equation}
 where \(H^g_R\) is generated by \(X_\hal{2}\) and \(H^g_\mathrm{mat} \subseteq H_\mathrm{mat}\) is generated by \(X_{\tal{2}''}\).
Note that even though the Killing vectors \(\cP_{\alpha_2^0}\) vanish in the background they can still generate a nontrivial group \(H^g\).
In particular, the equivariance condition \eqref{eq:equivariance} becomes
\begin{equation}
\bigl[\cQ_{\alpha_2^0}, \cQ_{\beta_2^0}\bigr] = {f_{\alpha_2^0 \beta_2^0}}^{\gamma_2^0} \cQ_{\gamma_2^0} \,,
\end{equation}
and therefore non-vanishing moment maps imply a non-trivial gauge group \(H^g \subseteq H\).
The fact that \(H^g\) is a subgroup of \(H\) and that it is generated by the moment maps \(\cQ_{\alpha_2^0}\) allows us to restrict \(H^g\) further.
The expression \eqref{eq:backgroundQ} for the generators \(X_{\alpha_{2}^0}\) of \(H^g\) in combination with the general property \eqref{eq:Hdelta} of every element of \(\h\) yields
\begin{equation}
{\bigl(X_{\alpha_2^0}\bigr)_{(\be{2}}}^{\delta_2} \delta_{\ga{2})\delta_2} = 0 \,.
\end{equation}
Therefore, an equivalent invariance property must hold true also for the structure constants \({f_{\alpha_2^0 \beta_2^0}}^{\gamma_2^0}\) of the Lie algebra  \(\h^g\) of \(H^g\), i.e.
\begin{equation}
{f_{\alpha_2^0 (\beta_2^0}}^{\delta_2^0} \delta_{\gamma_2^0)\delta_2^0} = 0 \,.
\end{equation}
The presence of the invariant symmetric positive-definite matrix \(\delta_{\al{2}\be{2}}\) implies that \(\h^g\) is reductive, i.e.~that it is the direct sum of an abelian Lie algebra and a semi-simple Lie algebra, and that the semi-simple factors in \(H^g\) are compact, see e.g.~\cite{Weinberg:1996kr} for a proof.

So far we have not included the first equation of \eqref{eq:AdSconditionsQP} into our analysis.
This condition completely determines the commutators \(\bigl[X_\hal{2}, X_\hbe{2}\bigr] = {X_{\hal{2}\hbe{2}}}^\hga{2} X_\hga{2}\) of the generators of \(H^g_R\) via
\begin{equation}\label{eq:HgRcomm}
{X_{\hal{2}\hbe{2}}}^\hga{2} = {\bigl(\cQ_\hal{2}\bigr)_\hbe{2}}^\hga{2} = {\bigl(\cQ^R_\hal{2}\bigr)_\hbe{2}}^\hga{2} \,.
\end{equation}
However, it still leaves some freedom for the embedding of \(H^g_R\) into \(H\) because it does not determine \({X_{\hal{2}\tbe{2}}}^\tga{2} = {\bigl(\cQ_\hal{2}\bigr)_\tbe{2}}^\tga{2}\).
Let us denote the subgroup of \(H_R\) which is generated by \(\cQ^R_\hal{2}\) by \(\hat H^g_R\).
It follows from the equivariance condition \eqref{eq:equivariance} that also \(\bigl[\cQ^R_\hal{2}, \cQ^R_\hbe{2}\bigr] = {X_{\hal{2}\hbe{2}}}^\hga{2} \cQ^R_\hga{2}\).
Therefore \(H^g_R\) and \(\hat H^g_R\) share the same commutator relations and are isomorphic (at least at the level of their Lie algebras).
Nonetheless, as subgroups of \(H\) they do not need to be identical since \(H^g_R\) is not necessarily a subgroup of just \(H_R\) but might be embedded diagonally into \(H = H_R \times H_\mathrm{mat}\).
This is the case if \({X_{\hal{2}\tbe{2}}}^\tga{2}\) is non-vanishing.

Given an explicit expression for the matrices \(B_\hal{2}\) we could now compute \(\cQ^R_\hal{2}\) from the prescription \eqref{eq:AdSconditionsQP} and thus determine \(\hat H^g_R\).
This calculation is demonstrated for a couple of examples in the next section.
However, without any reference to an explicit realization of \(B_\hal{2}\) we can already say a lot about \(H^g_R\) just from the general properties of \(B_\hal{2}\).
In appendix~\ref{app:representationtheory} we show that the \(\cQ^R_\hal{2}\) given by \eqref{eq:AdSconditionsQP} generate a subgroup \(\hat H^g_R \subseteq H_R\) under which \(A_0\) is invariant, i.e.~\(\bigl[\cQ^R_\al{2}, A_0\bigr] = 0\).
To be more specific, let us denote by \(\x\) the maximal subalgebra of \(\h_R\) such that \([\x, A_0] = 0\)
(i.e.~the stabilizer of \(A_0\))
and let us decompose the representation \(\mathbf v\) of \(\h_R\) which  corresponds to the index \(\hal{2}\) into irreducible representations of \(\x\).
The Lie algebra \(\hat\h^g_R\) of \(\hat H^g_R\) must be a subalgebra of \(\x\) such that the adjoint representation of \(\hat\h^g_R\) appears in the decomposition of \(\mathbf v\) into representations of \(\x\).

Let us finally discuss the spontaneous breaking of the gauge group \(G^g\) in the AdS vacuum.
In the background the gauged vielbeins \eqref{eq:gaugedP} read \(\hat \cP = \cP + A^{\tal{2}'} \cP_{\tal{2}'}\).
Inserting this expression into the scalar kinetic term \eqref{eq:scalarvielbeins} produces the mass term
\begin{equation}\label{eq:gaugemass}
\cL_\mathrm{mass} = \tfrac12 \delta_{\al{1}\be{2}}\cP^\al{1}_{\tal{2}'}\cP^\be{1}_{\tbe{2}'} A^{\tal{2}'} \wedge \ast A^{\tbe{2}'} \,.
\end{equation}
Because the \(\cP_{\tal{2}'}\) are linearly independent this generates mass terms for all gauge fields \(A^{\tal{2}'}\), while all the other gauge fields \(A^\hal{2}\) and \(A^{\tal{2}''}\) remain massless.
In other words the mass term \eqref{eq:gaugemass} breaks \(G^g\) spontaneously to \(H^g\), i.e.
\begin{equation}
G^g \rightarrow H^g_R \times H^g_\mathrm{mat} \,.
\end{equation}
This result is physically satisfactory as it shows that the gauge group must be broken to a product of abelian and compact semi-simple subgroups.
Moreover, 
%as discussed in the beginning of this section, 
as mentioned in the introduction,
we can interpret \(H^g_R\) as the R-symmetry group of the holographically dual SCFT and \(H_\mathrm{mat}\) as some additional flavor symmetry.

%The equivariance condition \eqref{eq:} implies that \(\cT_\hal{2}\) and \(\cT_{\tal{2}''}\) generate a subgroup \(H^g \subseteq G^g\).
%Moreover we see from \eqref{eq:} that 
%\begin{equation}\begin{aligned}
%\bigl[\cT_\hal{2}, \cT_\hbe{2}\bigr] &= {\bigl(T_\hal{2}\bigr)_\hbe{2}}^\ga{2} T_\ga{2} = 
%{\bigl(\cT_\hal{2}\bigr)_\hbe{2}}^\hga{2} \cT_\hga{2}  \,. \\
%\bigl[\cT_{\tal{2}''}, \cT_\hal{2}\bigr] &= {\bigl(T_{\tal{2}''}\bigr)_\hal{2}}^\be{2} T_\be{2} = 
%0 \,,
%\end{aligned}\end{equation}
%and therefore \(H^g\) factorises according to 
%\begin{equation}
%H^g = H^g_R \times H^g_\mathrm{mat} \subseteq{H} \,.
%\end{equation}

For theories where the scalar manifold is a symmetric space \(\cM = G/H\) the gauge group \(G^g\) must be a subgroup of \(G\).
The generators of \(G^g\) can be expressed in terms of the T-tensor \(\cT_\al{2}\) \eqref{eq:Ttensor}.
The AdS conditions \eqref{eq:AdSconditionsQP} dictate that they are of the general form
\begin{equation}\begin{aligned}
\cT_\hal{2} &= \cQ^R_\hal{2} + \cQ^\mathrm{mat}_\hal{2} \,, \\
\cT_{\tal{2}'} &=  \cP_{\tal{2}'} + \cQ^\mathrm{mat}_{\tal{2}'} \,, \\
\cT_{\tal{2}''} &=  \cQ^\mathrm{mat}_{\tal{2}''} \,,
\end{aligned}\end{equation}
where we employed our previous split of \(\tal{2}\) into \(\tal{2}'\) and \(\tal{2}''\).
The generators \(\cT_{\tal{2}'}\) can possibly lead to a non-compact or non-reductive gauge group \(G^g\), but according to our previous discussion they are spontaneously broken in the vacuum.

In the next section we will be especially interested in theories where the only multiplet is the gravitational multiplet.
For these theories there is no \(H_\mathrm{mat}\) and no gauge fields \(A^\tal{2}_M\).
Consequently the only generators of \(G^g\) are given by%
\footnote{Notice, that for the four-dimensional \(\cN = 6\) theory there could be in principle an additional generator \(\cT_0 = \cP_0\)
but we show in section~\ref{sec:D4} that \(\cP_0 = 0\).}
\begin{equation}\label{eq:maximalgenerators}
\cT_\hal{2} = \cQ^R_\hal{2} + \cP_\hal{2} = \cQ^R_\hal{2} \,,
\end{equation}
and therefore
\begin{equation}\label{eq:maximalGg}
G^g = H^g = H^g_R \,,
\end{equation}
i.e.~the complete gauge group must be reductive (and in the semi-simple case also compact) and is uniquely determined by the AdS conditions \eqref{eq:AdSconditionsQP}.

Let us finally mention that these results can be straightforwardly translated to maximally supersymmetric Minkowski solutions as well as to maximally supersymmetric solutions with non-trivial flux \cite{Louis:2016tnz}.
Both classes of solutions require not only \(A_1 = 0\) but also \(A_0 = 0\).
This in turn implies via \eqref{eq:AdSconditionsQP} that \(\cQ^R_\al{2} = 0\).
Hence here \(H^g_R\) is trivial.
%Hence all of the above results stay valid with the only exception that now \(H^g_R\) must be trivial.

\subsection{The moduli space}\label{sec:adsmoduli}

We now turn to the moduli spaces of AdS solutions, i.e.~we want to discuss if there are any directions in the scalar field space which are undetermined by the conditions \eqref{eq:adsconditions}.
Let us denote a point in the scalar manifold at which \eqref{eq:adsconditions} is satisfied by \(\left<\phi\right>\) and vary it according to
\begin{equation}\label{eq:phivariation}
\phi = \left<\phi\right> + \delta\phi \,,
\end{equation}
where \(\delta\phi\) is an infinitesimal variation or in other words an infinitesimal tangent vector, i.e.~\(\delta\phi \in T_{\left<\phi\right>}\cM\).
Our goal is to determine if there are any variations \(\delta\phi\) 
%\footnote{In mathematical terms \(\delta\phi\) is an infinitesimal tangent vector, i.e.~\(\delta\phi \in T_{\phi_0}\cM\).} 
under which the AdS conditions \eqref{eq:adsconditions} do not change, i.e.~we are looking for solutions of
\begin{equation}\label{eq:adsmoduli}
\bigl< \partial_{\delta\phi} A^2_0 \bigr> = \bigl< \partial_{\delta\phi} A_1 \bigr> = 0 \,.
\end{equation}
However, the vanishing of the first derivative with respect to \(\delta\phi\) is a priori only a necessary condition for \(\delta\phi\) to be a modulus.
For the existence of a true modulus, i.e.~a continuous deformation parameter of the AdS solution, \(A_0^2\) and \(A_1\) have to be invariant not only under an infinitesimal variation \eqref{eq:phivariation} but also under finite variations.
Equivalently, a modulus is characterized by the vanishing of not only the first derivative with respect to \(\delta\phi\) but also of all higher-order derivatives,
\begin{equation}\label{eq:allordervariation}
\bigl< \partial^n_{\delta\phi} A^2_0 \bigr> = \bigl< \partial^n_{\delta\phi} A_1 \bigr> = 0 \,, \qquad \forall n \geq 1 \,,
\end{equation}
assuming analyticity in \(\phi\).
This resembles the distinction between marginal and exactly marginal deformations of SCFTs.

As mentioned in the discussion below equation \eqref{eq:adsconditions}, the vanishing of \(A_1\) already implies \(A_0^2 \sim \id\).
%It is hence conceivable that the vanishing of the variations of \(A_1\) also implies that the variations of \(A_0^2\) vanish.
Hence it is conceivable that also the vanishing of the variations of \(A_0^2\) is guaranteed by the vanishing of \(A_1\) and its variations.
Indeed, there is a relation of the form \(\cD A_0 \sim A_1\), called gradient flow equation \cite{DAuria:2001rlt}, between the (covariant) derivative of \(A_0\) and the value of \(A_1\).
We rederive the precise form of the gradient flow equation, adopted to our notation, 
%in equation \eqref{eq:gradientflow}
in %the 
appendix~\ref{app:susycalculations}.
It reads
\begin{equation}\label{eq:gradientflow}
\cD_{\al{1}} A_0 = \tfrac{1}{2(D-2)} \bigl(A^\dagger_1 C_{\alpha_1} + C^\dagger_{\alpha_1} A_1 \bigr) \,,
\end{equation}
where \(C_\al{1}\) are the same matrices as in the supersymmetry variations \eqref{eq:cFhat} and \eqref{eq:cFtilde}.
At every point in the scalar manifold where \(A_1 = 0\) we therefore automatically have \(\cD_{\delta\phi} A_0 = 0\) for all variations \(\delta\phi \in T_{\left<\phi\right>} \cM\).
Thus
\begin{equation}
\partial_{\delta\phi} A^2_0 = \cD_{\delta\phi} A^2_0 = (\cD_{\delta\phi} A_0) A_0 + A_0 (\cD_{\delta\phi} A_0) = 0 \,,
\end{equation}
where the replacement of the ordinary derivative of \(A_0^2\) with its covariant derivative is allowed due to \(A_0^2 \sim \id\).
Analogously, the vanishing of all higher-order variations of \(A_1\) implies the vanishing of all higher-order variations of \(A_0^2\), i.e.
\begin{equation}
%\cD^n_{\delta\phi}  A_1 = 
\partial^n_{\delta\phi} A_1 = 0 \,,\quad \forall n \geq 0 \qquad \Rightarrow \qquad \partial^n_{\delta\phi} A_0^2 = 0 \,,\quad \forall n \geq 1 \,.
\end{equation}
It is therefore sufficient to study the variations of \(A_1\).

%Let us shortly
Note that the gradient flow equation \eqref{eq:gradientflow} together with \eqref{eq:generalpotential} also guarantees that every solution of \eqref{eq:adsconditions} is indeed a critical point of the potential \(V\), i.e.
\begin{equation}
\bigl<\partial_{\delta\phi} V \bigr> = 0 \,,\qquad \forall\,  \delta\phi \in T_{\left<\phi\right>}\cM \,,
\end{equation}
and therefore a solution of the equations of motion.

%From \eqref{eq:generalpotential} for the pwe can compute the derivative of \(V\) with respect to an arbitrary \(\delta\phi\),
%\begin{equation}
%\partial_{\delta\phi} V = - \tfrac{2(D-1)(D-2)}{\cN} \tr\left(\partial_{\delta\phi} A_0^2\right) + \tfrac{2}{\cN} \tr\left(A_1^\dagger \partial_{\delta\phi} A_1 \right)
%\end{equation}

Let us temporarily neglect the problem of finding exact solutions \(\delta\phi\) of \eqref{eq:allordervariation} at all orders, but let us for the moment only focus on the leading order variation.
This means we are looking for solutions of
\begin{equation}\label{eq:A1moduli}
\bigl<\cD_{\delta\phi} A_1 \bigr> = \bigl<\partial_{\delta\phi} A_1 \bigr> = 0 \,,
\end{equation}
where \(\cD\) and \(\partial\) can be identified due to \(A_1 = 0\).
If \(\delta\phi\) solves \eqref{eq:A1moduli} it is straightforward to show that
\begin{equation}
\bigl< \partial^2_{\delta\phi} V \bigr> = 0 \,,
\end{equation}
and therefore \(\delta\phi\) corresponds to a massless excitation.
%As discussed in the introduction 
A massless scalar fields gets mapped via the AdS/CFT correspondence to an operator of conformal dimension \(\Delta = d\) on the \(d\)-dimensional boundary SCFT \cite{Gubser:1998bc, Witten:1998qj}. %\Snote{More explanation / ref. / remove?}
This
%again
 illustrates that a solution of \eqref{eq:A1moduli} is dual to a supersymmetric marginal deformation.
On the other hand a solution of \eqref{eq:allordervariation} fulfills \(\bigl<\partial^n_{\delta\phi} V \bigr> = 0\) (\(\forall n \geq 1\)) and thus corresponds to an exactly marginal deformation.

From now on we assume that all derivatives are evaluated at \(\phi = \left<\phi\right>\) and stop indicating this explicitly to simplify the notation.
In the previous section we found that the general AdS conditions \eqref{eq:adsconditions} constrain the background values of the dressed moment maps \(\cQ^R_\al{2}\) and Killing vectors \(\cP_\al{2}\) to be of the form \eqref{eq:AdSconditionsQP}.
Therefore, a solution of \eqref{eq:A1moduli} must necessarily satisfy
\begin{equation}\label{eq:QPmoduli}
\cD_{\delta\phi} \cQ^R_\al{2} = \cD_{\delta\phi} \cP_\hal{2} = 0 \,.
\end{equation}
In many cases gaugings are the only possible deformations of a supergravity and \(A_0\) and \(A_1\) can be expressed exclusively in terms of \(\cQ^R_\al{2}\) and \(\cP_\hal{2}\).
Under these circumstances \eqref{eq:QPmoduli} is also a sufficient condition for \eqref{eq:A1moduli}.
In the remainder of this section we want to assume that this is indeed the case.
However, if there are other contributions to the shift matrices, e.g.~by a non-trivial superpotential, \eqref{eq:A1moduli} and \eqref{eq:QPmoduli} are not equivalent.
%By means of \eqref{eq:QRA0A1} and \eqref{eq:PA1} this in general only implies that
%\begin{equation}\begin{aligned}\label{eq:A1QPmoduli}
%\bigl(\nabla\!_{\delta\phi} A_1^\dagger\bigr) C_\al{1} +  C_\al{1}^\dagger \bigl(\nabla\!_{\delta\phi} A_1\bigr) &= 0 \,, \\
%\bigl(\nabla\!_{\delta\phi} A_1^\dagger\bigr) C_\al{2} +  C_\al{2}^\dagger \bigl(\nabla\!_{\delta\phi} A_1\bigr) &= 0 \,, 
%\end{aligned}\end{equation}
%and not necessarily that \(\nabla\!_{\delta\phi} A_1 = 0\).
%Therefore we can strictly speaking not deduce that \eqref{eq:A1moduli} and \eqref{eq:QPmoduli} are equivalent.
%It might be possible that a solution of \eqref{eq:QPmoduli} is not a solution \eqref{eq:A1moduli}.
%In the remainder we want however concentrate only on those theories for which gaugings are the only possible deformations and for which \(A_0\) and \(A_1\) can be entirely expressed in terms of \(\cQ^R_\al{2}\) and \(\cP_\hal{2}\).
%Under these circumstances \eqref{eq:A1moduli} and \eqref{eq:QPmoduli} are indeed equivalent and it is enough to study the variations of \(\cQ^R_\al{2}\) and \(\cP_\hal{2}\) to determine the moduli space.

In the previous section we have seen that  the gauge group \(G^g\) gets spontaneously broken if there are Killing vectors \(\cP_\al{2}\) with non-vanishing background values. 
According to Goldstones theorem we expect that for each broken generator there exists one massless scalar field, the Goldstone boson.
Indeed, a gauged supergravity theory is constructed in such a way that its action and hence also the potential \(V\) are \(G^g\)-invariant.
The shift matrices \(A_0\) and \(A_1\), however, since they couple to the fermions, are only gauge invariant up to a compensating \(H\)-transformation, described by the \(H\)-compensator \(W_I\) \eqref{eq:W}.
This \(H\)-transformation drops out in the expression for \(V\) in terms of \(A_0\) and \(A_1\) \eqref{eq:generalpotential} due to the involved trace.
Consequently an infinitesimal gauge transformation parametrized by \(\lambda^I\) which acts on the scalar fields as \eqref{eq:scalargaugetransf}
\begin{equation}\label{eq:goldstone}
\delta\phi = \lambda^\al{2} \cP_\al{2} \,,
\end{equation}
is expected to solve \eqref{eq:allordervariation}.
This variation describes one independent solution \(\lambda^{\alpha'_2}\) for each non-vanishing Killing vector \(\cP_{\alpha'_2}\).
Therefore there is one massless scalar field for each spontaneously broken generator of the gauge group \(G^g\).
Nonetheless, these fields cannot be counted as moduli.
As Goldstone bosons of a spontaneously broken gauge symmetry they describe the additional degrees of freedom of the massive gauge fields \(A_M^{\al{2}'}\) and get eaten by the St\"uckelberg mechanism.
%In other words the scalar modes \eqref{eq:goldstone} are pure gauge and therefore non-physical.

%Let us show that each non-vanishing Killing vector indeed gives a solution of \eqref{eq:QPmoduli}.
%We first want to show that for each spontaneously broken generator of the gauge symmetry group \(G^g\) there is a corresponding massless mode \(\delta\phi\) which satisfies \eqref{eq:QPmoduli}.
%The appropriate ansatz for \(\delta\phi\) reads
%\begin{equation}\label{eq:goldstone}
%\delta\phi = \lambda^\al{2} \cP_\al{2} \,,
%\end{equation}
%with \(\lambda^\al{2}\) an infinitesimal parameter.
Let us now explicitly show that \eqref{eq:goldstone} solves \eqref{eq:QPmoduli}.
Before we can compute the variations of \(\cQ^R_\al{2}\) and \(\cP_\al{2}\) with respect to \eqref{eq:goldstone},
we need to determine how the covariant derivative acts on the vielbein \(\cV_\al{2}^I\).
We denote the covariant derivative in a Killing direction by \(\cD_\al{2} = \cP^\al{1}_\al{2} \cD_\al{1}\) and
recall its definition in terms of connection form \(\theta\),
\begin{equation}
\cD_\al{2}\cV^I_\be{2}  = \cL_\al{2} \cV^I_\be{2} + \iota_\al{2} {\theta_\be{2}}^\ga{2} \cV^I_\ga{2} \,,
\end{equation}
where we used the fact that the Lie derivative acts on \(\cV_\al{2}^I\) as an ordinary derivative.
From \eqref{eq:generalvielbeinvariation} and the definition of the moment map \eqref{eq:genmomentmap} we obtain
\begin{equation}\label{eq:genvielbeincovderiv}
\cD_\al{2}\cV^I_\be{2} = \left[- {X_{\al{2}\be{2}}}^\ga{2} + {\bigl(\cQ_\al{2}\bigr)_\be{2}}^\ga{2} \right] \cV^I_\ga{2} \,.
\end{equation}
From this we can compute
\begin{equation}
\cD_\al{2} \cQ^R_\be{2} = \bigl(\cD_\al{2} \cV^I_\be{2}\bigr) \cQ^R_I + \cV^I_\be{2} \bigl(\cD_\al{2} \cQ^R_I\bigr) \,. \\
\end{equation}
Inserting \eqref{eq:genvielbeincovderiv} and the covariant derivative of the moment map \eqref{eq:Qderiv} gives
\begin{equation}\begin{aligned}\label{eq:Qgoldstone}
\cD_\al{2} \cQ^R_\be{2} &= - {X_{\al{2}\be{2}}}^\ga{2} \cQ^R_\ga{2} + {(\cQ_\al{2})_\be{2}}^\ga{2} \cQ^R_\ga{2} + \Omega(\cP_\al{2}, \cP_\be{2}) \\
&= {(\cQ_\al{2})_\be{2}}^\ga{2} \cQ^R_\ga{2} - \bigl[\cQ^R_\al{2}, \cQ^R_\be{2}\bigr] = 0 \,,
\end{aligned}\end{equation}
where we used the equivariance condition \eqref{eq:equivariance}.
In the last step we used that the \(\cQ^R_\al{2}\) span a subalgebra of \(H_R\) with generalized structure constants given by \({(\cQ^R_\al{2})_\be{2}}^\ga{2}\) (compare the discussion below \eqref{eq:HgRcomm}) and that \({(\cQ_\al{2})_\be{2}}^\ga{2} \cQ^R_\ga{2} %= (\cQ_\al{2})_\be{2}^\hga{2} \cQ^R_\hga{2}
 = {(\cQ^R_\al{2})_\be{2}}^\ga{2} \cQ^R_\ga{2}\).
%, which follows from \((\cQ^\mathrm{mat}_\al{2})_\be{2}^\hga{2} = 0\).
In a similar fashion we can also compute the covariant derivative of \(\cP_\hal{2}\) from the covariant derivative of \(\cP_I\) given in \eqref{eq:Pcovderiv},
\begin{equation}\begin{aligned}\label{eq:Pgoldstone}
\cD_\al{2} \cP^\al{1}_\hbe{2} &= \bigl(\cD_\al{2} \cV^I_\hbe{2}\bigr) \cP^\al{1}_I + \cV^I_\hbe{2} \bigl(\cD_\al{2} \cP^\al{1}_I\bigr) \\
%&= - {X_{\al{2}\hbe{2}}}^\ga{2} \cP_\ga{2} + \cV^I_\al{2} \cV^J_\hbe{2} \bigl(\cD_I \cP_J - \cD_J \cP_I\bigr) = 0
&= {(\cQ_\al{2})_\hbe{2}}^\hga{2} \cP^\al{1}_\hga{2} - {(\cQ_\al{2})_\be{1}}^\al{1} \cP^\be{1}_\hbe{2} = 0
 \,.
\end{aligned}\end{equation}
%where we used that \(\cV^J_\hbe{2} \cD_J \cP_I = \cP^\al{1}_\hbe{2} \cD_\al{1} \cP_I= 0\) and that \(\cD_I \cP_J - \cD_J \cP_I = \bigl[\cP_I, \cP_J\bigr] = {X_{IJ}}^K \cP_K\).
Together \eqref{eq:Qgoldstone} and \eqref{eq:Pgoldstone} show that the ansatz \eqref{eq:goldstone} indeed satisfies \eqref{eq:QPmoduli}.
By applying \eqref{eq:Qgoldstone} and \eqref{eq:Pgoldstone} recursively 
%to themselves
one can also show that all higher-order derivatives of \(\cQ^R_\al{2}\) and \(\cP_\hal{2}\) with respect to \eqref{eq:goldstone} vanish.
Note that we inserted the AdS conditions \eqref{eq:AdSconditionsQP} only in the very last step.

We have just seen that the Goldstone bosons appear generically as solutions of \eqref{eq:allordervariation}, however, they do not contribute to the moduli space.
%To find the true moduli space they have to be divided out of the space of solutions of \eqref{eq:allordervariation}.
Here, we do not attempt to find the remaining solutions of \eqref{eq:allordervariation}, which span the moduli space, in a similar general fashion.
This has been achieved explicitly for various theories in \cite{deAlwis:2013jaa,Louis:2014gxa,Louis:2015mka,Louis:2015dca,Louis:2016qca}.
Instead, we only consider theories where the scalar manifold is a symmetric %homogeneous
 space, as introduced in section~\ref{sec:coset}.

If the scalar manifold is a symmetric %homogeneous
space \(\cM = G/H\),
it is most convenient to parametrize the scalar variation \(\delta\phi\) in terms of the corresponding \(\k\) valued quantity \(\cP_{\delta\phi}\), defined as
\begin{equation}
\cP_{\delta\phi} = \iota_{\delta\phi} \cP \in \k \,.
\end{equation}
To compute the (covariant) variations of the general AdS conditions \eqref{eq:AdSconditionsQP} it is necessary to determine the variations of the moment maps \(\cQ_I\) and Killing vectors \(\cP_I\) as well as of the vielbeins \(\cV^I_\al{2}\).
From \eqref{eq:dressedgenerator} we infer that in the coset case \(\cQ_I\) and \(\cP_I\) are given by the \(\h\)-components and the \(\k\)-components of the dressed gauge group generators \(\cT_I\).
Applying \eqref{eq:cosetreprderiv} to the definition \eqref{eq:dressedgenerator} of \(\cT_I\) yields
\begin{equation}
\cD_{\delta\phi} \cT_I = \bigl[\cT_I, \cP_{\delta\phi}\bigr] \,,
\end{equation}
and after splitting this into an \(\h\)-part and a \(\k\)-part one obtains
\begin{equation}\label{eq:PQvariation}
\cD_{\delta\phi} \cQ_I = \bigl[\cP_I, \cP_{\delta\phi}\bigr] \,,\qquad \cD_{\delta\phi} \cP_I = \bigl[\cQ_I, \cP_{\delta\phi}\bigr] \,.
\end{equation}
On the other hand, as discussed in the last paragraph of section~\ref{sec:coset}, the vielbeins \(\cV^\al{2}_I\) are given by the coset representative \(L\) expressed in the appropriate representations.
Analogously \(\cV_\al{2}^I\) is given by the inverse vielbein \(L^{-1}\).
Hence its covariant derivative takes the same form as the covariant derivative of \(L^{-1}\) and is according to \eqref{eq:cosetreprderiv} given by
\begin{equation}\label{eq:cVvariation}
\cD_{\delta\phi} \cV^I_\al{2} = - {\bigl(\cP_{\delta\phi}\bigr)_\al{2}}^\be{2} \cV^I_\be{2} \,,
\end{equation}
where \(\bigl({\cP_{\delta\phi}\bigr)_\al{2}}^\be{2}\) denotes \(\bigl(\cP_{\delta\phi}\bigr)\) expressed in the \(\h\)-representation of the dressed vector fields (i.e.~the representation which is labeled by the index \(\al{2}\)).

After this preparation we are in the position to analyze the general conditions \eqref{eq:QPmoduli}.
With \eqref{eq:PQvariation} and \eqref{eq:cVvariation} they read
\begin{equation}\begin{aligned}\label{eq:cosetmoduli}
\cD_{\delta\phi} \cQ^R_\al{2} &= - {(\cP_{\delta\phi})_\al{2}}^\be{2} \cQ^R_\be{2} + \bigl[\cP_\al{2}, \cP_{\delta\phi}\bigr]\!^R = 0 \,, \\
\cD_{\delta\phi} \cP_\hal{2} &= 
- {(\cP_{\delta\phi})_\hal{2}}^\be{2} \cP_\be{2} + 
\bigl[\cQ_\hal{2}, \cP_{\delta\phi}\bigr] = 0 \,,
\end{aligned}\end{equation}
where
the superscript \(R\)
%\((\,\cdot\,)^R\) 
denotes the projection of an \(\h\)-valued quantity onto \(\h_R\).
To proceed we recall that it follows from \eqref{eq:reductive} that \(\k\) transforms in some representation of \(\h\) with respect to the adjoint action.
We can therefore decompose \(\k\) intro irreducible representations \(\k_i\) of the subalgebra \(\h^g_R\) of \(\h\), i.e.
\begin{equation}\label{eq:kdecomp}
\k = \bigoplus_{i = 1, \dots, N} \k_i \,,\qquad [\h^g_R, \k_i] \subseteq \k_i \,.
\end{equation}
Let us denote the set of all solutions of \eqref{eq:cosetmoduli} by \(\f\), i.e.
\begin{equation}
\f = \left\{\cP_{\delta\phi} \in \k : \cD_{\delta\phi} \cQ^R_\al{2} = \cD_{\delta\phi} \cP_\hal{2} = 0 \right\} \,.
\end{equation}
It follows directly from \eqref{eq:cosetmoduli} that for \(\cP_{\delta\phi} \in \f\) also \([\cQ_\hal{2}, \cP_{\delta\phi}] \in \f\) and therefore
\begin{equation}\label{eq:fdecomp}
\f = \bigoplus_{i \in I} \k_i \,,\qquad I \subseteq \left\{1,\dots,N\right\} \,,
\end{equation}
i.e.~if \eqref{eq:cosetmoduli} is satisfied by one element of some irreducible \(\h^g_R\)-representation, it holds for all elements of this representation.

Let us furthermore introduce 
\begin{equation}\label{eq:kg}
\k^g = \mathrm{span}(\cP_\al{2}) \,,
\end{equation}
i.e.~the projection of the Lie algebra \(\g^g\) of \(G^g\) onto \(\k\).
According to our previous considerations \(\k^g\) corresponds to the Goldstone bosons of the spontaneous symmetry breaking \(G^g \rightarrow H^g\).
Therefore, \(\k^g\) must always be contained in the set of solutions \(\f\), which can be seen directly by inserting \(\cP_{\delta\phi} = \cP_\al{2}\) into \eqref{eq:cosetmoduli}.
Also \(\k^g\) is a \(\h^g_R\) representation (not necessarily an irreducible one) in the above sense and hence
\begin{equation}\label{eq:fsplit}
\f = \k^g \oplus \k_{AdS} \,,
\end{equation}
where \(\k_{AdS}\) spans the non-trivial solutions of \eqref{eq:cosetmoduli} and therefore the candidates for supersymmetric moduli.
The second condition of \eqref{eq:cosetmoduli} implies \(\bigl[\cQ_\hal{2}, \cP_{\delta\phi}\bigr] \subseteq \k^g
\) or equivalently
\begin{equation}
[\h^g_R, \f] \subseteq \k^g \,.
\end{equation}
According to \eqref{eq:kdecomp} this is only possible for two \(\h^g_R\)-representations: \(\k^g_R\) itself and the singlets which commute with \(\h^g_R\).
Hence, we deduce
\begin{equation}\label{eq:singlets}
[\h^g_R, \k_{AdS}] = 0 \,.
\end{equation}
Consequently, all moduli must necessarily commute with \(\h^g_R\) or in other words they must be singlets with respect to the adjoint action of \(\h^g_R\).
%This is often a strong statement and can highly
Depending on the supergravity at hand, this condition can significantly
constrain the existence of a moduli space.
Moreover, finding singlets in the branching of a Lie algebra representation into irreducible representations of a subalgebra is a very well understood problem.

Using this result the conditions on supersymmetric moduli \eqref{eq:cosetmoduli} can be simplified even further.
In terms of the generators \(\cQ_\hal{2}\) of \(\h^g_R\) equation \eqref{eq:singlets} reads
\begin{equation}
\bigl[\cQ_\hal{2}, \cP_{\delta\phi}] = 0 \,.
\end{equation}
Inserting this back into \eqref{eq:cosetmoduli} gives
\begin{equation}
{(\cP_{\delta\phi})_\hal{2}}^\be{2} \cP_\be{2} = 0 \,,
\end{equation}
and using the split of the index \(\tal{2}\) into \(\tal{2}'\) and \(\tal{2}''\) introduced in section~\ref{sec:adsgaugings} we obtain
\begin{equation}\label{eq:Pdeltaphi1}
{(\cP_{\delta\phi})_\hal{2}}^{\tbe{2}'} = 0 \,.
\end{equation}
On the other hand, we infer from the first equation in \eqref{eq:cosetmoduli} that
\begin{equation}\label{eq:Pdeltaphi2}
{(\cP_{\delta\phi})_{\tal{2}''}}^\be{2} \cQ^R_\be{2} = 0 \,.
\end{equation}
We show in appendix~\ref{app:symP} that \({(\cP_{\delta\phi})_\hal{2}}^\tbe{2}\) is symmetric in its indices, i.e.
\begin{equation}
{(\cP_{\delta\phi})_\hal{2}}^\tbe{2} = \delta_{\hal{2}\hde{2}} \delta^{\tbe{2}\tga{2}} {(\cP_{\delta\phi})_\tga{2}}^\hde{2} \,.
\end{equation}
Applying this relation to \eqref{eq:Pdeltaphi1} and \eqref{eq:Pdeltaphi2} we find
\begin{equation}
{(\cP_{\delta\phi})_{\tal{2}}}^\be{2} \cQ^R_\be{2} = 0 \,.
\end{equation}
Therefore, we find the following set of conditions on the supersymmetric moduli \(\k_{AdS}\),
\begin{equation}\begin{gathered}\label{eq:cosetmoduli2}
{(\cP_{\delta\phi})_{\al{2}}}^\be{2} \cQ^R_\be{2} = {(\cP_{\delta\phi})_\hal{2}}^\be{2} \cP_\be{2} = 0 \,, \\
\bigl[\cQ_\hal{2}, \cP_{\delta\phi}] = \bigl[\cP_\al{2}, \cP_{\delta\phi}]^R = 0 \,.
\end{gathered}\end{equation}
These conditions are usually simpler to analyze than the original conditions \eqref{eq:cosetmoduli} and will serve as the starting point for most of our further discussions.

%However, we want to stress that \eqref{eq:singlets} is in general only a necessary condition.
%Not all \(\h^g_R\) singlets in the decomposition \eqref{eq:kdecomp} are necessarily elements of \(\k_{AdS}\), one must furthermore check that they satisfy the first equation in \eqref{eq:cosetmoduli} as well as \((\cP_{\delta\phi})_\hal{2}^\be{2} \cP_\be{2} = 0\).

However, a priori it is not clear that \(\k_{AdS}\) really describes the moduli space of the AdS solution, since we only checked for the vanishing of the first derivatives.
A simple sufficient condition for a solution \(\cP_{\delta\phi} \in \k_{AdS}\) of \eqref{eq:cosetmoduli} or \eqref{eq:cosetmoduli2} to be a true modulus is that it keeps all generators \(\cT_\al{2}\) of the gauge group \(G^g\) invariant, i.e.
\begin{equation}\label{eq:Tmoduli}
\cD_{\delta\phi} \cT_\al{2} = - {(\cP_{\delta\phi})_\al{2}}^\be{2} \cT_\be{2} + \bigl[\cT_\al{2}, \cP_{\delta\phi}\bigr] = 0 \,,
\end{equation}
and not only \(\cQ^R_\al{2}\) and \(\cP_\hal{2}\) as in \eqref{eq:cosetmoduli}.
Due to the linear action of the covariant derivative \(\cD_{\delta\phi}\) all higher-order covariant derivatives of \(\cT_\al{2}\) vanish if the first derivative \eqref{eq:Tmoduli} vanishes.
Moreover, we show in appendix~\ref{app:Tmoduli} that if all elements of \(\k_{AdS}\) satisfy \eqref{eq:Tmoduli} the moduli space is a symmetric %homogeneous
space as well.
This means, that we can find a subalgebra \(\h_{AdS}\) of \(\h\) such that \(\g_{AdS} = \h_{AdS} \oplus \k_{AdS}\) is a subalgebra of \(\g\).
\(\g_{AdS}\) and \(\h_{AdS}\) in turn generate subgroups \(G_{AdS} \subseteq G\) and \(H_{AdS} \subseteq H\) and the moduli space is given by
\begin{equation}\label{eq:symmetricmodulispace}
\cM_{AdS} = \frac{G_{AdS}}{H_{AdS}} \,,
\end{equation}
which is symmetric because \(\g_{AdS}\) inherits the properties \eqref{eq:reductive} and \eqref{eq:symmetric} from \(\g\).

Let us discuss the implications of the general conditions \eqref{eq:cosetmoduli2} for different theories with specific numbers of supersymmetries.
We begin with four and five-dimensional theories with \(q = 8\) real supercharges (i.e.~$\cN =2$ supergravities).
A general discussion of their AdS vacua and the corresponding moduli spaces can be found in \cite{deAlwis:2013jaa, Louis:2016qca}.
The scalar field manifold \(\cM\) of such theories factorizes into the product
\begin{equation}\label{eq:N2scalarmanifold}
\cM = \cM_V \times \cM_H \,,
\end{equation}
where \(\cM_V\) is spanned by the scalar fields in vector multiplets and \(\cM_H\) is spanned by the scalar fields in hyper multiplets.
We denote the former by \(\phi_V\) and the latter by \(\phi_H\).
The geometry of \(\cM_V\) depends on the space-time dimension, \(\cM_H\) on the other hand is in both cases a quaternionic K\"ahler manifold.
Generically \(\cM_V\) and \(\cM_H\) are not necessarily symmetric but there exist many symmetric manifolds of the form \(G/H\) which describe viable scalar geometries for such theories.
In these cases it is possible to use our previous results to determine the moduli space of an AdS solution.

Note that for \(\cN = 2\) theories the gauge fields \(A^\al{2}\) are non-trivial sections only over the first factor \(\cM_V\) in \eqref{eq:N2scalarmanifold} and do not depend on \(\cM_H\).
Therefore, also the variation matrix \({(\cP_{\delta\phi})_\al{2}}^\be{2}\) acting on \(\cV^\al{2}_I\) depends only on the variation of the vector multiplet scalars \(\delta\phi_V \in T_{\left<\phi\right>} \cM_V\).
This implies that the first line of \eqref{eq:cosetmoduli2} is completely independent of \(\cM_H\) and only restricts \(\delta\phi_V\).
In the following we analyze the condition \({(\cP_{\delta\phi})_{\al{2}}}^\be{2} \cQ^R_\be{2} = 0\) separately for the two cases \(D=4\) and \(D=5\) and show that it determines \(\delta\phi_V\) completely, irrespective of the specific choice of \(\cM_V\) or the gauge group \(G^g\).

In five dimensions there is one (real) graviphoton field \(A^{\hal{2} = 0}\).
According to \eqref{eq:AdSconditionsQP} the corresponding moment map \(\cQ^R_{\hal{2} = 0}\) needs to be non-vanishing and generates the gauged R-symmetry group \(H^g_R = \U(1)\), see also the discussion in the following section.
Therefore, \eqref{eq:cosetmoduli2} implies that
\begin{equation}
{(\cP_{\delta\phi})_{\al{2}}}^0 = 0 \,.
\end{equation}
Moreover, we compute in appendix~\ref{app:symP} that \({(\cP_{\delta\phi})_{\tal{2}}}^0\) can be expressed directly in terms of the variation \(\delta \phi^{\al{1}}_V\) of the scalar fields on \(\cM_V\), see \eqref{eq:D5Pdeltaphi},
\begin{equation}
{(\cP_{\delta\phi})_{\tal{2} = \al{1}}}^0 = - \sqrt{\tfrac23} \delta_{\al{1}\be{1}} \delta\phi^\be{1}_V \,,
\end{equation} 
and hence
\begin{equation}
\delta \phi^{\al{1}}_V = 0 \,.
\end{equation}

In four dimensions the situation is similar, however, here the dressed graviphoton \(A^0\) is complex due to electric-magnetic duality.
We denote its complex conjugate by \(A^{\bar 0}\) and let the index \(\hal{2}\) take the values \(0\) and \(\bar 0\).
Therefore, we only have
\begin{equation}\label{eq:D4variationV}
{(\cP_{\delta\phi})_{\al{2}}}^0 \cQ^R_0 + {(\cP_{\delta\phi})_{\al{2}}}^{\bar 0} \cQ^R_{\bar 0} = 0 \,,
\end{equation}
where \(\cQ^R_{\bar 0}\) denotes the complex conjugate of \(\cQ^R_0\), which -- as in five dimensions -- has to be non-vanishing.
Moreover, \(\cM_V\) is a complex Manifold (to be precise a special K\"ahler manifold), so it is possible to describe the variation \(\delta\phi_V\) by a complex vector \(\delta\phi_V^\al{1}\) and its complex conjugate \(\delta\bar\phi_V^{\bar\alpha_1}\).
Inserting the explicit expressions \eqref{eq:D4Pdeltaphi} for \({(\cP_{\delta\phi})_{\al{2}}}^0\) and \({(\cP_{\delta\phi})_{\al{2}}}^{\bar 0}\) into \eqref{eq:D4variationV} gives
\begin{equation}
\delta\phi^\al{1}_V \cQ^R_0 = \delta\bar\phi^{\bar\alpha_1}_V \cQ^R_{\bar 0} = 0 \,,
\end{equation}
which in turn implies the vanishing of \(\delta\phi_V\).

As well in four as in five dimensions the variations of the vector multiplet scalars \(\delta\phi_V\) must vanish.
Therefore the geometry of \(\cM_V\) is not directly relevant for the structure of the moduli space.
It only restricts the possible gauge groups  to be contained in the isometry group of \(\cM_V\).
Consequently, a non-trivial moduli space \(\cM_{AdS}\) can be spanned only by scalar fields in hyper multiplets, i.e.
\begin{equation}
\cM_{AdS} \subseteq \cM_H \,,
\end{equation}
and is determined by the conditions in the second line of \eqref{eq:cosetmoduli2}.
The details of this computation will depend on the choice of a symmetric quaternionic K\"ahler manifold \(\cM_H\) and the gauge group \(G^g\).

For half-maximal supergravities (\(q = 16\)) the scalar manifold is given by the symmetric coset space%
\footnote{This is not true for the chiral theories in six and ten dimensions. However, these theories do not allow for supersymmetric AdS solutions.}
\begin{equation}\label{eq:halfmaximalcoset}
\cM = \frac{G^\ast}{H^\ast} \times \frac{\SO(10-D, n)}{\SO(10-D) \times \SO(n)} \,,
\end{equation}
where \(n\) denotes the number of vector multiplets.
In most cases \(G^\ast\) is given by \(\SO(1,1)\), only in four dimensions it is given by \(\SU(1,1)\). %, while for the chiral theories in six and ten dimensions \(G^\ast\) is trivial.
\(H^\ast\) is the maximal compact subgroup of \(G^\ast\), so in four dimensions \(H^\ast = \U(1)\) and in all other cases it is trivial.
The gauge fields transform in the vector representation of \(\SO(10-D, n)\) and also non-trivially with respect to \(G^\ast\).
Only in five dimensions there is an additional gauge field transforming as a singlet with respect to \(\SO(10-D, n)\).
Moreover, all scalar fields are either part of the gravity multiplet or of vector multiplets.
Therefore \({(P_{\delta\phi})_\al{2}}^\be{2}\) depends on the variation of all scalar fields in \(\cM\), in contrast to theories with \(q = 8\) supercharges.
For this reason the first condition in \eqref{eq:cosetmoduli2},
\begin{equation}\label{eq:halfmaximalcondition}
{(\cP_{\delta\phi})_{\al{2}}}^\be{2} \cQ^R_\be{2} = 0 \,,
\end{equation}
is particularly strong and often constraints the existence of supersymmetric moduli considerably.
The group \(G^\ast\) does not mix fields from different multiplets, therefore variations in the first factor \(G^\ast / H^\ast\) of \eqref{eq:halfmaximalcoset} contribute only to \({(\cP_{\delta\phi})_\hal{2}}^\hbe{2}\) and \({(\cP_{\delta\phi})_\tal{2}}^\tbe{2}\).
On the other hand, variations in the second factor of \eqref{eq:halfmaximalcoset} give rise only to \({(\cP_{\delta\phi})_\hal{2}}^\tbe{2}\) and \({(\cP_{\delta\phi})_\tal{2}}^\hbe{2}\).
For this reason the condition \({(\cP_{\delta\phi})_{\hal{2}}}^\hbe{2} \cQ^R_\hbe{2}\) enforces all variations in \(G^\ast / H^\ast\) to vanish, as we will illustrate in the next section for a concrete example.
Consequently a possible moduli space can only be a submanifold of the second factor of \eqref{eq:halfmaximalcoset}.

%Let us denote the real (or in \(D=4\) complex) scalar field spanning \(G^\ast / H^\ast\) by \(\phi^\ast\).
%Since \(G^\ast / H^\ast\) is real (complex) one-dimensional \({(\cP_{\delta\phi})_\hal{2}}^\hbe{2}\) must be of the form
%\begin{equation}
%{(\cP_{\delta\phi})_\hal{2}}^\hbe{2} = \delta\phi^\ast \, {P_\hal{2}}^\hbe{2} \,,
%\end{equation}
%where \({P_\hal{2}}^\hbe{2}\) is a constant matrix.
%An equivalent relation holds for \({(\cP_{\delta\phi})_\tal{2}}^\tbe{2}\).

In four dimensions there is also a supergravity theory with \(q = 12\) real supercharges.
The scalar manifold of this theory is given by
\begin{equation}
\cM = \frac{\SU(3,n)}{\mathrm{S}[\U(3) \times \U(n)]} \,,
\end{equation}
where \(n\) again denotes the number of vector multiplets.
The gauge fields arrange themselves into the complex vector representation of \(\SU(3,n)\).
The analysis of the moduli space is very similar to the half-maximal case.
In the next section we show explicitly that \eqref{eq:halfmaximalcondition} enforces the moduli space to be trivial.

Let us finally turn our attention to supergravities with more than 16 real supercharges, which thus have the gravitational multiplet as their only supermultiplet.
%only the gravity multiplet and vector multiplets.
%We argue in appendix~\ref{app:} that for these theories the conditions \(\nabla\!_{\delta\phi} \cQ^R_{\tal{2}} = 0\) and \(\nabla\!_{\delta\phi} \cP_{\hal{2}} = 0\) are in fact equivalent.
%Therefore \eqref{eq:cosetmoduli} simplifies to  
For these theories the conditions \eqref{eq:cosetmoduli} simplify considerably and become
\begin{equation}\begin{aligned}\label{eq:maximalmoduli}
\cD_{\delta\phi} \cQ^R_\hal{2} &= - {(\cP_{\delta\phi})_\hal{2}}^\hbe{2} \cQ^R_\hbe{2} = 0 \,, \\
\cD_{\delta\phi} \cP_\hal{2} &= %- (\cP_{\delta\phi})_\hal{2}^\be{2} \cP_\be{2} + 
\bigl[\cQ^R_\hal{2}, \cP_{\delta\phi}\bigr] = 0 \,.
\end{aligned}\end{equation}
Moreover, here the only generators of the gauge group are \(\cT_\hal{2} = \cQ^R_\hal{2} + \cP_\hal{2}\), see \eqref{eq:maximalgenerators}.
Therefore, \eqref{eq:maximalmoduli} is equivalent to \eqref{eq:Tmoduli} which shows that all solutions of \eqref{eq:maximalmoduli} are moduli and that the moduli space is a symmetric 
%homogeneous
space of the form \eqref{eq:symmetricmodulispace}.

To make this a bit more specific we note that there is no spontaneous symmetric breaking due to the vanishing of all Killing vectors \(\cP_{\hal{2}}\) in the background.
This is consistent with the observation \eqref{eq:maximalGg} that the entire gauge group is only given by \(H^g_R\).
Therefore we do not have to worry about possible Goldstone bosons and \(\k_{AdS}\) comprises all solutions of \eqref{eq:maximalmoduli}, i.e.
\begin{equation}\label{eq:kads}
\k_{AdS} = \left\{\cP \in \k : \bigl[ \cP, \cQ^R_\hal{2}\bigr] = {\cP_\hal{2}}^\hbe{2} \cQ^R_\hbe{2} = 0 \right\} \,.
\end{equation}
To extend this to a proper subalgebra of \(\g\) we define
\begin{equation}\label{eq:hads}
\h_{AdS} = \left\{\cQ \in \h : \bigl[ \cQ, \cQ^R_\hal{2}\bigr] = {\cQ_\hal{2}}^\hbe{2} \cQ^R_\hbe{2} = 0 \right\} \,,
\end{equation}
and \(\g_{AdS} = \k_{AdS} \oplus \h_{AdS}\).
%Remember, that the absence of generators \(\cT_{\tal{2}}\) implies that the term \(\k^g\) in \eqref{eq:fsplit} is not present here, i.e.~since there is no spontaneous symmetry breaking there are also no Goldstone bosons.
It is straightforward to show that \(\h_{AdS}\) and \(\g_{AdS}\) are subalgebras of \(\h\) and \(\g\), respectively, i.e.~they are closed with respect to the Lie bracket.
Consequently \(\k_{AdS}\) corresponds to the tangent space of the coset manifold \(\cM_{AdS} = G_{AdS}/H_{AdS}\).
We illustrate our techniques in the next section and compute the AdS moduli spaces for all theories with more than 16 supercharges explicitly.

\section{AdS solutions with \texorpdfstring{$q > 16$}{q > 16} supercharges}\label{sec:examples}

In this section we apply our previous general results to theories with more than 16 supercharges.
These theories all have symmetric scalar field spaces and are characterized by the absence of any other multiplets than the gravity multiplet.

At first we need to determine which theories allow for maximally supersymmetric AdS solutions at all.
It is well-known from \cite{Nahm:1977tg} that the corresponding AdS superalgebras exist only in certain dimensions and for limited numbers of supercharges.
Consequently, one expects that only those theories where an AdS superalgebra exists can be gauged in such a way that a (maximally supersymmetric) AdS solution is possible.
Maximally supersymmetric AdS backgrounds are characterized by the general condition \eqref{eq:adsconditions}, this means that we have to identify theories which allow for \(A_0 \neq 0\) but \(A_1 = 0\). Note that \(A_1 = 0\) is already enough to ensure \((A_0)^2 \sim \id\), which is necessary for unbroken supersymmetry.

This task simplifies a lot if the scalar manifold is a symmetric 
%homogeneous
space of the form \(\cM = G/H\).
As described in section~\ref{sec:coset}, in this case the gaugings can be conveniently described in terms of the T-tensor \(\cT\) \eqref{eq:Ttensor}, which is a scalar field dependent object with a well-defined transformation behavior under \(H\).
Moreover, the shift matrices \(A_0\) and \(A_1\) are built from the appropriate \(H\)-irreducible components of \(\cT\).
%Moreover, when we decompose it into irreducible representations of \(H\) some of its components correspond to the shift matrices \(A_0\) and \(A_1\).
In table~\ref{tab:A0A1} we explicitly list which irreducible components of \(\cT\) correspond to \(A_0\) and \(A_1\). %for supergravities with \(q > 16\) and \(D \geq 4\).
Due to its \(H\)-invariance the condition \(A_1 = 0\) implies that every irreducible component of \(\cT\) which is present in \(A_1\) must vanish identically.
Therefore, \(A_0 \neq 0\) is only possible if there is an irreducible component of \(\cT\) which is part of \(A_0\) but not of \(A_1\).%
\footnote{The situation is slightly more subtle if there are two independent components of \(\cT\) which are both transforming in the same representation.
If both of them are part of \(A_1\) it is possible that only a certain linear combination of them is set equal to zero.
A second linear combination that might be part of \(A_0\) might still be non-vanishing.
However, if we consider these two different linear combinations as independent irreducible representations our argumentation is still valid.}
Inspection of table~\ref{tab:A0A1} shows that for more than 16 real supercharges this is only possible in dimensions \(D = 4, 5\) and \(7\).\footnote{Note that we restrict the discussion to \(D \geq 4\).}

\begin{table}[htb]
\centering
\begin{tabular}{|c|c|c|c|c|c|c|}
\hline
$D$ & $q$ & $H = H_R$ & $A_0$ & $A_1$ & Ref. &  AdS$_D$ \\
\hline
11 & 32 & - & - & - & & \\
\hline
10 & $(32,0)$ & $\U(1)$ & - & - & & \\
& $(16,16)$ & -  & $\mathbf{1}_m$ & $\mathbf{1}_m$ & \cite{Romans:1985tz} & \\
\hline
9 & 32 & $\U(1)$ & $\mathbf{0} \oplus \mathbf{1}_a$ & $\mathbf{0} \oplus \mathbf{1}_a \oplus \mathbf{1}_b$ & \cite{Bergshoeff:2002nv, FernandezMelgarejo:2011wx} &  \\
\hline 
8 & 32 & $\U(2)$ & $\mathbf{1_{+1}}$ & $\mathbf{1_{+1}} \oplus \mathbf{3_{+1}} \oplus \mathbf{5_{+1}}$ & \cite{Bergshoeff:2003ri, deRoo:2011fa} & \\ 
\hline
7 & 32 & $\USp(4)$ & $\mathbf 1 \oplus \mathbf 5$ & $\mathbf{5} \oplus \mathbf{14} \oplus \mathbf{35}$ & \cite{Samtleben:2005bp} & \checkmark \\
\hline
6 & (16,16) & $\USp(4) \times \USp(4)$ & $(\mathbf{4},\mathbf{4})$ & $(\mathbf{4},\mathbf{4}) \oplus (\mathbf{4},\mathbf{16}) \oplus (\mathbf{16},\mathbf{4})$ & \cite{Bergshoeff:2007ef} & \\
& (16,8) & $\USp(4) \times \USp(2)$ & $(\mathbf{4},\mathbf{2})_a$ & $(\mathbf{4},\mathbf{2})_a \oplus (\mathbf{4},\mathbf{2})_b \oplus (\mathbf{16},\mathbf{2})$ & \cite{Roest:2009sn} & \\
\hline
5 & 32 & $\USp(8)$ & $\mathbf{36}$ & $\mathbf{315}$ & \cite{deWit:2004nw} & \checkmark \\
& 24 & $\USp(6)$ & $\mathbf{21}_a$ & $\mathbf{14} \oplus \mathbf{21}_b \oplus \mathbf{70}$ & &  \checkmark \\
\hline
4 & 32 & $\SU(8)$ & $\mathbf{36}$ & $\mathbf{420}$ & \cite{deWit:2007kvg} & \checkmark \\
& 24 & $\U(6)$ & $\mathbf{21_{+1}}$ & $\mathbf{15_{+1}} \oplus \mathbf{35_{-3}} \oplus \mathbf{105_{+1}}$ & \cite{Andrianopoli:2008ea, Roest:2009sn} &  \checkmark \\
& 20 & $\U(5)$ & $\mathbf{15_{+1}}$ & $\mathbf{\overline{5}_{-3}} \oplus \mathbf{10_{+1}} \oplus \mathbf{\overline{40}_{+1}}$ & \cite{Trigiante:2016mnt} & \checkmark \\
\hline
\end{tabular}
\caption{Deformations of supergravities with \(q >16\) and \(D \geq 4\).
The last column indicates whether a maximally supersymmetric AdS solution is possible.
A subscript ``$m$'' denotes a massive deformation.
If there are multiple independent deformations transforming in the same \(H_R\) representation they are distinguished by the subscripts ``$a$'' and ``$b$''.
\(\mathbf{1}_m\) denotes the massive deformation of type IIA supergravity.
Integer subscripts denote the respective \(\U(1)\) charges.
%For $D=9$ and $D=6$, $q = (16,8)$ all components of \(A_1\) have to vanish independently.
}
\label{tab:A0A1}
\end{table}

In the following we want to analyze the gaugings which can lead to AdS solutions and the respective moduli spaces for the allowed theories from table~\ref{tab:A0A1} explicitly.
The first step consists in finding the subgroup \(H^g_R \subset H_R\) which is generated by the moment maps \(\cQ_\hal{2}\).
\(H^g_R\) is a subgroup of \(H_R\) which leaves \(A_0\) invariant and which is gaugeable by the graviphotons. 
We will see in the examples that it is always the maximal such subgroup of \(H_R\).
We determine \(H^g_R\) in a case-by-case analysis for the dimensions \(D = 4,5\) and \(7\) separately und verify the results using the explicit formula \eqref{eq:AdSconditionsQP} for the moment maps \(\cQ^R_\hal{2}\).
We want to stress that the results for \(H^g_R\) are universal and not restricted to theories with \(q > 16\).
However, if \(q > 16\) the only possible multiplet is the gravitational multiplet and there can be no other gauge fields than the graviphotons.
Therefore, as explained in section~\ref{sec:adsgaugings}, the gauge group \(G^g\) must be reductive and is uniquely fixed by \(G^g = H^g_R\).
 
The knowledge of the gauge group \(H^g_R\) finally allows us to determine the moduli spaces of the AdS solutions.
One of the key results of section~\ref{sec:adsmoduli} is that moduli must necessarily be uncharged with respect to \(H^g_R\).
As explained in section~\ref{sec:coset} the Lie algebra \(\g\) of \(G\) splits into the Lie algebra \(\h\) of \(H\) and its orthogonal complement \(\k\).
%Thus an analogous decomposition holds for the adjoint representation of \(\g\).
It is \(\k\) which corresponds to the non-compact directions of \(G\) and therefore to the physical scalar fields.
Moreover, \(\h\) and \(\k\) satisfy \(\left[\h,\k\right] \subseteq \k\) so \(\k\) transforms in an \(\h\)-representation with respect to the adjoint action.
As \(\h^g_R\) is a subalgebra of \(\h\) we can decompose \(\k\)
% (or better the respective \(\h\)-representation)
 into irreducible representations of \(\h^g_R\).
We have seen that only the singlets in this decomposition are candidates for moduli.

We summarize the results for \(H^g_R\) and the relevant decompositions in table~\ref{tab:adsdecomp}.
It shows that the only theory with \(\h^g_R\)-singlets in the decomposition of \(\k\) is the five-dimensional maximal (i.e. \(\cN = 8\) or \(q = 32\)) supergravity.
We argue in due course that the corresponding scalar fields are indeed moduli.
The absence of singlets shows that all the other theories cannot have a non-trivial moduli space.

\begin{table}[htb]
\centering
{\tabulinesep=0.7mm
\begin{tabu}{|c|c|c|c|c|c|}
\hline
$D$ & $q$ & $G/H$ & $H^g_R$ & $\g \rightarrow \h \oplus \k$ & $\k \rightarrow \bigoplus \k_i$  \\
\hline
7 & 32 & 
$\frac{\SL(5)}{\SO(5)}$ 
%$\SL(5)/\SO(5)$
& $\SO(5)$ &$\mathbf{24} \rightarrow \mathbf{10} \oplus \mathbf{14}$ & $\mathbf{14} \rightarrow 
\mathbf{14}$ \\
\hline
5 & 32 & $\frac{\E_{6(6)}}{\USp(8)}$ & $\SU(4)$ & $\mathbf{78} \rightarrow \mathbf{36} \oplus \mathbf{42}$ & $\mathbf{42} \rightarrow 
2 \cdot \mathbf{1} \oplus \mathbf{10} \oplus \mathbf{\overline{10}} \oplus \mathbf{20'}$\\
& 24 & $\frac{\SU^*(6)}{\USp(6)}$ & $\U(3)$ & $\mathbf{35} \rightarrow \mathbf{21} \oplus \mathbf{14}$ & $\mathbf{14} \rightarrow
\mathbf{3_{-1}} \oplus \mathbf{\overline{3}_{+1}} \oplus \mathbf{8_0}$ \\
\hline
4 & 32 & $\frac{\E_{7(7)}}{\SU(8)}$ & $\SO(8)$ & $\mathbf{133} \rightarrow \mathbf{63} \oplus \mathbf{70}$ & $\mathbf{70} \rightarrow \mathbf{70}$ \\
& 24  & $\frac{\SO^*(12)}{\U(6)}$ & $\SO(6)$ & $\mathbf{66} \rightarrow \mathbf{1_0} \oplus \mathbf{35_0} \oplus \mathbf{15_1} \oplus \mathbf{\overline{15}_{-1}}$ & $\mathbf{15_1} \oplus \mathbf{\overline{15}_{-1}} \rightarrow 
2 \cdot \mathbf{15}$ \\
& 20 & $\frac{\SU(5,1)}{\U(5)}$ & $\SO(5)$ & $\mathbf{35} \rightarrow \mathbf{1_0} \oplus \mathbf{24_0} \oplus \mathbf{5_1} \oplus \mathbf{\overline{5}_{-1}}$ & $\mathbf{5_1} \oplus \mathbf{\overline{5}_{-1}}\rightarrow 
2 \cdot \mathbf{5}$ \\
\hline
\end{tabu}}
\caption{Relevant representation theoretical decompositions for the determination of AdS moduli spaces.
Firstly, the branching of the adjoint representation of \(\g\) into \(\h\)-representations and secondly the branching of the \(\h\)-representation corresponding to \(\k\) into representations of \(\h^g_R\).}
\label{tab:adsdecomp}
\end{table}

In the following we discuss each of the three dimensions \(D=4,5,\) and \(7\) independently.
For each case we demonstrate how to explicitly compute the gauge group \(G^g = H^g_R\) using the general formula \eqref{eq:AdSconditionsQP}.
%The results are collected in table~\ref{tab:adsdecomp}.
Moreover, for the maximal five-dimensional theory we show that the two singlets in the decomposition of \(\k\) are indeed moduli and compute the corresponding moduli space.

Let us shortly outline our strategy for determining the gauge algebra \(\h^g_R\) and the moduli space \(\cM_\mathrm{AdS}\):
\begin{enumerate}
\item Find the maximal subalgebra \(\x \subseteq \h_R\) such that \(\bigl[\x, A_0\bigr] = 0\), i.e.~\(\x\) is the stabilizer of \(A_0\) in \(\h_R\), and decompose the graviphotons \(A^\hal{2}\) into irreducible representations with respect to \(\x\).
%\item Decompose the graviphotons \(A^\hal{2}\) into irreducible representations with respect to \(\x\).
\item The adjoint representation of the gauge algebra \(\h^g_R\) must be contained in this decomposition.
The result can be confirmed explicitly using \eqref{eq:AdSconditionsQP}.
\item Decompose the scalar fields \(\k\) into representations of \(\h^g_R\) (see table~\ref{tab:adsdecomp}). The singlets are candidates for moduli.
\end{enumerate}

\subsection{Four-dimensional AdS solutions}\label{sec:D4}

The R-symmetry group of a four-dimensional supergravity with \(q = 4 \cN\) real supercharges is
\begin{equation}\label{eq:d4ralgebra}
H_R = \begin{cases}
\U(\cN) & \text{if}\; \cN \neq 8 \\
\SU(\cN) & \text{if}\; \cN = 8 
\end{cases} \,,
\end{equation}
where \(\cN\) is the number of chiral supersymmetry parameters \(\epsilon^i_+ = \Gamma_\ast \epsilon^i_+\).
Their charge conjugates \(\epsilon_{-i} = (\epsilon^i_+)^C\) have opposite chirality, i.e.~\(\epsilon_{-i} = -\Gamma_\ast \epsilon_{-i}\).\footnote{Our spinor conventions are outlined in appendix~\ref{app:conventions}.}
\(\mathrm{(S)}\U(\cN)\) indices are raised and lowered by complex conjugation.
We summarize some properties of four-dimensional supergravities in appendix~\ref{app:sugras}.

%The chiral and antichiral gravitini as well as therefore infinitesimal supersymmetry parameters \(\epsilon\) transform under this group in the \(\mathbf N \oplus \mathbf{\overline N}\) representation.
%It is convenient to arrange them in column vectors as 
%\begin{equation}
%\psi_M = \begin{pmatrix} \psi^i_M \\ \psi_{M i} \end{pmatrix} \qquad\text{and}\qquad \epsilon = \begin{pmatrix}\epsilon^i \\ \epsilon_i \end{pmatrix} \,.
%\end{equation}
%In this notation the matrix gravitino shift matrix \(A_0\) takes the form
%\begin{equation}\label{eq:D4A}
%A_0 = \begin{pmatrix} 0 & A^{ij} \\ A_{ij} & 0 \end{pmatrix} \,,
%\end{equation}
%where \(A_{ij}\) is symmetric and \(A^{ij} = (A_{ij})^\ast\). \Snote{Add note on fermions and complex conjugation.}

As outlined there, the shift matrix \((A_0)_{ij} = \bigl((A_0)^{ij}\bigr)^\ast\) is a symmetric matrix. %, i.e.~\((A_0)_{ij} = (A_0)_{(ij)}\).
The condition \eqref{eq:adsconditions} on maximally supersymmetric AdS vacua reads
\begin{equation}
(A_0)_{ik} (A_0)^{kj} = - \frac{\Lambda}{12} \delta_i^j \,.
\end{equation}
It implies that all eigenvalues \(\lambda_i\) of \(A_{ij}\) satisfy \(\left|\lambda_i\right| = \sqrt{\frac{\Lambda}{12}}\) , but they can in principle differ by a complex phase.
As outlined above we need to find the stabilizer algebra of \(A_0\) in \(\h^g_R\), i.e.~the maximal subalgebra \(\x \in \h^g_R\) commuting with \(A_0\).
As explained for example in \cite{ORaifeartaigh:1986agb}, there is always an element \(U \in \SU(\cN)\) such that
%The maximal subgroup of \(H_R\) under which \(A_{ij}\) (and therefore also \(A_0\)) is invariant is  \(\SO(\cN)\), as for example shown in \cite{}.
%To see this one notes that there is always a \(U \in \SU(\cN)\) such that
\begin{equation}\label{eq:orthogonalA}
(A_0)_{kl} U^k_i U^l_j = e^{i\omega} \sqrt{\frac{\Lambda}{12}} \delta_{ij} \,,
\end{equation}
i.e.~it is possible to align the phases of all eigenvalues of \(A_0\) by a special unitary transformation.
If \(H_R = \U(\cN)\) we can perform an additional \(\U(1)\) rotation to remove the overall phase factor \(e^{i\omega}\) as well.
However, this is not possible if \(H_R\) is only \(\SU(\cN)\).
\eqref{eq:orthogonalA} is invariant with respect to orthogonal transformations and therefore
\begin{equation}
\x = \so(\cN) \,.
\end{equation}

Next we decompose the dressed graviphotons \(A^\hal{2}\) into irreducible representations of \(\x\).
They are given by \(A^{[ij]}_M\) and their complex conjugates \(A_{M[ij]} = (A^{[ij]}_M)^\ast\).
Both transform in the same way with respect to \(\x = \so(\cN)\), namely in the antisymmetric tensor representation.
This is at the same time also the adjoint representation of \(\so(\cN)\),
so we expect the gauged R-symmetry algebra to be given by \(\h^g_R = \so(\cN)\).
For the \(\cN = 6\) theory there is an additional graviphoton \(A^0_M\), transforming as an R-symmetry singlet.
However, there is no generator of \(\x\) left which could be gauged by \(A^0_M\).
%\footnote{Nonetheless, \(A^0\) can still generate an independent \(U(1)\)-gauge symmetry which does not correspond to an isometry of \(\cM\) \cite{Borghese:2014gfa}.}

%To finally determine the gauged R-symmetry group \(H^g_R\) we also have to understand the branching of the \(\h_g\) representation \(\mathbf v\) of the dressed graviphotons \(A^\hal{2}_M\) into irreducible representations of \(\so(\cN)\).
%For every four-dimensional supergravity the dressed graviphotons \(A^\hal{2}_M\) are given by \(A^{[ij]}_M\) and \(A_{M[ij]}\), satisfying the pseudo-reality condition \((A_{M[ij]})^\ast = A^{[ij]}_M\).
%The presence of both of these representations is due to electric-magnetic duality. \Snote{Formulate more precisely...}.
%Morevoer, for \(\cN = 6\) there is an additional R-singlet \(A^0_M\), which is not present in all the other theories.

To compute the generators \(\cQ^R_\hal{2}\) of \(\h^g_R\) explicitly, using the general formula \eqref{eq:AdSconditionsQP}, it is necessary to combine the fundamental and anti-fundamental representation of \(\su(\cN)\) into a column vector, e.g.~\(\epsilon^i = (\epsilon^i_+, \epsilon_{-i})^T\), see also appendix~\ref{app:sugras}.
Analogously we arrange \((A_0)_{ij}\) and \((A^0)^{ij}\) into a \((2\cN) \times (2\cN)\) matrix as
\begin{equation}\label{eq:D4A0}
A_0 = \begin{pmatrix} 0 & (A_0)^{ij} \\ (A_0)_{ij} & 0 \end{pmatrix} \,.
\end{equation}
Inserting \eqref{eq:D4A0} together with the explicit expression for \(B_\hal{2}\) given in \eqref{eq:appD4B} and \eqref{eq:appD4Bb}
% know the explicit form of the matrices \(B_\hal{2}\).
%In four dimensions they are given by
%\begin{equation}\label{eq:D4B}
%B_{ij} = \begin{pmatrix} 0 & (B_{ij})^{kl} \ \\ 0 & 0 \end{pmatrix} \qquad \text{and} \qquad
%B^{ij} = \begin{pmatrix} 0 & 0 \ \\ -(B^{ij})_{kl} & 0 \end{pmatrix} \,,
%\end{equation}
%with 
%\begin{equation}\label{eq:D4Bb}
%(B_{ij})^{kl} = \tfrac{1}{\sqrt{2}} \delta^{kl}_{ij} \,.
%\end{equation}
%The additional matrix \(B_{0}\) which we need to formulate the \(\cN = 6\) theory vanishes identically,% 
%\footnote{This can be e.g. seen from the truncation of the maximal \(\cN = 8\) theory to \(\cN = 6\).
%To perform this  truncation one first breaks \(\SU(8)\) to \(\U(6) \times \SU(2)\) and then keeps only those fields which are singlets under \(SU(2)\).
%The additional vector field \(A^0_M\) corresponds then to \(A^{[78]}_M\) and therefore \((B_0)^{ij} = (B_{78})^{ij} = 0\) for \(i,j = 1, \dots, 6\).
%In a similar fashion one can also determine the numerical coefficient in \eqref{eq:D4Bb}.
%It has to be the same for all \(\cN\) and can therefore be fixed for \(\cN =2\) by \eqref{eq:BCanticom} since here \(C_\hal{2} = 0\).} i.e.
%\begin{equation}
%B_{0} = 0 \,.
%\end{equation}
into \eqref{eq:AdSconditionsQP} yields
\begin{equation}\label{eq:D4Q}
\cQ^R_{ij} \equiv \begin{pmatrix}{\bigl(\cQ^R_{ij}\bigr)^k}_l & \bigl(\cQ^R_{ij}\bigr)^{kl} \\ \bigl(\cQ^R_{ij}\bigr)_{kl} & {\bigl(\cQ^R_{ij}\bigr)_k}^l \end{pmatrix} 
= \frac{1}{\sqrt{2}}\begin{pmatrix}\delta^k_{[i} (A_0)_{j]l} & 0 \\ 0 & - \delta^k_{[i} (A_0)_{j]l} \end{pmatrix} \,,
\end{equation}
and an analogous result for \(\cQ^{R\,ij}\).
%Moreover, for \(\cN = 6\) there is the additional generator \(\cQ_0\) coupling to \(A^0_M\), but inserting  \eqref{eq:appD4B0} into \eqref{eq:AdSconditionsQP} gives \(\cQ_0 = 0\).
After diagonalizing \((A_0)_{ij}\) and \((A_0)^{ij}\) by an \(\SU(\cN)\) transformation \eqref{eq:orthogonalA}
we find from \eqref{eq:D4Q} the following generators of the gauged R-symmetry \(\h^g_R\),
\begin{equation}\label{eq:D4Qresult}
{(\cQ^R_{ij})^k}_l = -{(\cQ^R_{ij})_l}^k = e^{i \omega} \sqrt{\frac{|\Lambda|}{24}} \delta^k_{[i} \delta_{j]l} \,,\qquad \cQ^{R\,ij} = \bigl(\cQ^R_{ij}\bigr)^\ast \,.
\end{equation} 
We recognize the generators of \(\so(\cN)\). Therefore, for all four-dimensional theories the gauged R-symmetry is indeed given by
\begin{equation}
H^g_R = \SO(\cN) \,.
\end{equation}
We want to point out again that for \(\cN \neq 8\) we can use the left-over \(\U(1)\) freedom to remove the complex phase \(e^{i\omega}\).
For \(\cN = 8\), however, this is not possible and \(\omega\) parametrizes a family of inequivalent \(\SO(8)\)-gaugings, known as \(\omega\)-deformations \cite{DallAgata:2012mfj}.%
\footnote{See \cite{Borghese:2014gfa} for a discussion of \(\omega\)-deformations in \(\cN = 6\) supergravity.}

Let us finally discuss the role of the additional gauge field \(A^\tal{2}_M = A^0_M\) in the \(\cN = 6\) theory, which could in principle gauge another isometry
generated by
\begin{equation}
\cT_0 = \cQ^R_0 + \cP_0 \,.
\end{equation}
Since \(B_0 = 0\) \eqref{eq:appD4B0} it follows directly from \eqref{eq:AdSconditionsQP} that
\begin{equation}
\cQ^R_0 = 0 \,.
\end{equation}
However, for the same reason \eqref{eq:AdSconditionsQP} a priori does not require \(\cP_0 = 0\),
but if we evaluate the commutator between a generator \(\cQ^R_\hal{2}\) of \(H^g_R\) and \(\cP_0\) we find
\begin{equation}\label{eq:N6QP}
\bigl[\cQ^R_\hal{2}, \cP_0\bigr] = \bigl[\cT_\hal{2}, \cT_0\bigr] = {\bigl(\cT_\hal{2}\bigr)_0}^\al{2} \cT_\al{2} = 0 \,,
\end{equation}
since \(A^0\) is uncharged with respect to \(G\).
Moreover, we can read of from table~\ref{tab:adsdecomp} that there are no \(\h^g_R\) singlets in \(\k\).
Therefore, \eqref{eq:N6QP} implies
\begin{equation}
\cP_0 = 0 \,,
\end{equation}
and \(A^0_M\) cannot gauge an isometry of \(\cM\)
Nonetheless, \(A^0_M\) can still generate an independent \(U(1)\)-gauge symmetry which does not correspond to an isometry \cite{Borghese:2014gfa}.

As mentioned above and summarized in table~\ref{tab:adsdecomp} none of the four-dimensional solutions with \(q \geq 16\) admits for \(\h^g_R\) singlets in the decomposition of \(\k\).
Therefore for all three cases the moduli space is trivial.

\subsection{Five-dimensional AdS solutions}\label{sec:D5}

The R-symmetry group of a five-dimensional supergravity with \(q = 4 \cN\) real supercharges is given by
\begin{equation}
H_R = \USp(\cN) \,,
\end{equation}
where \(\cN\) is the number of supersymmetry parameters \(\epsilon^i\) satisfying the symplectic Majorana condition \eqref{eq:symplmajorana}.
The relevant properties of five-dimensional supergravities are summarized in appendix~\ref{app:sugras}.

Firstly, note that in five dimensions the shift matrix \(\left(A_0\right)_{ij} =  \Omega_{ki} A^k_{0\,j}\) is symmetric, see \eqref{eq:appD5A09}.
\(\Omega_{ij}\) is the \(\usp(\cN)\) invariant tensor introduced in \eqref{eq:symplmajorana} which can be used to raise and lower indices.
Moreover, we require \(A_0\) to satisfy the condition \eqref{eq:adsconditions} on maximally supersymmetric AdS vacua which reads
\begin{equation}\label{eq:D5AdS}
A^i_{0\,k} A^k_{0\,j} = \lambda^2 \delta^i_j \,,\qquad \lambda^2 = \frac{\left|\Lambda\right|}{24} \,.
\end{equation}
Let us determine the maximal subalgebra \(\x \subseteq \usp(\cN)\) which commutes with \(A_0\).
For this purpose we note that \eqref{eq:D5AdS} implies together with \(A^i_{0\, i} = A_{0\,ij} \Omega^{ij} = 0\) that
the eigenvalues of \(A^i_{0\,j}\) are given by \(\pm \lambda\), with multiplicity \(\cN/2\) each.
We denote the respective eigenvectors by \(e^i_\alpha\) and \(e^i_{\bar\alpha}\) and introduce \(A_{\alpha\beta} =  e^i_\alpha A_{0\, ij} e^j_\beta\), \(\Omega_{\alpha\beta} = e^i_\alpha \Omega_{ij} e^j_\beta\), ... .
The symmetry of \(A_0\) requires that
\begin{equation}\begin{aligned}
A_{\alpha\beta} &= - \lambda \Omega_{\alpha\beta} = 0 \,, \\
A_{\bar\alpha\bar\beta} &= \lambda \Omega_{\bar\alpha\bar\beta} = 0 \,, \\
A_{\alpha\bar\beta} &= A_{\bar\beta\alpha} = \lambda \Omega_{\alpha\bar\beta} \,.
\end{aligned}\end{equation}
Expressed in this basis \(A_0\) has the form of a hermitian metric which is invariant with respect to unitary transformations and therefore
\begin{equation}\label{eq:D5x}
\x = \u(\cN/2) = \u(1) \oplus \su(\cN/2) \,.
\end{equation}
Working in the eigenbasis of \(A_0\) corresponds to splitting the fundamental representation of \(\usp(\cN)\) labeled by \(i\) into the fundamental and anti-fundamental representation of \(\x\), labeled by \(\alpha\) and \(\bar\alpha\).
Note that it is moreover possible to choose a convenient basis of eigenvectors in which
\begin{equation}
\Omega_{\alpha\bar\beta} = \delta_{\alpha\bar\beta} \,.
\end{equation}

In the next step we have to look at the dressed graviphoton fields \(A^\hal{2}\) given in \eqref{eq:appD5graviphotons}.
In five-dimensional supergravities there generically exist the graviphoton fields \(A^{[ij]}_M\) constrained by the condition \(A^{ij}_M \Omega_{[ij]} = 0\), i.e.~transforming in the traceless antisymmetric tensor representation of \(\usp(\cN)\).
Moreover, for theories with \(\cN \neq 8\) there is an additional graviphoton \(A^0_M\), transforming under \(\usp(\cN)\) as a singlet. 
To understand how these representations branch into representations of \(\x\) we express them in the eigenbasis of \(A_0\).
The \(\usp(\cN)\) singlet \(A^0_M\) stays of course inert under \(\u(\cN)\) and therefore transforms in the adjoint representation of \(\u(1)\).
On the other hand, the vector fields \(A^{[ij]}_M\) decompose as
\begin{equation}\label{eq:ialphabaralpha}
A^{[ij]}_M \rightarrow A^{[\alpha\beta]}_M \oplus A^{[\bar\alpha\bar\beta]}_M \oplus A^{\alpha\bar\beta}_M \,,
\end{equation}
where the last term satisfies \(A_M^{\alpha\bar\beta} \delta_{\alpha\bar\beta} = 0\).
Therefore, the \(A_M^{\alpha\bar\beta}\) transform in the adjoint representation of \(\su(\cN)\).
Consequently, we expect that the gauged R-symmetry algebra \(\h^g_R\) is given by \(\h^g_R = \u(\cN/2)\) if the singlet \(A^0_M\) is present (i.e.~for \(\cN \neq 8\)) and otherwise by \(\h^g_R = \su(\cN/2)\) (i.e.~for \(\cN = 8\)).
%According to the arguments presented in appendix~\ref{app:} we therefore expect \(\h^g_R\) to be \(\u(\cN/2)\) whenever a singlet \(A^0_M\) is present (i.e. for \(\cN \neq 8\)) or \(\h^g_R = \su(\cN/2)\) otherwise (i.e. for \(\cN = 8\)).
%Note that \eqref{eq:ialphabaralpha} indeed satisfies indeed \eqref{eq:maxcriterion}.

Let us explicitly verify this result.
%The matrices \(B_\hal{2}\) are given in \eqref{eq:appD5B}.
%\begin{equation}
%\bigl(B_0\bigr)^k_l = \tfrac i2\sqrt{\tfrac{8-\cN}{2\cN}}\delta^k_l \,,\qquad \bigl(B_{ij}\bigr)^k_l = i \delta^k_{[i} %\Omega_{j]l} + \tfrac i\cN \Omega_{ij} \delta^k_l \,,
%\end{equation}
Inserting the expression \eqref{eq:appD5B} for the matrices \(B_\hal{2}\)  into \eqref{eq:AdSconditionsQP} yields
\begin{equation}\begin{aligned}\label{eq:D5Rgen}
(\cQ^R_0)^k_l &=  2i \sqrt{\tfrac{8-\cN}{2\cN}} (A_0)^k_l \,,\\
 (\cQ^R_{ij})^k_l &= 2i \left((A_0)^k_{[i} \Omega_{j]l} - \delta^k_{[i} (A_0)_{j]l} + \tfrac2\cN \Omega_{ij} (A_0)^k_l \right) \,.
\end{aligned}\end{equation}
These are the generators of \(\h^g_R\).
We want to express them in the basis of eigenvectors of \(A_0\).
The result reads
\begin{equation}\begin{aligned}
(\cQ^R_0)^\gamma_\delta &= 2 i \lambda \sqrt{\tfrac{8-\cN}{2\cN}} \delta^\gamma_\delta \,, \\
(\cQ^R_{\alpha\bar\beta})^\gamma_\delta &= - 2 i \lambda \left(\delta^\gamma_\alpha \delta_{\bar\beta\delta} - \tfrac2\cN \delta_{\alpha\bar\beta} \delta^\gamma_\delta\right) \,,
\end{aligned}\end{equation}
and similarly for \((\cQ^R_0)^{\bar\gamma}_{\bar\delta}\) and \((\cQ^R_{\alpha\bar\beta})^{\bar\gamma}_{\bar\delta}\).
All other components are either determined by antisymmetry or vanish identically.
We recognize that \(\cQ^R_0\) commutes with all other generators and thus spans the abelian algebra \(\u(1)\).
The \(\cQ^R_{\alpha\bar\beta}\) on the other hand are hermitian and traceless and therefore are
 the generators of \(\su(\cN/2)\).
This confirms that the gauged R-symmetry is given by
\begin{equation}
H^g_R = \begin{cases}
\U(\cN/2) & \text{if}\; \cN \neq 8 \\
\SU(\cN/2) & \text{if}\; \cN = 8 
\end{cases} \,.
\end{equation}

The next step is the determination of the moduli space.
The relevant decompositions of the representation \(\k\) of the scalar fields into irreducible representations  of \(\h^g_R\) are summarized in table~\ref{tab:adsdecomp}.
Only for the maximal theory with \(\h^g_R = \su(4)\) there are singlets in the decomposition, which thus is the only theory where a non-trivial moduli space can exist.

%For this purpose we need to find the singlets in the decomposition of \(\k\) into \(\su_4\) irreps.
%Anticipating the result we decompose \(\h\) as well and find:
Let us check that these singlets are indeed moduli and determine the geometry of the manifold they span.
From table~\ref{tab:adsdecomp} we read off that the scalar manifold of the maximal theory is given by 
\begin{equation}
\cM = \frac{\E_{6(6)}}{\USp(8)} \,,
\end{equation}
and that the decomposition of the adjoint representation of \(\mathfrak{e}_{6(6)}\) into representations of \(\usp(8)\) reads \(\mathbf{78} \rightarrow \mathbf{36} \oplus \mathbf{42}\).
The \(\mathbf{36}\) is the adjoint representation of \(\h = \usp(8)\) and the \(\mathbf{42}\) corresponds to \(\k\).
To determine the geometry of the moduli space \(\cM_{AdS}\) (which is a submanifold of \(\cM\)) we decompose both into representations of \(\h^g_R = \su(4)\) and find
%\Snote{check $\mathbf{20}$}
\begin{equation}\begin{aligned}\label{eq:d5n8modulidecomp}
\h &\colon\quad \mathbf{36} \rightarrow \mathbf{1} + \mathbf{10} + \mathbf{\overline{10}} + \mathbf{15}\,, \\
\k &\colon\quad \mathbf{42} \rightarrow 2 \cdot \mathbf{1} + \mathbf{10} + \mathbf{\overline{10}} + \mathbf{20'} \,.
\end{aligned}\end{equation}
%There are two singlets in the decomposition of \(\k\) and therefore the corresponding AdS solution has a two-dimensional moduli space.
Next we determine the algebra \(\g_{AdS}\) spanned by the three singlets in \eqref{eq:d5n8modulidecomp}.
For this purpose we note that the \(\mathbf{36}\) corresponds to a symmetric \(\usp(8)\)-tensor \(\Lambda_{(ij)}\), and that the \(\mathbf{42}\) is given by a completely antisymmetric \(\usp(8)\) 4-tensor \(\Sigma_{[ijkl]}\), constrained by the tracelessness condition \(\Omega^{ij}\Sigma_{ijkl} = 0\) \cite{deWit:2004nw}.
Together \(\Lambda_{ij}\) and \(\Sigma_{ijkl}\) span the adjoint representation of \(\mathfrak{e}_{6(6)}\) and satisfy the commutator relations
\begin{equation}\begin{aligned}\label{eq:e6}
\bigl[\Lambda_{ij}, \Lambda_{kl}\bigr] &= \Omega_{ik} \Lambda_{jl} + \dots \,, \\
\bigl[\Lambda_{ij}, \Sigma_{klmn}\bigr] &= \Omega_{ik} \Sigma_{jlmn} + \dots \,, \\
\bigl[\Sigma_{ijkl}, \Sigma_{mnop}\bigr] &= \Omega_{im} \Omega_{jn} \Omega_{ko} \Lambda_{lp} + \dots \,, 
\end{aligned}\end{equation} 
where the ellipses stand for all terms which need to be added to obtain the correct (anti-) symmetry on the right-hand side.
Moreover, a generator \(T = \lambda^{ij} \Lambda_{ij} + \sigma^{ijkl} \Sigma_{ijkl}\) of \(\mathfrak{e}_{6(6)}\) acts on a tensor \(X_{[ij]}\) in the antisymmetric traceless representation (i.e.~the \(\mathbf{27}\)) of \(\usp(8)\) as \cite{deWit:2004nw}
\begin{equation}\label{eq:usp8action}
(T X)_{ij} = - 2 {\lambda_{[i}}^k X_{j]k} + {\sigma_{ij}}^{kl} X_{kl} \,.
\end{equation}
To reproduce the decomposition \eqref{eq:d5n8modulidecomp} we express \(\Lambda_{ij}\) and \(\Sigma_{ijkl}\) in the eigenbasis of \(A_0\) which was constructed above.
The three singlets are given by
\begin{equation}
\Lambda^0 = \tfrac{1}{4}\delta^{\alpha\bar\beta} \Lambda_{\alpha\bar\beta} \,,\quad
\Sigma^- = \tfrac{1}{4!}\epsilon^{\alpha\beta\gamma\delta} \Sigma_{\alpha\beta\gamma\delta} \,,\quad
\Sigma^+ = \tfrac{1}{4!}\epsilon^{\bar\alpha\bar\beta\bar\gamma\bar\delta} \Sigma_{\bar\alpha\bar\beta\bar\gamma\bar\delta} \,. 
\end{equation}
From \eqref{eq:e6} we find
\begin{equation}
\left[\Lambda^0, \Sigma^\pm\right] = \pm \Sigma^\pm \,,\qquad \left[\Sigma^-, \Sigma^+\right] = \Lambda_0 \,,
\end{equation}
These are the well-known commutator relations of \(\su(1,1)\).
Moreover, from \eqref{eq:usp8action} it follows that \(\Lambda^0\) and \(\Sigma^{\pm}\) indeed satisfy the conditions \eqref{eq:kads} and \eqref{eq:hads} on supersymmetric moduli and therefore
\begin{equation}
\h_{AdS} = \mathrm{span}\bigl(\{\Lambda^0\}\bigr) \,,\qquad \k_{AdS} = \mathrm{span}\bigl(\{\Sigma^-, \Sigma^+\}\bigr) \,,
\end{equation}
and \(\g_{AdS} = \h_{AdS} \oplus \k_{AdS} = \su(1,1)\).
Consequently, the moduli space is given by the coset space
\begin{equation}
\cM_\mathrm{AdS} = \frac{\SU(1,1)}{\U(1)} \,.
\end{equation}

\subsection{Seven-dimensional AdS solutions}\label{sec:D7}

The R-symmetry group of a seven-dimensional supergravity theory with \(q = 8 \cN\) real supercharges is given by
\begin{equation}
H_R = \USp(\cN) \,,
\end{equation}
where \(\cN\) is the number of supersymmetry parameters \(\epsilon^i\) satisfying the symplectic Majorana condition \eqref{eq:symplmajorana}.
We summarize the essential properties of seven-dimensional supergravities in appendix~\ref{app:sugras}.

In seven dimensions the shift matrix \(\left(A_0\right)_{ij}\) is antisymmetric \eqref{eq:appD7A0}.
Hence, in general it decomposes into two irreducible \(\usp_\cN\) representations: The singlet representation (proportional to \(\Omega_{ij}\)) and the antisymmetric traceless representation.
However, as we see from table~\ref{tab:A0A1} the second AdS condition \(A_1 = 0\) enforces the antisymmetric traceless part to vanish\footnote{In \(D=7\) there only exist the \(\cN=4\) and the \(\cN = 2\) theories.
The case \(\cN = 2\) is not contained in table~\ref{tab:A0A1} but an antisymmetric traceless representation of \(\USp(2)\) does not exist.} and therefore
\begin{equation}\label{eq:D7A0AdS}
(A_0)_{ij} = \pm \sqrt\frac{|\Lambda|}{60}\Omega_{ij} \,.
\end{equation}
Consequently, the maximal subalgebra \(\x\) of \(\h_R = \usp(\cN)\) commuting with \(A_0\) is \(\usp(\cN)\) itself, i.e.
\begin{equation}
\x = \usp(\cN)\,.
\end{equation}
Therefore the decomposition of the dressed graviphotons \(A^\hal{2}_M\) into representations of \(\x\) is trivial.
As stated in \eqref{eq:appD7graviphotons} the graviphotons are given by \(A^{(ij)}_M\), i.e.~they transform in the symmetric tensor representation of \(\usp(\cN)\).
This is also its adjoint representation,
so we expect the gauged R-symmetry algebra to be given by \(\h^g_R = \usp(\cN)\).
%To argue that this is indeed the gauged R-symmetry algebra \(h^g_R\) we have to verify that the graviphotons transform in the correct \(\usp(\cN)\) representation.
%Indeed, in seven dimensions the dressed graviphotons \(A^\hal{2}_M\) are given by \(A^{(ij)}_M\), i.e. they transform in the symmetric tensor representation which is the adjoint of \(\usp(\cN)\).
%As moreover the criterion \eqref{eq:} is satisfied we can conclude that
%\begin{equation}
%\h^g_R = \h_g = \usp(\cN) \,.
%\end{equation}
Let us verify this explicitly.
The matrices \(B_\hal{2}\) are given in \eqref{eq:appD7B}.
%\begin{equation}\label{eq:D7B}
%(B_{ij})^k_l = \sqrt{2} \delta^k_{(i} \Omega_{j)l} \,.
%\end{equation} 
%and are therefore nothing but the generators of \(\usp(\cN)\) in the fundamental representation.
Inserting the expression stated there as well as \eqref{eq:D7A0AdS} into \eqref{eq:AdSconditionsQP} gives
\begin{equation}
(\cQ^R_{ij})^k_l = 6 \sqrt\frac{\left|\Lambda\right|}{30} \delta^k_{(i} \Omega_{j)l} \,.
\end{equation}
These indeed are the generators of \(\usp(\cN)\) in the fundamental representation, which
confirms our above result and hence
\begin{equation}
H^g_R = \USp(\cN) \,.
\end{equation}

In seven-dimensions the only supergravity with \(q > 16\) is the maximal \(\cN = 4\) theory.
Also here the decomposition of \(\k\) into irreducible representations of \(\h^g_R\) does not contain any singlets, see table~\ref{tab:adsdecomp}.
This shows that the AdS moduli space is trivial.

\section{AdS solutions in four-dimensional \texorpdfstring{$\cN = 3$}{N = 3} supergravity}\label{sec:N=3}

In this section we discuss the maximally supersymmetric AdS solutions of four-dimensional $\cN = 3$ supergravity.%
\footnote{Aspects of AdS solutions and gaugings of four-dimensional $\cN = 3$ supergravities have also been discussed in \cite{Karndumri:2016fix, Karndumri:2016miq, Karndumri:2016tpf}.}
Contrary to the previously discussed supergravities this theory can be coupled to an arbitrary number of vector multiplets, which makes the situation slightly more complicated.
In particular, the additional vector fields can be used to gauge additional symmetries and thus the gauge group \(G^g\) can be larger than \(H^g_R\) and possibly non-compact.

The discussion of half-maximal supergravities could be performed along similar lines.
Their field content is also given only by the gravity multiplet and vector multiplets.
An explicit analysis of AdS solutions in half-maximal supergravities in dimensions \(D=4,5,6\) and \(7\) can be found in \cite{Louis:2014gxa, Louis:2015dca, Karndumri:2016ruc, Louis:2015mka}.

The scalar manifold of for four-dimensional \(\cN = 3\) supergravity is given by the coset space \cite{Castellani:1985ka}
\begin{equation}
\cM = \frac{\SU(3,n)}{\mathrm{S}[\U(3) \times \U(n)]} \,,
\end{equation}
where \(n\) denotes the number of vector multiplets.
The gauge fields \(A^I_M\) transform in the \((\mathbf{3 + n}) \oplus \overline{(\mathbf{3 + n})}\) representation of \(G = \SU(3,n)\).
Consequently, the dressed gauge fields \(A^\al{2}_M\) transform in the \((\mathbf{3}, \mathbf{1})_{-1} \oplus (\mathbf{1},\overline{\mathbf{n}})_{-3/n}\) representation of \(H = \U(3) \times \SU(n)\), where the subscripts denote the \(\U(1)\) charge.
Moreover, we denote the complex conjugate of \(A^\al{2}_M\) by \(A^{\bar\alpha_2}_M\).

To determine if maximally supersymmetric AdS solutions exist we need to know which of the irreducible \(H\)-representations of the T-tensor appear in the shift matrices \(A_0\) and \(A_1\).
We can read them of from \cite{Trigiante:2016mnt},
\begin{equation}\begin{aligned}
A_0 &\colon (\mathbf{6}, \mathbf{1})_{+1} \,, \\
A_1 &\colon (\overline{\mathbf{3}}, \mathbf{1})_{+1} \oplus (\mathbf{1},\mathbf{n})_{3/n} \oplus (\mathbf{3},\mathbf{n})_{2+3/n} \oplus (\mathbf{8},\mathbf{n})_{3/n} \,.
\end{aligned}\end{equation}
We observe that the \(H\)-representation of \(A_0\) does not appear in \(A_1\) and therefore \(A_0 \neq 0\) and \(A_1 = 0\) is possible.
Consequently, the conditions \eqref{eq:adsconditions} can be solved and maximally supersymmetric AdS solutions exist.
Moreover, the \((\mathbf{6}, \mathbf{1})_{+1}\) representation of \(A_0\) is the symmetric tensor representation of \(\U(3)\), which agrees precisely with our general considerations in section~\ref{sec:D4}.
There, we have determined that this form of \(A_0\) implies that the three moment maps \(\cQ^R_\hal{2}\) generate the gauged R-symmetry group
\begin{equation}
H^g_R = \SO(3) \,.
\end{equation}
Of course, the general gauge group \(G^g \subset \SU(3,n)\) can be much more complicated, however, its precise form is not relevant for our further analysis.

Let us now discuss the moduli spaces of such solutions.
We denote the generators of \(\SU(3,n)\) by \(t_{I \bar J}\).
In the fundamental representation they read
\begin{equation}
{(t_{I \bar J})_K}^L = \eta_{K \bar J} \delta^L_I - \tfrac1{3+n} \eta_{I \bar J} \delta^L_K \,,
\end{equation}
where \(\eta_{I\bar J} = \mathrm{diag}(-1,-1,-1,+1, \dots, +1)\).
According to the splitting \(\su(3,n) \rightarrow \u(3) \oplus \su(n)\) they decompose as
\begin{equation}
t_{I \bar J} \rightarrow t_{\hal{2} \bar{\hat \beta}_2} \oplus t_{\tal{2} \bar{\tilde \beta}_2} \oplus t_{\hal{2} \bar{\tilde \beta}_2} \oplus t_{\tal{2} \bar{\hat \beta}_2} \,.
\end{equation}
The first two terms span the maximally compact subalgebra \(\h =  \u(3) \oplus \su(n)\) and the second two terms span the non-compact part \(\k\) which corresponds to the tangent space of \(\cM\).
Therefore, we can expand the variation matrix \(\cP_{\delta\phi} \in \k\) as
\begin{equation}
\cP_{\delta\phi} = \delta\phi^{\hal{2} \bar{\tilde \beta}_2} t_{\hal{2} \bar{\tilde \beta}_2} +  \delta\phi^{\tal{2} \bar{\hat \beta}_2} t_{\tal{2} \bar{\hat \beta}_2} \,.
\end{equation}
%In particular \( \delta\phi^{\tal{2} \bar{\hat \beta}_2}\) is the complex conjugate of \(\delta\phi^{\hal{2} \bar{\tilde \beta}_2}\).
Inserting this parametrization into \eqref{eq:halfmaximalcondition} yields
\begin{equation}
{(\cP_{\delta\phi})_{\tal{2}}}^\hbe{2} \cQ^R_\hbe{2} = \delta_{\tal{2} \bar{\tilde\gamma}_2 } \delta\phi^{\hbe{2} \bar{\tilde\gamma}_2} \cQ^R_\hbe{2} = 0 \,.
\end{equation}
Moreover, since \(H^g_R = \SO(3)\), all three moment maps \(\cQ^R_\hal{2}\) are non-vanishing and linearly independent, c.f.~also \eqref{eq:D4Qresult}.
Therefore,
\begin{equation}
\delta\phi^{\hal{2} \bar{\tilde\beta}_2} = 0 \,.
\end{equation}
In the same way we infer from \({(\cP_{\delta\phi})_{\bar{\tilde\alpha}_2}}^{\bar{\hat\beta}_2}\cQ^R_{\bar{\hat\beta}_2} = 0\) the vanishing of \(\delta\phi^{\tal{2} \bar{\hat \beta}_2}\).
This shows that the moduli space is trivial.

\section{Conclusions}

In this paper we studied maximally supersymmetric AdS$_D$ vacua of gauged supergravities in dimensions $D\geq4$.
We performed a model independent analysis and focused on properties which are independent of a possible higher-dimensional origin.
Moreover, we described supergravity as much as possible in a universal language which is formally independent of the number of space-time dimensions $D$ and supersymmetries $\cN$.
This allowed us to develop generic properties of such vacua.

Unbroken supersymmetry imposes algebraic conditions \eqref{eq:adsconditions} on the shift matrices \(A_0\) and \(A_1\) which in turn restrict the admissible gauge groups.
We found that the gauge group -- after a possible spontaneous symmetry breaking -- is always of the form \(H^g_R \times H^g_\mathrm{mat}\), where \(H^g_R\) is unambiguously determined by the conditions on \(A_0\) and \(A_1\).
\(H^g_\mathrm{mat}\), on the other hand, is only constrained by the general prescriptions on gauging supergravities and can only exist in the presence of vector multiplets.
Both factors are direct products of abelian and non-compact semi-simple Lie groups, i.e.~reductive.
This is agrees with the structure of the global symmetry groups of SCFTs, where \(H^g_R\) corresponds to the R-symmetry group and \(H^g_\mathrm{mat}\) to a possible flavor symmetry.

Moreover, the conditions on \(A_0\) and \(A_1\) determine at which points of the scalar field space AdS$_D$ solutions exist.
A continuous family of such points corresponds to a non-trivial moduli space of solutions.
Focusing on the special case where the scalar manifold is a symmetric space of the form \(\cM = G/H\) we derived general group theoretical conditions \eqref{eq:cosetmoduli2} on the existence of such a moduli space.

We used these results to discuss the maximally supersymmetric AdS$_D$ solutions of all gauged supergravities with more than 16 real supercharges
and -- as a less supersymmetric example -- of gauged \(\cN = 3\) supergravity in four dimensions.
We explicitly determined their gauge groups and showed that almost all of them do not allow for non-trivial moduli spaces.
The only exception occurs for maximal supergravity in five dimension where the moduli space is given by \(\SU(1,1)/\U(1)\).
These results are in one-to-one agreement with predictions from the AdS/CFT correspondence.
It has been shown in \cite{Cordova:2016xhm} that the dual SCFTs do not admit supersymmetric marginal deformations as well and thus do not have conformal manifolds.
Moreover, the \(\SU(1,1)/\U(1)\) moduli space in five dimensions corresponds to the complex gauge coupling of the dual four-dimensional \(\cN = 4\) super Yang-Mills theory.

Similarly to our analysis of the $\cN =3$ case in four dimensions it should be possible to apply the results of this paper to all gauged half-maximal supergravities. 
These theories are structurally very similar because their field content is given by the gravity multiplet and an arbitrary number of vector multiplets.
However, a detailed discussion of maximally supersymmetric AdS$_D$ solutions of such theories has already been performed in \cite{Louis:2014gxa, Louis:2015dca, Karndumri:2016ruc, Louis:2015mka}.
In addition, there are many examples for symmetric scalar field spaces in gauged supergravities with less supersymmetry, i.e.~for the $\cN = 2$ theories in four and five dimensions \cite{deWit:1995tf,Andrianopoli:1996cm,Bergshoeff:2004kh}.
General results on the moduli spaces of AdS$_D$ solutions of these theories have been obtained in \cite{deAlwis:2013jaa,Louis:2016qca}, but it would be very interesting to work out explicit examples using our results.

Moreover, there are various gauged supergravities in three dimensions which admit maximally supersymmetric AdS$_D$ solutions \cite{deWit:2003ja,Fischbacher:2002fx}.
Unfortunately, these theories are outside the scope of our discussion.
The relevant formulae on the Killing vectors and moment maps \eqref{eq:AdSconditionsQP}  are only valid in dimensions $D \geq 4$.
Therefore, an independent analysis of their AdS$_D$ solutions and the corresponding moduli spaces would be highly desirable.

%%%%%%%%%%%%%%%%%%%%%%%%%%%%%%%%%%%%
\section*{Acknowledgments}
%%%%%%%%%%%%%%%%%%%%%%%%%%%%%%%%%%%%

This work was supported by the German Science Foundation (DFG) under
the Collaborative Research Center (SFB) 676 ``Particles, Strings and the Early
Universe'' and the Research Training Group (RTG) 1670 ``Mathematics
inspired by String Theory and Quantum Field Theory''.
The work of SL was also supported by the ANR grant Black-dS-String.
The work of PR was also supported by a James Watt Scholarship.

We have benefited from conversations with Constantin Muranaka and Marco Zagermann.

%%%%%%%%%%%%%%%%%%%%%%%%%%%%%%%%%%%%%%%%%%%%%%%%%%%%%%%%%%%
\newpage

\appendix
\noindent
{\bf\Huge Appendix}

\section{Conventions and Notations}\label{app:conventions}

In this appendix we summarize the conventions and notations used in this paper.
We mostly follow the sign and spinor conventions of \cite{Freedman:2012zz}. 

\subsection*{Metric}
The space-time metric is mostly positive, i.e.
\(
\eta_{MN} = \mathrm{diag}(-, +, \dots, +)
\).

\subsection*{Indices}
In our description of supergravities we use the following indices:
\vspace{0.3em}
\begin{itemize}
\setlength{\itemsep}{0.3em}
\item space-time: \(M, N,  \ldots \in \{0,1, \dots, D-1\} \)
\item gravitini (R-symmetry): \(i,j, \ldots \in \{1,\dots, \cN\} \)
\item spin-1/2 fermions: \(a,b,\ldots\)
\item scalars: \(r, s, \ldots\)
\item gauge fields: \(I, J, \ldots\)
\item $p$-form field strengths: \(I_p, J_p, \ldots\)
\item dressed $p$-form field strengths: \(\al{p}, \be{p}, \ldots\)
\end{itemize}
\vspace{0.3em}
Moreover, we use hated or a tilded indices to indicate whether a field belongs to the gravity multiplet or any other multiplet, i.e.~the fermions \(\chi^{\hat a}\) belong to the gravity multiplet and \(\chi^{\tilde a}\) to matter multiplets.

%\subsection*{Covariant Derivatives}
%
%\begin{itemize}
%\setlength{\itemsep}{0.3em}
%\item space-time: \(\nabla = \dd - \omega\)
%\item H-covariant derivative: \(\cD\)
%\item gauged H-covariant derivative: \(\hat\cD\)
%\end{itemize}

\subsection*{$\Gamma$-matrices}

The \(D\)-dimensional gamma matrices \(\Gamma^M\) span a Clifford algebra and are defined via their anti-commutation relation
%\Snote{Flat metric \(\eta^{MN}\) vs. curved \(g^{MN}\)?}
\begin{equation}\label{eq:gammadef}
\Gamma^M \Gamma^N + \Gamma^N \Gamma^M = 2 g^{MN} \id \,.
\end{equation}
%In the main text 
Their antisymmetric products 
appear frequently and we abbreviate
\begin{equation}\label{eq:gammaprod}
\Gamma^{M_1\dots M_p} = \Gamma^{[M_1} \dots \Gamma^{M_p]} \,,
\end{equation}
where the antisymmetrization \([\dots]\) is with total weight 1, i.e. \(\Gamma^{MN} = \frac{1}{2}\left(\Gamma^M\Gamma^N - \Gamma^N \Gamma^M\right)\).
In even dimensions \(D = 2m\) we additionally have the chirality operator 
\(\Gamma_\ast\) defined by 
\begin{equation}\label{eq:gamma5}
\Gamma_\ast = (-i)^{m+1} \Gamma_0 \Gamma_1 \dots \Gamma_{D-1} \,,
\end{equation}
which allows us to define projection operators 
\begin{equation}\label{eq:Ppm}
P_{\pm} = \tfrac{1}{2} (\id \pm \Gamma_{\ast}) \,.
\end{equation}
Moreover, one introduces the charge conjugation matrix \(C\) which is defined by the properties
\begin{equation}\label{eq:C}
C^T = - t_{0} C \,,\qquad (\Gamma^M)^T = t_0 t_1 C \Gamma^M C^{-1} \,,
\end{equation}
where \(t_0\) and \(t_1\) are sign factors collected in Table~\ref{tab:spinors}.

\subsection*{Spinors}

For a set of complex spinors \(\epsilon^i\) transforming as a vector in the fundamental representation of the R-symmetry group \(H_R\) we denote the (Dirac) conjugates by \(\bar\epsilon_i\) with a lowered index, i.e.
\begin{equation}
\bar\epsilon_i \equiv (\epsilon^i)^\dagger i \Gamma^0 \,.
\end{equation}
It is convenient to introduce the spinor \(\epsilon_i\) with lowered index as the charge conjugate of \(\epsilon^i\), i.e. \(\epsilon_i = (\epsilon^i)^C\), defined by the relation
\begin{equation}\label{eq:chargeconjugate}
\bar\epsilon_i = (-t_0 t_1) \epsilon_i^T C \,,
\end{equation}
where \(\epsilon_i^T C\) is called the Majorana conjugate of \(\epsilon_i\).
With this notation bilinears of spinors \(\epsilon^i\) and \(\eta^j\) satisfy \cite{Freedman:2012zz}
\begin{equation}\label{eq:bilinearswap}
\bar\epsilon_i \Gamma^{M_1\dots M_p} \eta^j = t_p\, \bar\eta^j \Gamma^{M_1\dots M_p} \epsilon_i \,,
\end{equation}
where \(t_2 = -t_0\), \(t_3 = -t_1\), \(t_{p+4} = t_p\) and \(\bar\eta^j \equiv (-t_1) \eta_i^\dagger i \Gamma^0 = (-t_0 t_1) (\eta^i)^T C\).
This relation is particularly useful if there is a relation between \(\epsilon^i\) and \(\epsilon_i\), i.e.~if the spinors satisfy a (symplectic) Majorana condition.

Applying charge conjugation twice yields \(\bigl((\epsilon^i)^C\bigr){}^C = (- t_1) \epsilon^i\) and according to the sign of \((- t_1)\) we can introduce Majorana or symplectic Majorana spinors.
If \(t_1 = - 1\), the charge conjugation is a strict involution and it is consistent to impose the reality constraint
\begin{equation}\label{eq:majorana}
\epsilon^i = \delta^{ij} \epsilon_j \,,
\end{equation}
with \(\delta^{ij}\) the identity matrix.
A spinor satisfying \eqref{eq:majorana} is called a Majorana spinor and has half as many real degrees of freedom compared with an unconstrained spinor.

If \(t_1 = + 1\), the above Majorana condition would be inconsistent but we can instead impose the symplectic Majorana condition,
\begin{equation}\label{eq:symplmajorana}
\epsilon^i = \Omega^{ij} \epsilon_j \,,
\end{equation}
where \(\Omega^{ij} = (\Omega_{ij})^\ast\) is a non-degenerate antisymmetric matrix satisfying \(\Omega_{ik}\Omega^{jk} = \delta_i^j\).
Note that this condition is only consistent for an even number of spinors \(\epsilon^i\) because otherwise a matrix \(\Omega_{ij}\) with the required properties does not exist.

If not denoted otherwise we always assume spinors to fulfill the (symplectic) Majorana conditions \eqref{eq:majorana} or \eqref{eq:symplmajorana}, respectively.
%in particular if we are not referring to a specific dimension.
The benefit of this choice is that it gives spinor bilinears well-defined reality properties.
For example, symplectic Majorana spinors \(\epsilon^i\) and \(\eta^i\) satisfy
\begin{equation}\label{eq:reality}
\left(\bar \epsilon_i \Gamma^{M_1\dots M_p} \eta^j\right)^\ast = (-t_0 t_1)^{(p+1)} \Omega^{ik} \Omega_{jl} \bar \epsilon_k \Gamma^{M_1\dots M_p} \eta^l \,.
\end{equation}
By replacing \(\Omega_{ij}\) with \(\delta_{ij}\) one obtains the analogous relation for Majorana spinors.
This allows us to easily construct real Lagrangians.
We illustrate this with the example of the gravitino mass term.
% \eqref{eq:appfermionmass}.
Up to a prefactor it is given by
\begin{equation}\label{eq:majoranagravmass}
{(A_0)^i}_j \bar\psi_{iM} \Gamma^{MN} \psi^j_N \,,
\end{equation}
and is real if
\begin{equation}
\left({(A_0)^i}_j \right)^\ast = (-t_0 t_1) {(A_0)_i}^j = (-t_0 t_1) \Omega^{ik} \Omega_{jl} {(A_0)^l}_k \,.
\end{equation}
Consequently, we assume all objects with indices \(i,j,\dots\) to be pseudo real or pseudo imaginary, which means that indices can be raised or lowered by complex conjugation (up to a sign factor).

However, using (symplectic) Majorana conditions can sometimes obscure the action of the R-symmetry, especially in even dimensions where we furthermore can distinguish between left- and right-handed spinors.

If \(D\) is odd (symplectic) Majorana spinors are the only minimal spinor representations.
Note that the Majorana condition \eqref{eq:majorana} is invariant under \(H_R = \SO(\cN)\) transformations, where \(\cN\) denotes the number of spinors \(\epsilon^i\).
The symplectic Majorana condition \eqref{eq:symplmajorana}, on the other hand, is invariant under \(H_R = \USp(\cN)\).

In even dimensions \(D\) the situation is slightly more complicated since here the projectors \(P_{\pm}\) given in \eqref{eq:Ppm} can be used to define chiral or Weyl spinors.
We need to distinguish between two different cases.
Let us first consider the situation where \((\Gamma_\ast \epsilon^i)^C = - \Gamma_\ast (\epsilon^i)^C\) (as well as \(t_1 = - 1\)), which implies that the charge conjugate of a left-handed spinor is right-handed and vice versa.
Therefore, a Majorana spinor cannot have a definite chirality.
Nonetheless, we can decompose \(\epsilon^i\) into its left and right handed component, i.e.
\begin{equation}
\epsilon^i = \epsilon^i_+ + \epsilon_{-i} \,,
\end{equation}
with 
\begin{equation}
\epsilon^i_+ \equiv P_+ \epsilon^i \,,\qquad \epsilon_{-i} \equiv P_- \epsilon^i \,.
\end{equation}
Note that \((\epsilon^i_+)^C = \epsilon_{-i}\), i.e.~the positioning of the indices is consistent with \eqref{eq:chargeconjugate}.
On the other hand this also implies that \(\epsilon^i_+\) and \(\epsilon_{-i}\) do not satisfy the Majorana condition \eqref{eq:majorana} individually.
Consequently, we loose the ability to raise and lower indices with \(\delta_{ij}\).
 %such that \(\epsilon^i_+\) and \(\epsilon_{-i}\) are transforming in the \(\mathbf N\) and \(\mathbf{\overline N}\) representations, respectively.
Moreover, Weyl spinors \(\epsilon^i_+\) do not satisfy the reality property \eqref{eq:reality} anymore.
However, we can still write down a relation similar to the Majorana condition \eqref{eq:majorana} if we replace \(\epsilon^i\) by a column vector \(\epsilon^I\) consisting of \(\epsilon^i_+\) and \(\epsilon_{-i}\), i.e.
\begin{equation}
\epsilon^i \rightarrow \epsilon^I \equiv \begin{pmatrix} \epsilon^i_+ \\ \epsilon_{-i} \end{pmatrix} \,,\qquad \text{and} \qquad \epsilon_i \rightarrow \epsilon_I \equiv (\epsilon^I)^\cc = \begin{pmatrix} \epsilon_{-i} \\ \epsilon^i_+ \end{pmatrix} \,.
\end{equation}
With this notation we have
\begin{equation}
\epsilon^I = \Delta^{IJ} \epsilon_J \,,\qquad\text{where}\qquad \Delta^{IJ} = \begin{pmatrix} 0 & \delta^j_i \\
\delta^i_j & 0 \end{pmatrix} \,,
\end{equation}
which formally resembles \eqref{eq:majorana} or \eqref{eq:symplmajorana}.
The formal replacement of \(\epsilon^i\) by \(\epsilon^I\) (and analogously for all other involved spinors) enables us to convert our general formulae (collected in appendix~\ref{app:susy}) from (symplectic) Majorana spinors to Weyl spinors.

Let us illustrate this with the gravitino mass term.
Using chiral spinors \(\psi^i_{+M}\) and \(\psi_{M-i}\) it reads
\begin{equation}\label{eq:weylgravmass}
(A_0)_{ij} \, \bar\psi^i_{+M} \Gamma^{MN} \psi^j_{+N} + \mathrm{h.c.} =  (A_0)_{ij} \, \bar\psi^i_{M+} \Gamma^{MN} \psi^j_{N+} + (A_0)^{ij} \, \bar\psi_{M-i} \Gamma^{MN} \psi_{N-j} \,,
\end{equation}
where \((A_0)^{ij} = \left((A_0)_{ij}\right)^\ast\).
Note that we stick to our convention that raising and lowering indices is related to complex conjugation.
\eqref{eq:weylgravmass} can be cast into a form equivalent to \eqref{eq:majoranagravmass} by combining \(\psi^i_{M+}\) and \(\psi_{M-i}\) into a column vector, i.e.~\(\psi^I_M = \left(\psi^i_{M+}, \psi_{M-i}\right)^T\) and by introducing
\begin{equation}
%A^i_{0\,j} \rightarrow
 A^I_{0\,J} = \begin{pmatrix} 0 & (A_0)^{ij} \\ (A_0)_{ij} & 0 \end{pmatrix} \,.
\end{equation}
With this notation \eqref{eq:weylgravmass} reads
\begin{equation}
{(A_0)^J}_J \bar\psi_{IM} \Gamma^{MN} \psi^J_N \,,
\end{equation}
which is (after the replacements \(\psi^i_M \rightarrow \psi^I_M\) and \(A^i_{0\,j} \rightarrow A^I_{0\,J}\)) of the same form as \eqref{eq:majoranagravmass}.
We finally want to mention that the Weyl condition \(\epsilon^i_+ = P_+ \epsilon^i_+\) is invariant with respect to \(H_R = (\mathrm{S})\U(\cN)\).

Now we turn to the second case where \((\Gamma_\ast \epsilon^i)^C = \Gamma_\ast (\epsilon^i)^C\). Here one can consistently define (symplectic) Majorana-Weyl spinors.
This means we can have two independent sets of spinors \(\epsilon^i_+\) and \(\epsilon^{i'}_{+}\),
\begin{equation}
P_{\pm} \epsilon^i_\pm = \epsilon^{i'}_\pm \,,
\end{equation}
which individually satisfy \eqref{eq:majorana} or \eqref{eq:symplmajorana}, respectively.
Analogously to the odd-dimensional case we find \(H_R = \SO(\cN_+) \times \SO(\cN_-)\) or \(H_R = \USp(\cN_+) \times \USp(\cN_-)\), where \(\cN_+\) denotes the number of chiral spinors \(\epsilon^i_+\) and \(\cN_-\) the number of anti-chiral spinors \(\epsilon^{i'}_-\).%
\footnote{The notation \(\cN = (\cN_+, \cN_-)\) is also common.}
In this case a sum over the index \(i\) in a general formula is implicitly understood to run over \(i'\) as well, unless stated otherwise.%
\footnote{This prescription can be formalized by replacing \(\epsilon^i\) with \(\epsilon^I = (\epsilon^i_+, \epsilon^{i'}_+)^T\), similarly as in our previous discussion.}
We summarize the irreducible spinor representations together with the compatible R-symmetry groups \(H_R\) for various dimensions in table~\ref{tab:spinors}.

\begin{table}[htb]
\centering
\begin{tabular}{|c|cc|c|c|}
\hline
$D$  (mod 8) & $t_0$ & $t_1$ & irrep. & $H_R$ \\
\hline
3 & $+$ & $-$ & M & $\SO(\cN)$ \\
4 & $+$ & $-$ & M / W & $(\mathrm{S})\U(\cN)$ \\
5 & $+$ & $+$ & S & $\USp(\cN)$ \\
6 & $-$ & $+$ & SW & $\USp(\cN_+) \times \USp(\cN_-)$ \\
7 & $-$ & $+$ & S & $\USp(\cN)$ \\
8 & $-$ & $-$ & M / W & $(\mathrm{S})\U(\cN)$ \\
9 & $-$ & $-$ & M & $\SO(\cN)$ \\
10 & $+$ & $-$ & MW & $\SO(\cN_+) \times \SO(\cN_-)$ \\
\hline
\end{tabular}
\caption{Spinor conventions in various dimensions \cite{Freedman:2012zz}. \(t_0\) and \(t_1\) are the sign factors introduced in \eqref{eq:C}. ``M'' stands for Majorana spinors, ``S'' for symplectic Majorana spinors and ``W'' for Weyl spinors.
In four and eight dimensions one can have either Majorana or Weyl spinors (but not both), while in six and ten dimensions (symplectic) Majorana-Weyl spinors are possible.}
\label{tab:spinors}
\end{table}

\section{Supersymmetry variations}\label{app:susy}

In this appendix we summarize the general form of the supergravity Lagrangian and supersymmetry variations and derive some important relations between the fermionic shift matrices and the Killing vectors and moment maps.

In appendix~\ref{app:susyvariations} we summarize the supersymmetry variations of the fermions and bosons in a general supergravity theory and comment on some of the properties of the involved objects.
In appendix~\ref{app:lagrangian} we review the general form of a supergravity Lagrangian.
In appendix~\ref{app:susycalculations} we compute the Killing vectors \(\cP_I\) and their moment maps \(\cQ^R_I\) in terms of the shift matrices \(A_0\) and \(A_1\).
In appendix~\ref{app:sugras} we give explicit expressions for some of the previously introduced objects in dimensions \(D = 4,5,7\).

\subsection{Supersymmetry variations}\label{app:susyvariations}

In this appendix we collectively present the general form of the supersymmetry variations of the fields present in a (gauged) supergravity theory.
These expressions are universal and not restricted to a specific dimension or number of supercharges.
Moreover, we assume all spinors to satisfy to be (symplectic) Majorana.
See appendix~\ref{app:conventions} for our spinor conventions and for the conversion from Majorana to chiral spinors.

The supersymmetry variations of the bosonic fields read
\begin{subequations}\begin{align}
\delta e^A_M &= \tfrac12 \bar\epsilon_i \Gamma^A \psi^i_M \,,  \label{eq:vielbeinvariation} \\
\delta A^{I_p}_{N_1\dots N_{p-1}} &= \tfrac{p!}{2} \cV^{I_p}_{\alpha_p} 
%\delta^{\alpha_p \beta_p}
\left[\bigl(B^{\alpha_p}\bigr)^i_j \bar \psi_{i[N_1} \Gamma_{N_2 \dots N_{p-1}]} \epsilon^j + \bigl(C^\al{p}\bigr)^a_i \bar\chi_a \Gamma_{N_1 \cdots N_{p-1}} \epsilon^i\right] + \dots \,, \label{eq:Avariation}
\end{align}\end{subequations}
where we have omitted possible terms that depend on the other $p$-form fields and their supersymmetry variations.
The supersymmetry variations of the fermionic fields up to terms of higher order in the fermionic fields are given by
\begin{subequations}\begin{align}
\delta \psi^i_M &= \hat\cD_M  \epsilon^i  + \left(\cF_M\right)^i_j \epsilon^j + A^i_{0\,j} \Gamma_M \epsilon^j + \dots \,, \label{eq:appgravitinovariation} \\
\delta \chi^a &= \cF^a_i \epsilon^i + A^a_{1\,i} \epsilon^i + \dots \,, \label{eq:appspin12variation}
\end{align}\end{subequations}
where \(\hat\cD\) is the covariant derivative introduced in \eqref{eq:gaugedQ}.
The shift matrices \(A_0\) and \(A_1\) generically depend on the scalar fields.
Moreover, we have defined the abbreviations
\begin{equation}\label{eq:appFM}
\big(\cF_M\big)^i_j  =  \tfrac{1}{2(D-2)} \sum_{p \geq 2} 
%\sum_{\hat\alpha_p}
\big(B_{ \hat\alpha_p}\big)^i_j\,
F^{\hat\alpha_p}_{{N_1}\dots {N_p}} {T^{{N_1}\dots {N_p}}}{}_M \,,
\end{equation}
with
\begin{equation}\label{eq:T}
{T^{{N_1}\dots {N_p}}}{}_M = {\Gamma^{{N_1}\dots {N_p}}}{}_M + p\tfrac{D-p-1}{p-1} \Gamma^{[N_1\dots N_{p-1}} \delta^{N_p]}_M \,,
\end{equation}
as well as
\begin{equation}
\cF^{a}_i = \tfrac12 \sum_{p \geq 1} \big(C_{\alpha_p}\big)^{a}_i \, F^{\alpha_p}_{N_1\dots N_p} \Gamma^{N_1\dots N_p} \epsilon^i \,.
\end{equation}
The matrices \(B_\al{p}\) and \(C_\al{p}\) are constant and mediate between the different representations of \(H\) that occur in the theory.
To be more specific, we denote the generators of \(H\) in the respective representations by \({(J_A)_i}^j\), \({(J_A)_a}^b\) and \({(J_A)_\al{p}}^\be{p}\) and demand
\begin{equation}\begin{aligned}\label{eq:appBCproperty}
{\bigl(J_A\bigr)_\al{p}}^\be{p} B_\be{p} &= \bigl[J_A, B_\al{p}\bigr]  \,, \\
{\bigl(J_A\bigr)_\al{p}}^\be{p} \bigl(C_\be{p}\bigr)_i^a &= \bigl(J_A\bigr)_i^j \bigl(C_\al{p}\bigr)^a_j -  \bigl(C_\al{p}\bigr)^b_i {\bigl(J_A\bigr)_b}^a \,.
\end{aligned}\end{equation}
To keep the notation compact we defined
\begin{equation}
B_{\tilde \alpha_p} = B_{\alpha_1} = 0 \,.
\end{equation}
The closure of the supersymmetry algebra imposes a Clifford algebra like condition on \(B_\al{p}\) and \(C_\al{p}\),
%\Snote{Check sign factor.}
\begin{equation}\label{eq:BCanticom}
\frac{(p!)^2}{D-2}\frac{D-p-1}{p-1} \left(B^\dagger_{\alpha_p} B_{\beta_p} + B^\dagger_{\beta_p} B_{\alpha_p}\right) +  (p!)^2 \left(C^\dagger_{\alpha_p} C_{\beta_p} + C^\dagger_{\beta_p} C_{\alpha_p}\right)  = 2 \delta_{\alpha_p \beta_p} \id \,.
\end{equation}

\subsection{The general Lagrangian}\label{app:lagrangian}

In this appendix we state the general Lagrangian of a (gauged) supergravity theory at the two derivative level.
The Lagrangian can be split into a purely bosonic part and a part that also depends on the fermionic fields, i.e.
\begin{equation}
\cL = \cL_B + \cL_F \,,
\end{equation}
The bosonic Lagrangian is already given in \eqref{eq:bosonicaction}, we restate it here for the sake of completeness
\begin{equation}\begin{aligned}\label{eq:appbosoniclagrangian}
e^{-1} \cL_B &= -\frac{R}{2} 
%+ \frac{e}{2} g_{rs}(\phi) D\phi^r \wedge \ast D\phi^s
- \frac{1}{2} \sum_{p \geq 1} M^{(p)}_{I_{p} J_{p}}\!\left(\phi\right)\, F^{I_p} \wedge \ast F^{J_p} - V + e^{-1} \cL_\mathrm{top} \\
&= -\frac{R}{2} 
%+ \frac{e}{2} g_{rs}(\phi) D\phi^r \wedge \ast D\phi^s
- \frac{1}{2} \sum_{p \geq 1} \delta_{\al{p}\be{p}} \, F^\al{p} \wedge \ast F^\be{p}  - V+ e^{-1} \cL_\mathrm{top} \,.
\end{aligned}\end{equation}
Note that we often denote the dressed scalar field strengths \(\hat\cP^\al{1} \equiv F^\al{1}\). 
The scalar potential reads \eqref{eq:generalpotential}
\begin{equation}\label{eq:appgeneralpotential}
V = - \tfrac{2(D-1)(D-2)}{\cN} \tr(A_0^\dagger A_0) + \tfrac{2}{\cN}\tr(A_1^\dagger A_1) \,,
%\delta^i_j V = - 2(D-1)(D-2) \bigl(A^\dagger_0\bigr)^i_k A^k_{0\,j} + 2 \bigl(A_1^\dagger\bigr)^i_a A^a_{1\,j} \,,
\end{equation}
where \(A_0\) and \(A_1\) are the fermionic shift matrices from \eqref{eq:appgravitinovariation} and \eqref{eq:appspin12variation}.
Moreover, there can be a topological term \(\cL_\mathrm{top}\) which does not depend on the space-time metric.

The fermionic Lagrangian (which despite its name in general also depends on the bosonic fields) is of the general form
\begin{equation}
\cL_F = \cL_\mathrm{kin,f} + \cL_\mathrm{pauli} + \cL_\mathrm{mass} + \cO(f^4) \,.
\end{equation}
The kinetic terms of the fermions read
\begin{equation}\label{eq:appkinferm}
e^{-1} \cL_\mathrm{kin,f} = - \frac{1}{2} \bar \psi_{iM} \Gamma^{MNP} \hat\cD_N \psi^i_P - \frac{1}{2} \bar \chi_{a} \Gamma^M \hat\cD_M \chi^a \,,
\end{equation}
where \(\hat\cD\) denotes the gauge covariant derivative given in \eqref{eq:gaugedQ} and \eqref{eq:chigaugedcovderiv}.
Local supersymmetry requires the existence of Pauli-like interaction terms between the $p$-form field strengths \(F^\al{2}\) and the fermions.
They are of the form
\begin{equation}\label{eq:pauli}
\cL_\mathrm{pauli} = \sum_{p \geq 1} \left(\cL^{(p)}_{F\bar\psi\psi} + \cL^{(p)}_{F\bar\chi\psi} + \cL^{(p)}_{F\bar\chi\chi} \right) \,,
\end{equation}
where
\begin{subequations}\begin{align}
e^{-1}\cL^{(p)}_{F\bar\psi\psi} &= - \frac{1}{4(p-1)} F^{\alpha_p}_{M_1 \dots M_p}   \left(B_{\alpha_p}\right)^i_j \bar\psi^N_i \Gamma_{[N} \Gamma^{M_1 \dots M_p} \Gamma_{P]} \psi^{P\,j} \,, \label{eq:apppaulia} \\
e^{-1}\cL^{(p)}_{F\bar\chi\psi} &= \frac12 F^{\alpha_p}_{M_1 \dots M_p} \left(C_{\alpha_p}\right)^a_i \bar\chi_a \Gamma^N \Gamma^{M_1\dots M_p} \psi^i_N \,, \label{eq:apppaulib} \\
e^{-1} \cL^{(p)}_{F\bar\chi\chi} &= \frac12 F^{\alpha_p}_{M_1 \dots M_p} \left(D_{\alpha_p}\right)^a_b \bar\chi_a \Gamma^{M_1\dots M_p} \chi^b \,.
\end{align}\end{subequations}
\(B_\al{p}\) and \(C_\al{p}\) are the same matrices as in the supersymmetry variations \eqref{eq:appgravitinovariation} and \eqref{eq:appspin12variation}.
The matrices \(D_\al{p}\) have similar properties. Their precise form, however, is not relevant for our discussion.
If the theory is gauged (or otherwise deformed) the Lagrangian also includes mass terms for the fermions which read
\begin{equation}\label{eq:appfermionmass}
e^{-1} \cL_\mathrm{mass} = \frac{D-2}{2} A^i_{0\,j} \bar\psi_{iM} \Gamma^{MN} \psi^j_N + A^a_{1\,i} \bar\chi_a \Gamma^M \psi^i_M + M^a_b \bar\chi_a \chi^b \,,
\end{equation}
where \(A_0\) and \(A_1\) are the same matrices as in \eqref{eq:appgravitinovariation} and \eqref{eq:appspin12variation}.
The third mass matrix \(M^a_b\) also depends on the scalar fields and the gaugings/deformations, but it is not relevant for our discussion.
Moreover, the supersymmetric completion of the Lagrangian requires terms of higher order in the fermions which we do not give here.

\subsection{Killing vectors and moment maps}\label{app:susycalculations}

In a supergravity theory the variation of the vielbein \(e^A_M\) of the space-time metric \eqref{eq:vielbeinvariation} induces additional terms in the variation of the sigma model kinetic term in \eqref{eq:appbosoniclagrangian} which are not present in global supersymmetry.
They read
\begin{equation}\label{eq:deltaPP}
\delta \cL_{\hat\cP\hat\cP} = -\frac{e}{2} \delta_{\alpha_1\beta_1} \hat\cP_M^{\alpha_1} \hat\cP_N^{\beta_1} \bar\epsilon_i \left(\tfrac12 g^{MN} \Gamma^P - \Gamma^{(M} g^{N)P} \right) \psi^i_P+ \dots \,.
\end{equation}
These terms are canceled by the Pauli term \eqref{eq:apppaulib} for \(p = 1\).
Indeed, inserting the variation \eqref{eq:appspin12variation} of the spin-$\frac12$ fermions \(\chi^a\) into \eqref{eq:apppaulib} yields
\begin{equation}\label{eq:deltaPchipsi}
e^{-1} \delta\cL^{(1)}_{\hat \cP\bar\chi\psi} = - \frac{1}{2} \hat\cP^{\alpha_1}_M \hat\cP_N^{\beta_1} \bigl(C_{\alpha_1}\bigr)^a_i \bigl(C^\dagger_{\beta_1}\bigr)_a^j \bar \epsilon_j \left(\tfrac12\Gamma^{MNP} - \tfrac12g^{MN} \Gamma^P + \Gamma^{(M} g^{N)P}\right) \psi^i_P  + \dots \,,
\end{equation}
which cancels \eqref{eq:deltaPP} due to \eqref{eq:BCanticom}.
Only the term cubic in the \(\Gamma\)-matrices does not have a counterpart in \eqref{eq:deltaPP}.
This term, however, is canceled by the kinetic term of the gravitini in \eqref{eq:appkinferm}.
From the gravitino variation \eqref{eq:appgravitinovariation} and \eqref{eq:Dcommgauged} we see that its variation contains
\begin{equation}\begin{aligned}\label{eq:psikinvariation}
e^{-1} \delta \cL_{\bar\psi\hat\cD\psi} &= \frac{1}{2}\bigl(\hat\cH^R_{MN}\bigr)^i_j \bar \psi_{iP} \Gamma^{MNP} \epsilon^j +\ldots \,, \\
&= \frac{1}{2}\bigl(\cH^R_{MN} + F^\al{2}_{MN} Q^R_\al{2}\bigr)^i_j \bar \psi_{iP} \Gamma^{MNP} \epsilon^j +\ldots \,,
\end{aligned}\end{equation}
where \(\cH^R_{MN}\) is the field strength of the R-connection \(\cQ^R_M\) \eqref{eq:Dcomm} and \(\cQ^R_\al{2}\) are the generalized moment maps defined in \eqref{eq:genmomentmap} and \eqref{eq:Qderiv}.
Comparing \eqref{eq:psikinvariation} with \eqref{eq:deltaPchipsi} requires
\begin{equation}
\cH^R_{MN} = -\tfrac12 C^\dagger_{\alpha_1} C_{\beta_1} \hat\cP^{\alpha_1}_{[M} \hat\cP^{\beta_1}_{N]} \,.
\end{equation}
However, in a gauged theory we still need to take care of the second term in \eqref{eq:psikinvariation} which contains the 2-form field strengths \(F^\al{2}_{MN}\).
For this purpose we vary the \(p=2\) Pauli terms \eqref{eq:apppaulia} and \eqref{eq:apppaulib} as well as the fermionic mass terms \eqref{eq:appfermionmass}.
The relevant terms in their variations are given by
\begin{equation}\begin{aligned}
e^{-1}\delta \cL_{\bar\psi\psi} &= \tfrac12 F^{\hat\alpha_2}_{MN} A^i_{0\,k}
\left(B_{\hat\alpha_2}\right)^k_j \bar\psi^P_i \left(-(D-3)
   {\Gamma^{MN}}_P + 2 \delta_P^{[M}\Gamma^{N]}\right) \epsilon^j
+\ldots \,, \\
e^{-1} \delta \cL^{(2)}_{F\bar\psi\psi} &= \tfrac12 F^{\hat\alpha_2}_{MN}  \left(B_{\hat\alpha_2}\right)^i_k A^k_{0\,j} \bar\psi^P_i \left( -(D-3) {\Gamma^{MN}}_P - 2 \delta_P^{[M}\Gamma^{N]}\right) \epsilon^j +\ldots\,, \\
e^{-1} \delta \cL_{\bar\chi\psi} &= \tfrac12 F^\al{2}_{MN} \bigl(A^\dagger_1\bigr)^i_a \left(C_\al{2}\right)^a_j  \bar\psi_j^P \left(- {\Gamma^{MN}}_P - 2 \delta_P^{[M}\Gamma^{N]}\right) \epsilon^j + \dots \,, \\
e^{-1} \delta \cL^{(2)}_{F\bar\chi\psi} &= \tfrac12 F^\al{2}_{MN}  \bigl(C^\dagger_\al{1} \bigr)^i_a A^a_{1\,j} \bar\psi_j^P \left({\Gamma^{MN}}_P - 2 \delta_P^{[M}\Gamma^{N]}\right) \epsilon^j + \dots \,.
\end{aligned}\end{equation}
The terms cubic in the \(\Gamma\)-matrices have to cancel \eqref{eq:psikinvariation} so we determine that the moment maps \(\cQ^R_\al{2}\) are given by
\begin{equation}\label{eq:QRA0A1}
\cQ^R_{\alpha_2} = (D-3) \bigl\{A_0, B_{\alpha_2}\bigr\} + \bigl(A^\dagger_1 C_{\alpha_2} - C^\dagger_{\alpha_2} A_1 \bigr) \,. \\
\end{equation}
Let us also derive a similar condition on the Killing vectors \(\cP_\al{2}\) \eqref{eq:gaugedP}.
Similar relations have first been obtained for \(D=4\) in \cite{DAuria:2001rlt}.
% \(= \cP^\al{1}_\al{2} e_\al{1}\), where \(e_\al{1}\) is the vielbein of the sigma-model metric, defined in \eqref{eq:}.
For this purpose we compute another term in the supersymmetry variation of the kinetic term of the scalar fields.
The gauged \(\hat\cP^\al{1}\) depend on the gauge fields \(A_M^\al{2}\) via \(\hat\cP^\al{1}_M = \cP_M^\al{1} + \cP^\al{1}_\al{2} A_M^\al{2}\),
therefore inserting the variation of \(A_M^\al{2}\) \eqref{eq:Avariation} into \eqref{eq:appbosoniclagrangian} gives
\begin{equation}\label{eq:scalarkinvar}
e^{-1} \delta \cL_{\hat \cP \hat \cP} = \delta_{\al{1}\be{1}}\hat \cP^\al{1}_M \cP^\be{1}_\al{2} \bigl(B^\al{2}\bigr)^i_j \bar\psi^M_i \epsilon^j + \dots \,.
\end{equation}
Similar to the above analysis we compute the relevant terms in the variations of the \(p=1\) Pauli terms \eqref{eq:apppaulib} and of the fermionic mass terms \eqref{eq:appfermionmass}.
The result reads
\begin{equation}\begin{aligned}\label{eq:appfermpaulip1var}
e^{-1} \delta \cL_{\bar\chi_\psi} &= \tfrac12 \hat \cP^\al{1}_M \bigl(A^\dagger_1\bigr)^i_a \left(C_\al{1}\right)^a_j \bar\psi^N_i \left({\Gamma^M}_N - \delta^M_N \right) \epsilon^j + \dots \,, \\
e^{-1} \delta \cL^{(1)}_{\hat \cP \bar\chi_\psi} &= \tfrac12 \hat \cP^\al{1}_M \bigl(C^\dagger_\al{1} \bigr)^i_a A^a_{1\,j}  \bar\psi^N_i \left({\Gamma^M}_N + \delta^M_N \right) \epsilon^j + \dots \,.
\end{aligned}\end{equation}
Comparing this with \eqref{eq:scalarkinvar} yields
\begin{equation}\label{eq:PA1}
\delta_{\al{1}\be{1}}\cP^\be{1}_\al{2} B^\al{2} = \tfrac12 \bigl(A^\dagger_1 C_{\alpha_1} - C^\dagger_{\alpha_1} A_1 \bigr) \,.
\end{equation}
We finally want to cancel also the terms quadratic in \(\Gamma\) in \eqref{eq:appfermpaulip1var}.
This gives rise to a gradient flow equation for \(A_0\) \cite{DAuria:2001rlt}.
Inserting \eqref{eq:appgravitinovariation} into the kinetic term of the gravitini \eqref{eq:appkinferm} gives
\begin{equation}\begin{aligned}
e^{-1} \delta \cL_{\bar\psi\hat\cD\psi} &= - (D-2)\bigl(\cD_M A_0\bigr)^i_j \bar\psi_{iN} \Gamma^{MN} \epsilon^j + \dots \\
&= - (D-2) \cP^\al{1}_M \bigl(\cD_\al{1} A_0\bigr)^i_j \bar\psi_{iN} \Gamma^{MN} \epsilon^j + \dots \,,
\end{aligned}\end{equation}
The comparison with 
%the terms quadratic in \(\Gamma\) in 
\eqref{eq:appfermpaulip1var} yields
\begin{equation}\label{eq:appgradientflow}
\cD_{\al{1}} A_0 = \tfrac{1}{2(D-2)} \bigl(A^\dagger_1 C_{\alpha_1} + C^\dagger_{\alpha_1} A_1 \bigr) \,.
\end{equation}
Note that it is possible to derive a similar relation expressing \(\cD_\al{1} A_1\) in terms of \(A_0\) and the third fermion mass matrix \(M^a_b\) \cite{DAuria:2001rlt}.

\subsection{Supersymmetry variations in various dimensions}\label{app:sugras}

In this appendix we give some explicit expressions for the general formulae collected above.
In particular we state the properties of \(A_0\) and give expressions for \(B_\hal{2}\).
We only consider the dimensions \(D=4,5,7\) which are the relevant cases for section~\ref{sec:examples}.

\subsection*{$D=4$}

In four dimensions we have, according to table~\ref{tab:spinors}, the choice between Majorana or Weyl spinors.
However, as explained in appendix~\ref{app:conventions} the R-symmetry is manifest only if we select the latter.
Accordingly, we choose the gravitini \(\psi^i_{M+}\) to be chiral, i.e. \(\Gamma_\ast \psi^i_{M+} = \psi^i_{M+}\).
Therefore, their charge conjugates \(\psi_{M-i} = (\psi^i_{M+})^C\) are antichiral.
The gravitini transform in the fundamental (or antifundamental representation, respectively) with respect to the R-symmetry group \(H_R\), given by
\begin{equation}\label{eq:appd4ralgebra}
H_R = \begin{cases}
\U(\cN) & \text{if}\; \cN \neq 8 \\
\SU(\cN) & \text{if}\; \cN = 8 
\end{cases} \,,
\end{equation}
where \(i = 1, \dots, \cN\).

To apply the results of the previous section we arrange \(\psi^i_{M+}\) and \(\psi_{M-i}\) in a combined column vector, and similarly for the supersymmetry parameters \(\epsilon^i_+\) and \(\epsilon_{-i}\),
\begin{equation}
\psi^i_M \rightarrow \begin{pmatrix} \psi^i_{M+} \\ \psi_{M- i} \end{pmatrix} \qquad\text{and}\qquad \epsilon^i \rightarrow \begin{pmatrix}\epsilon^i_+ \\ \epsilon_{-i} \end{pmatrix} \,,
\end{equation}
see also the discussion in appendix~\ref{app:conventions}.
Sticking to this notation, the gravitino shift matrix \(A_0\) in \eqref{eq:appgravitinovariation} reads
%Since the multiplication with one \(\Gamma\)-matrix inverts the chirality of spinors we infer from \eqref{eq:appfermionvariations} that the shift matrix \(A_0\) is of the general form
\begin{equation}
A_0 = \begin{pmatrix} 0 & (A_0)^{ij} \\ (A_0)_{ij} & 0 \end{pmatrix} \,,
\end{equation}
where \((A_0)^{ij} = \left((A_0)_{ij}\right)^\ast\).
This form is due to the fact that the multiplication with one \(\Gamma\)-matrix inverts the chirality of a spinor.
Moreover, the formula \eqref{eq:bilinearswap} applied to the gravitino mass term in \eqref{eq:appfermionmass} shows that \((A_0)_{ij}\) is symmetric.
In combination these properties imply that \(A_0\) is a hermitian matrix.

The dressed vector fields \(A^{\hal{2}}\) from the gravity multiplet  (i.e.~the graviphotons) are given by
\begin{equation}\label{eq:appD4graviphotons}
A^\hal{2}_M = %\begin{cases}
%\bigl(A^{[ij]}_M,A_{M[ij]}\bigr)  & \text{if}\; \cN \neq 6 \\
%\bigl(A^{[ij]}_M, A_{M[ij]}, A^0_M\bigr) & \text{if}\; \cN = 6 
%\end{cases} \,,
\bigl(A^{[ij]}_M,A_{M[ij]}\bigr) \,,
\end{equation} 
where \(A_{M[ij]} = \bigl(A^{[ij]}_M\bigr)^\ast\).
Only for the \(\cN = 6\) there is an additional gauge field in the gravity multiplet which is a singlet with respect to the global symmetry group of the theory and hence also with respect to \(H_R\).
We denote it by
\begin{equation}
A_M^\tal{2} = \bigl(A^0_M\bigr) \,,
\end{equation}
as if it would belong to an additional vector multiplet.
This is consistent since -- as we will see below -- the corresponding field strength \(F^0\) does not enter the supersymmetry variation of the gravitini.
%\footnote{\Snote{Add reference}}
This field content is constructed easiest by starting with the maximal \(\cN = 8\) theory \cite{Cremmer:1978ds,Cremmer:1979up,deWit:1982bul} and then decomposing the R-symmetry according to \(\SU(8) \rightarrow \U(\cN) \times \SU(8 - \cN)\) (see e.g.~\cite{Andrianopoli:2008ea}).
The spectrum of a theory with \(\cN\) supersymmetries is obtained by keeping only those fields which transform as singlets with respect to the second factor \(\SU(8 - \cN)\).
This also explains the appearance of the additional vector field \(A^0_M\) in the \(\cN = 6\) theory, where \(A^0_M = A^{[78]}_M\) is indeed invariant under \(\SU(2)\).
Note that \eqref{eq:appD4graviphotons} implies that there are no graviphotons for \(\cN = 1\).

In the same spirit one can determine the general form of the matrices \((B_\hal{2})\) in the supersymmetry variations of the gravitini \eqref{eq:appFM}.
For the \(\cN = 8\) theory they can be read off from \cite{deWit:1982bul} and are given by
\begin{equation}\label{eq:appD4B}
B_{ij} = \begin{pmatrix} 0 & (B_{ij})^{kl} \ \\ 0 & 0 \end{pmatrix} \qquad \text{and} \qquad
B^{ij} = \begin{pmatrix} 0 & 0 \ \\ -(B^{ij})_{kl} & 0 \end{pmatrix} \,,
\end{equation}
with 
\begin{equation}\label{eq:appD4Bb}
(B_{ij})^{kl} = \tfrac{1}{\sqrt{2}} \delta^{kl}_{ij} \,.
\end{equation}
Following the above argument, these expressions also hold for all other theories with \(\cN \neq 8\).
For \(\cN = 6\) there could in principle also be a matrix \(B_0\), but 
\begin{equation}\label{eq:appD4B0}
B_0 = 0 \,,
\end{equation}
since \((B_0)^{ij} = (B_{[78]})^{ij} = 0\) for \(i,j = 1, \dots 6\) and analogously for \((B_0)_{ij}\).
This justifies to treat \(A^0\) formally not as a graviphoton \(A^\hal{2}\).
The general structure of \eqref{eq:appD4B} and \eqref{eq:appD4Bb} is determined by the requirement that they transform invariantly with respect to \(H_R\).
Moreover, we can use the \(\cN = 2\) case to fix the numerical prefactor in \eqref{eq:appD4Bb}.
For \(\cN = 2\) theories there are no spin-1/2 fermions \(\chi^{\hat a}\) in the gravity multiplet and thus the matrices \(C_\hal{2}\) do not exist.
Therefore \eqref{eq:BCanticom} uniquely fixes the factor in \(B_\hal{2}\).

\subsection*{$D=5$}

In five dimensions we are using symplectic Majorana spinors, accordingly the R-symmetry group is given by
\begin{equation}
H_R = \USp(\cN) \,,
\end{equation}
where \(\cN\) denotes the number of gravitini \(\psi^i_M\), \(i = 1, \dots \cN\), satisfying the symplectic Majorana constraint \eqref{eq:symplmajorana}.
Every pair of symplectic Majorana spinors has 8 independent real components, hence the admissible values for \(\cN\) are \(2, 4, 6\) and \(8\).
In particular, we use the \(\USp(\cN)\)-invariant tensor \(\Omega_{ij} = (\Omega^{ij})^\ast\) to raise or lower indices.

Applying \eqref{eq:bilinearswap} and \eqref{eq:reality} to the gravitino mass term in \eqref{eq:appfermionmass} shows that the shift matrix \((A_0)_{ij}\) is symmetric and that
\begin{equation}\label{eq:appD5A09}
(A_0)_{ij} = (A_0)_{(ij)} = -(A_0^{ij})^\ast \,.
\end{equation}
In combination with the symmetry of \(A_0\), \eqref{eq:appD5A09} implies that \(A^i_{0\,j} = \Omega^{ik} (A_0)_{kj}\) is a hermitian matrix.
%where \(\Omega^{ij}\) is the \(\USp(\cN)\)-invariant tensor and was introduced below \eqref{eq:symplmajorana}.

For the graviphotons \(A^\hal{2}_M\) we follow a similar strategy as in four dimensions and start with the maximal theory with \(\cN = 8\), where \cite{Cremmer:1980gs,Gunaydin:1984qu,Pernici:1985ju,Gunaydin:1985cu}
\begin{equation}\label{eq:D5N8graviphotons}
A^\hal{2}_M = A^{[ij]}_M \,,\qquad A^{ij}_M \Omega_{ij} = 0 \,.
\end{equation}
To obtain the theories with \(\cN < 8\) we decompose \(\USp(8) \rightarrow \USp(\cN) \times \USp(8-\cN)\) and keep only those fields in \eqref{eq:D5N8graviphotons} which are singlets with respect to the second factor \(\USp(8-\cN)\).
This yields
\begin{equation}\label{eq:appD5graviphotons}
A^\hal{2}_M = \bigl(A^{[ij]}_M, A^0_M\bigr) \,,\qquad A^{ij}_M \Omega_{ij} = 0 \,,
\end{equation}
so for \(\cN \neq 8\) there is an additional vector field \(A^0_M\) in the gravity multiplet which is a singlet with respect to \(H_R\).
Note that for \(\cN = 2\) there is only \(A^0_M\). %since \(A^{[ij]}_M = 0\) due to the tracelessness condition.
Analogously we obtain \(B_0\) and \(B_{[ij]}\) for all \(\cN\) from starting with the expression for \(B_{[ij]}\) for the \(\cN = 8\) case.
The result reads
\begin{equation}\label{eq:appD5B}
\bigl(B_0\bigr)^k_l = \tfrac i2\sqrt{\tfrac{8-\cN}{2\cN}}\delta^k_l \,,\qquad \bigl(B_{ij}\bigr)^k_l = i \delta^k_{[i} \Omega_{j]l} + \tfrac i\cN \Omega_{ij} \delta^k_l \,.
\end{equation}
The general structure of these matrices is determined by \(\USp(\cN)\) invariance, and
as in four dimensions we can use \eqref{eq:BCanticom} to fix the numerical prefactor in \(B_0\) for \(\cN = 2\), which in turn determines the prefactors in \(B_0\) as well as in \(B_{[ij]}\) for all \(\cN\).

\subsection*{$D=7$}

In seven dimensions we are using symplectic Majorana spinors and the R-symmetry group is given by \(H_R = \USp(\cN)\), exactly as in five dimensions.
Here every pair of symplectic Majorana spinors carries 16 independent real components, so there is only the half-maximal theory with \(\cN = 2\) and the maximal theory with \(\cN = 4\).
The remaining discussion is very similar to the five-dimensional case, so let us only state the essential differences.

The gravitino shift matrix \(A_0\) satisfies
\begin{equation}\label{eq:appD7A0}
(A_0)_{ij} = (A_0)_{[ij]} = (A_0^{ij})^\ast \,.
\end{equation}
Both conditions in combinations imply that \(A^i_{0\,j}\) is hermitian.

The graviphotons \(A^\hal{2}\) as well as the matrices \(B_\hal{2}\) can be obtained from the maximal \(\cN = 4\) theory \cite{Sezgin:1982gi,Pernici:1984xx}.
The graviphotons are given by
\begin{equation}\label{eq:appD7graviphotons}
A^\hal{2} = A^{(ij)}_M \,,
\end{equation}
which is valid for all values of \(\cN\), since with respect to \(\USp(4) \rightarrow \USp(\cN) \times \USp(4-\cN)\) there cannot arise any additional \(\USp(4-\cN)\) singlets from the symmetric representation.
The matrices \(B_\hal{2}\) finally read
\begin{equation}\label{eq:appD7B}
(B_{ij})^k_l = \sqrt{2} \delta^k_{(i} \Omega_{j)l} \,.
\end{equation}
Note that locally
\begin{equation}
\USp(2) = \SU(2) \cong \SO(3) \,,\qquad \USp(4) \cong \SO(5) \,.
\end{equation}
Moreover, the graviphotons transform in the respective adjoint representations, and \eqref{eq:appD7B} is an explicit expression for the generators of \(\USp(\cN)\) in the fundamental representation.

\section{Properties of the gauged R-symmetry group \texorpdfstring{$H^g_R$}{Hg\_R}}\label{app:representationtheory}

In this appendix we discuss the implications of the formula \eqref{eq:AdSconditionsQP} on the gauged subalgebra \(\h^g_R\) of the R-symmetry algebra \(\h_R\).%
%It is self contained and can in principle be read independently from the rest of this thesis.%
\footnote{To keep the notation simple we deviate slightly from the notation used in the main part, e.g. we use \(\alpha\) instead \(\hal{2}\) and \(A\) instead of \(A_0\).}
Let \(\h_R\) a reductive Lie-algebra and let \(\{J_A\}\), \(A= 1,\dots,\dim(\h_R)\) be its generators.
Let \(\mathbf s\) and \(\mathbf v\) be two matrix representations of \(\h_R\), such that the generators in these representations read \((J_A)_i^j\) and \((J_A)_\alpha^\beta\), with \(i,j=1,\dots, \dim({\mathbf s})\) and \(\alpha,\beta = 1, \dots, \dim({\mathbf v})\).
We furthermore demand the existence of \(\dim({\mathbf v})\) linearly independent matrices \((B_\alpha)^j_i\) satisfying
\begin{equation}\label{eq:JB}
(J_A)_\alpha^\beta B_\beta = \bigl[J_A, B_\alpha\bigr] \,,
\end{equation}
where we suppressed the indices \(i\) and \(j\).
This condition implies that \(\mathbf v\) is contained in the tensor product decomposition of \(\mathbf s \otimes \mathbf s^\ast\), where \(\mathbf s^\ast\) denotes the dual representation of \(\mathbf s\).

Let us now assume that there is a matrix \((A)_i^j\) such that \(A^2 = \id\) and such that the matrices \((\cQ_\alpha)_i^j\), defined by
\begin{equation}\label{eq:Qdef}
\cQ_{\alpha} = \bigl\{A, B_\alpha\bigr\} \,,
\end{equation}
are elements of \(\h_R\).
It follows directly from the definition that
\begin{equation}\label{eq:QA}
\bigl[\cQ_\alpha, A\bigr] = 0 \,.
\end{equation}
Moreover, the condition \(\cQ_\alpha \in \h_R\) implies that there is a matrix \(\theta_\alpha^A\) -- usually called the embedding tensor, cf.~section~\ref{sec:gauging} -- such that \(\cQ_\alpha = \theta_\alpha^A J_A\).
This yields in combination with \eqref{eq:JB} and \eqref{eq:QA} that
\begin{equation}
\bigl[\cQ_\alpha, \cQ_\beta\bigr] = (\cQ_\alpha)_\beta^\gamma \cQ_\gamma \,,
\end{equation}
and therefore the \(\cQ_\alpha\) span a subalgebra \(\h^g_R \subseteq \h_R\).

Let \(\x\) be the maximal subalgebra of \(\h_R\) such that \([\x,A] = 0\) and let \(\cX_\a\), \(\a = 1, \dots \dim(\x)\), be the generators of \(\x\).
We now decompose the \(\h_R\)-representations \(\mathbf s\) and \(\mathbf v\) into irreducible representations of \(\x\), i.e.
\begin{equation}\label{eq:srdecomp}
{\mathbf s} = \bigoplus_{p = 1}^N {\mathbf s}_p \,,\qquad\text{and}\qquad {\mathbf v} = \bigoplus_{s = 1}^M {\mathbf v}_s \,.
\end{equation}
Analogously, we split the indices \(i\) into \((i_p)\) and \(\alpha\) into \((\alpha_s)\).
In this frame the generators \(\cX_\a\) become block-diagonal and
\begin{equation}
(\cX_\a)_{\alpha_s}^{\beta_s} \cQ_{\beta_s} = \bigl[\cX_\a, \cQ_{\alpha_s}\bigr] \,,
\end{equation}
for every \(s \in {1, \dots, M}\).
This implies that within each irreducible representation \({\mathbf v}_s\) either all the \(\cQ_{\alpha_s}\) vanish or are all non-vanishing and linearly independent.
Therefore \(\h^g_R\) must be a subalgebra of \(\x\) such that its adjoint representation is contained in the decomposition \eqref{eq:srdecomp}.
In other words, if the adjoint representation of the maximal subalgebra \(\mathfrak{z} \subseteq \x\) which satisfies this criterion is given by
\begin{equation}\label{eq:ydecomp}
\mathrm{ad}_\mathfrak{z} = \bigoplus_{s \in Z} {\mathbf v}_s \,,\qquad Z \subseteq \{1, \dots, M\} \,,
\end{equation}
we have
\begin{equation}
\mathrm{ad}_{\h^g_R} = \bigoplus_{s \in H} {\mathbf v}_s \,,\qquad \text{for some}\; H \subseteq Z \,.
\end{equation}

Under certain conditions it is possible to argue that an element \(s \in Z\) is also necessarily in \(H\).
Let \({\mathbf v}_{s}\) be one of the summands in \eqref{eq:ydecomp} (i.e.~\(s \in Z\)) such that 
\begin{equation}\label{eq:maxcriterion}
{\mathbf v}_{s} \notin {\mathbf s}_p \otimes \mathbf {\overline s}_q \,,\qquad\text{for}\qquad p \neq q \,,
\end{equation}
where \(\mathbf{ \overline s}_q\) denotes the dual representation
and therefore
\begin{equation}\label{eq:Bdecomp1}
(B_{\alpha_{s}})_{i_p}^{j_q} = 0 \,,\qquad\text{for}\qquad p \neq q \,.
\end{equation}
On the other hand we must have
\begin{equation}\label{eq:Bdecomp2}
(B_{\alpha_{s}})_{i_{p'}}^{j_{p'}} \neq 0 \,,
\end{equation}
for at least one \(p' \in \{1, \dots N\}\), since we demand all \(B_\alpha\) to be non-vanishing.
Moreover, the condition \([\cX_\a, A] = 0\) enforces (after a possible change of \(i\)-basis)
\begin{equation}\label{eq:Adecomp}
A_{i_p}^{j_q} = \begin{cases}
a_p\, \delta_{i_p}^{i_q} & \text{if}\; p = q \\
0 & \text{if}\; p \neq q
\end{cases} \,,
\end{equation}
where \((a_p)^2 = 1\) for all \(p\).
Inserting \eqref{eq:Bdecomp1}, \eqref{eq:Bdecomp2} and \eqref{eq:Adecomp} into \eqref{eq:Qdef} finally yields
\begin{equation}
\cQ_{\alpha_{s}} \neq 0 \,,
\end{equation}
and therefore \(s \in H\).
Note that \eqref{eq:maxcriterion} is a sufficient criterion for \(s \in H\) but not necessary.

\section{Variation of the vielbeins}\label{app:symP}

In this appendix we show that the variation matrix \({(\cP_{\delta\phi})_\al{2}}^\be{2}\) appearing in the variation \eqref{eq:cVvariation} of the vielbeins \(\cV^\al{2}_I\), i.e.
\begin{equation}\label{eq:appcVvariation}
\cD_{\delta\phi} \cV^\al{2}_I = \cV^\be{2}_I {(\cP_{\delta\phi})_\be{2}}^\al{2} ,
\end{equation}
always satisfies the property
\begin{equation}\label{eq:appsymP}
\bigl(\cP_{\delta\phi}\bigr)_{\hal{2}\tbe{2}} = \bigl(\cP_{\delta\phi}\bigr)_{\tbe{2}\hal{2}} \,,
\end{equation}
where \((\cP_{\delta\phi})_{\hal{2}\tbe{2}} = {(\cP_{\delta\phi})_\hal{2}}^\tga{2} \delta_{\tga{2}\tbe{2}}\) and \((\cP_{\delta\phi})_{\tbe{2}\hal{2}} = {(\cP_{\delta\phi})_\tbe{2}}^\hga{2} \delta_{\hga{2}\tal{2}}\).
To show \eqref{eq:appsymP} we perform a case-by-case analysis and discuss theories with different numbers \(q\) of real supercharges separately.

\subsection*{$q > 16$}

For these theories we do not have any vector multiplets and thus
\begin{equation}
{\bigl(\cP_{\delta\phi}\bigr)_\hal{2}}^\tbe{2} = {\bigl(\cP_{\delta\phi}\bigr)_\tbe{2}}^\hal{2} = 0 \,.
\end{equation}
Therefore \eqref{eq:appsymP} is satisfied trivially.

\subsection*{$q = 16$}

For half-maximal supergravities the duality group \(G\) is of the form
\begin{equation}
G = G^* \times \SO(10-D,n) \,,
\end{equation}
where \(n\) denotes the number of vector multiplets. In most cases the first factor \(G^\ast\) is given by \(\SO(1,1)\) while in \(D=4\) dimensions it is given by \(\SU(1,1)\), due to electric-magnetic duality.
Moreover, the gauge fields transform in the vector representation of \(\SO(10-D,n)\).

As explained in section~\ref{sec:adsmoduli} the variation \({(\cP_{\delta\phi})_\al{2}}^\be{2}\) corresponds to a non-compact generator of \(G\).
However, the group \(G^\ast\) does not mix fields from different multiplets, hence it can only give rise to \({(\cP_{\delta\phi})_\hal{2}}^\hbe{2}\) and \({(\cP_{\delta\phi})_\tal{2}}^\tbe{2}\).
This in turns means that the variations \({(\cP_{\delta\phi})_\hal{2}}^\tbe{2}\) and \({(\cP_{\delta\phi})_\tal{2}}^\hbe{2}\), in which we are interested, are elements of \(\so(10-D,n)\).
Therefore the split-signature metric
\begin{equation}
\eta_{\al{2}\be{2}} = \begin{pmatrix}
-\delta_{\hal{2}\hbe{2}} & 0 \\
0 & \delta_{\tal{2}\tbe{2}} \\
\end{pmatrix}
\end{equation}
is invariant with respect to \({(\cP_{\delta\phi})_\hal{2}}^\tbe{2}\) and \({(\cP_{\delta\phi})_\tal{2}}^\hbe{2}\), i.e.
\begin{equation}
- {(\cP_{\delta\phi})_\tbe{2}}^\hga{2} \delta_{\hga{2}\hal{2}} + {(\cP_{\delta\phi})_\hal{2}}^\tga{2} \delta_{\tga{2}\tbe{2}} = 0 \,,
\end{equation}
which shows \eqref{eq:appsymP}.

\subsection*{$q = 12$}

Such a theory exists only in $D=4$ dimensions (remember that we restrict our analysis to $D \geq 4$).
The duality group of the four-dimensional \(\cN = 3\) supergravity is given by
\begin{equation}
G = \SU(3,n) \,.
\end{equation}
Since \(\SU(3,n)\) is a subgroup of \(\SO(6, 2n)\) the above arguments also apply here.

\subsection*{$q = 8$}

These theories exist in dimensions \(D = 4,5\) and \(6\).
In six dimensions, however, the vector multiplets do not contain any scalar fields, moreover, the theory does not allow for supersymmetric AdS vacua.
Therefore, it is enough to consider only the cases \(D=4\) and \(D=5\).
We discuss them separately.

In four and five-dimensional \(\cN = 2\) supergravity the scalar field manifold \(\cM\) takes the form of a product
\begin{equation}
\cM = \cM_V \times \cM_H \,,
\end{equation}
where \(\cM_V\) is spanned by the scalar fields in vector multiplets and \(\cM_H\) is spanned by the scalars in hyper multiplets.
The gauge fields \(A^\al{2}_M\) are non-trivial sections only over \(\cM_V\), we can therefore restrict our attention to this space.

In five-dimensions \(\cM_V\) is a very special real manifold and can be described as a hypersurface of a \((n + 1)\)-dimensional real space with coordinates \(h^{I}\), \(I = 0, \dots, n\).%
\footnote{Our presentation follows \cite{Bergshoeff:2004kh}.} 
It is defined as the solution of the cubic polynomial equation
\begin{equation}
C_{IJK} h^{I} h^{J} h^{K} = 1 \,,
\end{equation}
where \(C_{IJK}\) is symmetric and constant.
This construction yields a metric \(M_{IJ}\) on the ambient space,
\begin{equation}
M_{IJ} = -2 C_{IJK} h^{K} + 3 h_{I} h_{J} \,,
\end{equation}
where \(h_{I} = C_{IJK} h^{J} h^{K}\).
This metric appears also as gauge kinetic metric \(M^{(2)}_{IJ}\) in \eqref{eq:bosonicaction}.
Moreover it induces a metric \(g_{rs}\) on \(\cM_V\) via
\begin{equation}
g_{rs} = h^{I}_r h^{I}_s M_{IJ} \,,
\end{equation}
where \(h^{I}_r\) is defined as the derivatives of \(h^{I}\), i.e.
\begin{equation}\label{eq:hIr}
h^{I}_r = - \sqrt{\tfrac32} \partial_r h^{I} \,.
\end{equation}
The covariant derivatives of \(h^{I}_r\) in turn satisfy
\begin{equation}\label{eq:hIrderiv}
\nabla_r h^{I}_s = - \sqrt{\tfrac32} \left(g_{rs} h^I + T_{rst} h^{I\, t}\right) \,,
\end{equation}
with \(T_{rst} = C_{IJK} h^{I}_r h^{J}_s h^{K}_t\).
We also need the relation
\begin{equation}
M_{I J} = h_{I} h_{J} + g_{rs}  h_{I}^r h_{J}^s \,,
\end{equation}
from which it follows that we can identify the vielbeins \(\cV^\al{2}_I\) introduced in \eqref{eq:kinmatrix} with \(h_{I}\) and \(h^{I}_r\), i.e.
\begin{equation}
\cV^{\hal{2} = 0}_{I} = h_{I} \,,\qquad \cV^{\tal{2} = \al{1}}_{I} = e^{\al{1}}_r h^r_{I} \,,
\end{equation}
where \(e^{\al{1}}_r\) are the vielbeins of the metric \(g_{rs}\) \eqref{eq:scalarvielbeins}.
Notice, that  we can identify the indices \(\tal{2}\) and \(\al{1}\) since there is precisely one scalar field per vector multiplet.
Finally, comparing \eqref{eq:appcVvariation} with \eqref{eq:hIr} and \eqref{eq:hIrderiv} yields
\begin{equation}
{\bigl(\cP_{\delta\phi}\bigr)_{\hal{2} = 0}}^{\tal{2} = \al{1}} = - \sqrt{\tfrac23} \delta\phi^\al{1} \,,
\end{equation}
as well as 
\begin{equation}\label{eq:D5Pdeltaphi}
{\bigl(\cP_{\delta\phi}\bigr)_{\tal{2} = \al{1}}}^{\hal{2} = 0} = - \sqrt{\tfrac23} \delta_{\al{1}\be{1}} \delta\phi^\be{1} \,.
\end{equation}
From this \eqref{eq:appsymP} follows directly.

In four dimensions \(\cM_V\) is a special K\"ahler manifold of complex dimension \(n\).
It is spanned by the complex scalars \((\phi^r, \bar\phi^{\bar r})\) and we denote its K\"ahler metric by \(g_{r\bar s}\).
A special K\"ahler manifold is characterized by the existence of a symplectic vector bundle over \(\cM_V\) and an holomorphic section \(\Omega\) on this vector bundle,%
\footnote{We follow the presentation and conventions from \cite{Andrianopoli:1996cm}.}
\begin{equation}
\Omega = \begin{pmatrix}
X^{I} \\
F_{I}
\end{pmatrix} \,,
\end{equation}
such that the K\"ahler potential \(\cK\) can be expressed as
\begin{equation}
\cK = - \ln\Bigl[i \left(\bar X^{I} F_{I} - \bar F_{I}  X^{I}\right)\Bigr] \,.
\end{equation}
Moreover one introduces 
\begin{equation}
V = \begin{pmatrix}
L^{I} \\
M_{I}
\end{pmatrix} = e^{\cK/2} \Omega = e^{\cK/2} \begin{pmatrix}
X^{I} \\
F_{I}
\end{pmatrix} \,,
\end{equation}
which satisfies
\begin{equation}\label{eq:appDbarrV}
\cD_{\bar r} V \equiv \Bigl(\partial_{\bar r} - \tfrac12 \partial_{\bar r} \cK \Bigr) V = 0 \,.
\end{equation}
The holomorphic covariant derivatives of \(V\), on the other hand, are not vanishing and one can define
\begin{equation}\label{eq:appUr}
U_r = \cD_r V = \begin{pmatrix}
f^{I}_r \\
h_{I\,r} \,.
\end{pmatrix}
\end{equation}
These objects in turn satisfy
\begin{equation}\label{eq:appDU}
\cD_r U_s = i C_{rst} g^{t \bar u} \bar U_{\bar u} \,,\qquad \cD_{\bar r} U_s = g_{\bar r s} V \,,
\end{equation}
where the precise properties of the completely symmetric tensor \(C_{rst}\) are not relevant for our further discussion.
Moreover, we need to introduce a complex, symmetric matrix \(\cN_{I J}\) which is defined by
\begin{equation}
M_{I} = \cN_{I J} L^{J} \,,\quad h_{I\,r} = \bar \cN_{I J} f^{I}_r \,.
\end{equation}
This matrix is related to the gauge kinetic matrix \(M^{(2)}_{I J}\) \eqref{eq:bosonicaction} via
\begin{equation}
M^{(2)}_{I J} = - \mathrm{Im}\, \cN_{I J} \,.
\end{equation}
The inverse of \(\mathrm{Im}\, \cN_{IJ}\) satisfies
\begin{equation}
- \frac12 \left(\mathrm{Im}\, \cN\right)^{IJ} = \bar L^{I} L^{J} + g^{r\bar s} f^{I}_r \bar f^{J}_{\bar s} \,,
\end{equation}
so we find for the (complex) inverse vielbeins \(\cV^I_\al{2}\),
\begin{equation}
\cV^{I}_{\hal{2} = 0} = \sqrt{2} \bar L^{I} \,,\qquad \cV^{I}_{\hal{2} = \bar 0} = \sqrt{2} L^{I}
\end{equation}
and
\begin{equation}
\cV^{I}_{\tal{2} = \al{1}} = \sqrt{2} e^r_{\al{1}} f^{I}_r \,,\qquad \cV^{I}_{\tal{2} = \bar \alpha_1} = \sqrt{2} \bar e^{\bar r}_{\bar\alpha_1} \bar f^{I}_{\bar r} \,,
\end{equation}
where \(e^r_{\al{1}}\) is a complex vielbein of the inverse metric \(g^{r\bar s}\), i.e.~\(g^{r\bar s} = \delta^{\al{1}\bar\beta_1} e^r_\al{1} \bar e^{\bar s}_{\bar \beta_1}\).
Thus we determine be comparing \eqref{eq:cVvariation} with \eqref{eq:appDbarrV} - \eqref{eq:appDU} that
\begin{equation}
{\bigl(\cP_{\delta\phi}\bigr)_{\hal{2} = \bar 0}}^{\tal{2} = \al{1}} = \delta\phi^\al{1} \,,\qquad {\bigl(\cP_{\delta\phi}\bigr)_{\hal{2} = \bar 0}}^{\tal{2} = \bar\alpha_1} = 0 \,,
\end{equation}
and
\begin{equation}\label{eq:D4Pdeltaphi}
{\bigl(\cP_{\delta\phi}\bigr)_{\tal{2} = \al{1}}}^{\hal{2} = 0} = 0 \,,\qquad {\bigl(\cP_{\delta\phi}\bigr)_{\tal{2} = \bar\alpha_1}}^{\hal{2} = 0} = \delta_{\bar\alpha_1 \be{1}} \delta\phi^{\be{1}} \,,
\end{equation}
as well as the respective relations for the complex conjugates.
This shows \eqref{eq:appsymP}.

\section{Symmetric Moduli Spaces}\label{app:Tmoduli}

In this appendix we show that the solutions of \eqref{eq:Tmoduli} span a symmetric
%homogeneous
space, even after dividing out possible Goldstone directions.
If the scalar field space is a symmetric space \(\cM = G/H\), the candidates for moduli (denoted by \(\k_{AdS}\) \eqref{eq:fsplit}) of a maximally supersymmetric AdS solution are characterized by the conditions \eqref{eq:cosetmoduli}.
In many examples all elements of \(\k_{AdS}\) satisfy also the stronger condition \eqref{eq:Tmoduli} which in turn guarantees that they are indeed moduli.
However, a priori not every solution of \eqref{eq:cosetmoduli} is necessarily a solution of \eqref{eq:Tmoduli}, in particular the Goldstone bosons \(\k^g\) which all solve \eqref{eq:cosetmoduli} might not all be solutions of \eqref{eq:Tmoduli}.
In the following we show how to divide the space of solutions of \eqref{eq:Tmoduli} by the remaining Goldstone directions and argue that the result corresponds to a symmetric submanifold \(\cM_{AdS} \subseteq \cM\).

Let us denote the set of all solutions of \eqref{eq:Tmoduli} by \(\k^f \subseteq \k\),
\begin{equation}
\k^f = \left\{\cP \in \k : - \cP_\alpha^\beta \cT_\beta + \bigl[\cT_\alpha, \cP\bigr] = 0 \right\} \,,
\end{equation}
where \(\cT_\alpha \in \g^g\) are the generators of the gauge group \(G^g\).
Analogously we define
\begin{equation}
\h^f = \left\{\cQ \in \h : - \cQ_\alpha^\beta \cT_\beta + \bigl[\cT_\alpha, \cQ\bigr] = 0 \right\} \,,
\end{equation}
and
\begin{equation}
\g^f = \h^f \oplus \k^f \,,
\end{equation}
where the direct sum is understood only as a direct sum on the level of vector spaces.
It follows readily from their definitions that \(\h^f\) as well as \(\g^f\) are both closed with respect to the Lie bracket, i.e.~they are subalgebras of \(\h\) and \(\g\) respectively.
(Note that \(\k^f\) itself cannot be a Lie algebra (unless it is abelian) due to \([\k,\k] \subseteq \h\).)
%\begin{equation}
%\bigl[\cQ, \cQ'\bigl]_\alpha^\beta \cT_\beta = \bigl( \cQ_\alpha^\gamma \cQ'{}_\gamma^\beta - \cQ'{}_\alpha^\gamma \cQ_\gamma^\beta\bigr) \cT_\beta = \Bigl[\bigl[\cT_\alpha, \cQ\bigr], \cQ' \Bigr] - \Bigl[\bigl[\cT_\alpha, \cQ'\bigr], \cQ \Bigr] = \Bigl[\cT_\alpha, \bigl[\cQ, \cQ'\bigr] \Bigr]
%\end{equation}

As in \eqref{eq:kg} we define
\begin{equation}
\k^g = \mathrm{span}(\cP_\alpha) \,,\quad \h^g = \mathrm{span}(\cQ_\alpha) \,,
\end{equation}
so \(\k^g\) and \(\h^g\) are the projections of the gauge algebra \(\g^g\) onto \(\k\) and \(\h\).
Note that in general \(\h^g \oplus \k^g\) can be larger than \(\g^g\).
Moreover, as noted in the discussion below \eqref{eq:kg}, \(\k^g\) corresponds to possible Goldstone bosons, so every element in \(\k^g\) which is at the same time also an element of \(\k^f\) must not be counted as a physical modulus and therefore has to be divided out.
Remember that we argued in section~\ref{sec:adsmoduli} that every element of \(\k^g\) is a solution of \eqref{eq:cosetmoduli}.
However, the condition \eqref{eq:Tmoduli} is stronger than \eqref{eq:cosetmoduli} and therefore it is possible that not every element of \(\k^g\) is contained in \(\k^f\).
For this reason we furthermore define
\begin{equation}
\k^{fg} = \k^f \cap \k^g \,,\qquad \h^{fg} = \h^f \cap \h^g \,,
\end{equation}
as well as
\begin{equation}
\g^{fg} = \h^{fg} \oplus \k^{fg} \,,
\end{equation}
i.e.~\(\k^{fg}\) corresponds to those Goldstone bosons which are also solutions of \eqref{eq:Tmoduli}.
In the next step we want to show that \(\g^{fg}\) is an ideal of \(\g^f\) and thus can be safely divided out.

Let \(\cP \in \k^f\) and \(\cP' \in \k^{fg}\).
This implies that there is a \(\cQ' \in \h^g\) such that
\begin{equation}
\cT' = \cQ' + \cP' \in \g^g \,.
\end{equation}
It follows from the definition of \(\k^f\) that 
\begin{equation}
\cT'' = \bigl[\cP, \cT'\bigr] \in \g^g \,.
\end{equation}
We split \(\cT''\) according to
\begin{equation}
\cT'' = \cQ'' + \cP'' \,,\quad\mathrm{s.t.}\quad \cQ'' \in \h \,,\; \cP'' \in \k \,.
\end{equation}
Therefore
\begin{equation}
\cQ'' = \bigl[\cP, \cP' \bigr] \in \h^g \,.
\end{equation}
Moreover, \(\cP\) and \(\cP'\) are both elements of \(\g^f\) and thus \(\cQ'' \in \h^{fg}\).
This shows that
\begin{equation}
\bigl[\k^f, \k^{fg}\bigr] \subseteq \h^{fg} \,. 
\end{equation}
Analogously one can show that \(\bigl[\h^f, \k^{fg}\bigr] \subseteq \k^{fg}\), \(\bigl[\k^f, \h^{fg}\bigr] \subseteq \k^{fg}\) and \(\bigl[\h^f, \h^{fg}\bigr] \subseteq \h^{fg}\).
Therefore \(\g^{fg}\) is an ideal of \(\g^f\) and \(\h^{fg}\) is an ideal of \(\h^f\), so we can define
\begin{equation}
\g_{AdS} = \g^f / \g^{fg} \qquad\mathrm{and}\qquad \h_{AdS} = \h^f / \h^{fg} \,.
\end{equation}
If we denote the Lie groups generated by \(\g_{AdS}\) and \(\h_{AdS}\) by \(G_{AdS} \subseteq G\) and \(H_{AdS} \subseteq H\) we find that
\begin{equation}
\cM_{AdS} = \frac{G_{AdS}}{H_{AdS}} 
\end{equation}
is a symmetric space.

%%%%%%%%%%%%%%%%%%%%%%%%%%%%%%%%%%%%%%%%%%%%%%%%%%%%%%%%%%%
%\newpage

\providecommand{\href}[2]{#2}\begingroup\raggedright\endgroup

\end{document}